%% file: main.tex
\documentclass[pdflatex,sn-mathphys-num]{sn-jnl}
\newcommand{\comment}[1]{}
\usepackage{graphicx}%
\usepackage{multirow}%
\usepackage{amsmath,amssymb,amsfonts}%
\usepackage{amsthm}%
\usepackage{bm, bbm}
\usepackage{mathrsfs, mathtools}%
\usepackage[title]{appendix}%
\usepackage{xcolor}%
\usepackage{textcomp}%
\usepackage{manyfoot}%
\usepackage{booktabs}%
\usepackage{algorithm}%
\usepackage{algorithmicx}%
\usepackage{algpseudocode}%
\usepackage{listings}%
\usepackage{siunitx}%
\usepackage{braket}
\usepackage{subcaption}
\usepackage{epstopdf}

\AppendGraphicsExtensions{.tga}

\usepackage{xr}
\begin{document}

\title[Quantum Computing for aGPCR autoproteolysis]{Multiscale Quantum-Classical Simulation of Protein Autoproteolysis in a Million-Atom Membrane Environment}

\input{authors.tex}

\abstract{
Predicting chemical reactions in biomolecules is challenging because reactive centers are embedded within complex environments that strongly influence their energetics and mechanisms. Here, we present a hybrid quantum-classical workflow that combines molecular dynamics, density functional theory, and correlated wavefunction methods, tightly integrating CPUs, GPUs, and quantum processing units (QPUs) to enable simulations across multiple length and accuracy scales. We apply this approach to ADGRG1 autoproteolysis, a maturation reaction in an adhesion G protein-coupled receptor that has not previously been studied using correlated electronic-structure methods. Our simulations treat a biologically realistic system containing more than one million atoms, including the full-length uncleaved receptor embedded in an endoplasmic-reticulum membrane, while incorporating a 54-qubit IQM quantum processor orchestrated via CUDA-Q and up to 1,200 NVIDIA H100 GPUs. These results demonstrate how quantum computers can serve as accelerators within scalable scientific workflows, extending correlated electronic-structure methods to biologically relevant systems.
}


\maketitle

\section{Introduction}
\label{sec:intro}


G protein-coupled receptors (GPCRs) are among the most pharmaceutically important protein families, targeted by approximately 36\% of approved drugs, with 516 approved agents and 337 candidates currently in clinical trials \cite{sanchezlorente2025}. Adhesion GPCRs (aGPCRs) constitute the second largest family within the GPCR superfamily \cite{hamann2015}, yet no aGPCR has been targeted for drug development, despite an association between several aGPCRs and human disease \cite{sanchezlorente2025}. Distinguished by their exceptionally large extracellular regions, most aGPCRs undergo a process called cis-autoproteolysis, in which the protein cuts its own backbone at the GPCR proteolysis site (GPS) located within the GAIN (GPCR autoproteolysis-inducing domain), without the help of an external enzyme \cite{arac2012, hamann2015}. A precise understanding of the molecular mechanism of this cleavage may provide the basis for targeting this receptor family therapeutically.

ADGRG1 is an aGPCR that undergoes autoproteolysis and has been implicated in several human diseases. Autoproteolysis is a self-cleavage process in which a protein catalyzes the breaking of its own peptide bonds. The autoproteolytically cleaved N-terminal fragment (NTF) of ADGRG1 has been detected in human blood plasma, and ADGRG1 knockout mice undergo prolonged bleeding, defective platelet plug formation, and delayed thrombotic occlusion, suggesting an important role for ADGRG1 in human platelet physiology \cite{yeung2020}. Mutations in ADGRG1 have been found to cause bilateral frontoparietal polymicrogyria (BFPP), a rare genetic disorder that causes cortical malformation \cite{bahibuisson2010} and, recently, high ADGRG1 levels in breast cancer patients have been shown to be significantly associated with an aggressive pathology in triple-negative breast cancer. In breast adipose tissue, ADGRG1 acts as a signaling hub to create an immunosuppressive and pro-metastatic niche in the tumor microenvironment (TME), making ADGRG1 a compelling therapeutic target with which to reprogram the TME and overcome resistance to immunotherapy in breast cancer \cite{zhou2026}.

Understanding ADGRG1 autoproteolysis requires modeling a chemical reaction occurring within a complex biomolecular environment. The quantities that govern the mechanism are the energy barriers separating reactants, intermediates, and products, as these determine reaction rates and reaction pathways. Precise prediction of such barriers demands an accurate description of electronic structure, yet a fully quantum-mechanical treatment of systems of this scale remains computationally intractable, even for large fault-tolerant quantum computers.

Multiscale electronic structure methods address this challenge by applying different levels of theory to different regions of a system. Approaches such as projection-based embedding (PBE) \cite{manby2012,goodpaster2014,ralli2024} and quantum mechanics/molecular mechanics (QM/MM) \cite{warshel1976,senn2009} enable the chemically active region to be treated at high accuracy while representing the surrounding environment at reduced cost. PBE has previously been combined with QM/MM to study enzymatic reactions through a correlated wavefunction (WF) description embedded within density functional theory (DFT), itself embedded in a classical environment \cite{bennie2016}. In this work, we adopt a related hierarchical WF-in-DFT-in-MM framework, in which the reactive center is treated explicitly at the wavefunction level, the local environment is described by DFT, and the wider biomolecular surroundings are represented through electrostatic embedding derived from a larger QM/MM model.

Although related approaches have been applied to aGPCR autoproteolysis, they have relied on substantially simplified structural models. Previous QM/MM studies of uncleaved rat ADGRL1 employed constructs containing the isolated N-terminal fragment and part of the tethered agonist solvated in water, without the transmembrane region \cite{seufert2025}. However, cleavage is believed to occur during biosynthesis in the endoplasmic reticulum, and studies of ADGRE2 suggest that the full seven-transmembrane architecture and membrane environment facilitate GAIN-domain autoproteolysis \cite{chung_self-cleavage_2025}. Furthermore, significant differences exist between human and rodent receptors \cite{salzman2016}, so mechanistic conclusions do not transfer directly between them.

Quantum computers are promising accelerators for the electronic structure components of such multiscale simulations. While a rigorous quantum advantage remains unproven, algorithms such as quantum phase estimation (QPE) \cite{aspuru2005simulated} are widely expected to exhibit more favorable scaling than classical approaches. Hybrid quantum-classical workflows are particularly attractive because they restrict quantum resources to the chemically most challenging regions of a system. Recent fragment-based studies have demonstrated calculations on increasingly large biomolecular systems. A subset of authors in this work developed a three-layer WF-in-DFT-in-MM framework incorporating quantum computing to demonstrate proton transfer in a solvated Zundel cation \cite{bickley2025}. Other work on multiscale QPU-HPC biochemistry has included the 303-atom Trp-cage miniprotein \cite{shajan2026} and protein-ligand complexes containing up to 12,635 atoms while needing quantum circuits of fewer than 100 qubits \cite{merz2026}.

For current and near-term hardware, quantum-selected configuration interaction (QSCI), also known as sample-based quantum diagonalization (SQD), represents a particularly promising approach. In QSCI, a quantum processor identifies important electronic configurations within the ground-state wavefunction, after which the Hamiltonian is diagonalized classically within the resulting reduced configuration space to recover electron correlation \cite{kanno2026,robledo2025}. This division of labor between quantum and classical resources exemplifies the emerging paradigm of quantum-centric supercomputing \cite{alexeev2024quantum,robledo2025}. Here, we introduce an iterative QSCI framework that uses measurements of stochastically time-evolved qubit populations to improve configuration recovery and incorporates a configuration-expansion stage inspired by previous selected-CI approaches \cite{weaving2026selected}. Unlike conventional non-iterative QSCI implementations, information from the classically diagonalized wavefunction is fed back into subsequent quantum sampling rounds. Specifically, we iteratively refine a biasing vector used in stochastic Hamiltonian sparsification during circuit compilation \cite{ouyang2020compilation,campbell2019}, preferentially sampling excitations expected to carry significant wavefunction weight outside the current variational space. In addition, the available circuit budget is distributed across multiple dominant determinants to promote broader exploration of configuration space.

Using this framework, we study the autoproteolysis mechanism of human ADGRG1 through a workflow combining structural modeling, molecular dynamics, reaction-path reconstruction, and quantum-computing-assisted reaction-barrier calculations. We construct a model of the full-length uncleaved receptor in an explicit membrane-solvent environment exceeding one million atoms and reconstruct a continuous cleavage pathway using QM/MM calculations. Energetics along this pathway are evaluated within a three-layer WF-in-DFT-in-MM embedding scheme, in which the reactive center is treated using iterative QSCI, embedded via projection-based embedding within a DFT description of the surrounding quantum region, itself embedded electrostatically within the larger biomolecular environment. Despite the substantially more realistic model employed here, our results are broadly consistent with earlier studies on the isolated GAIN domain of rat ADGRL1 \cite{seufert2025}. We recover reaction barriers of comparable magnitude and find evidence supporting a one-time autoproteolytic cleavage event, for which the receptor need not continuously maintain a reactive geometry. The observed reaction sequence is consistent with the previously proposed autoproteolysis pathway \cite{seufert2025}. While the reaction coordinate and QM-region selection in the present work were informed by the same prior study, the agreement across models spanning different receptor homologues and levels of structural complexity suggests that the key chemical features of the proposed mechanism are robust to these variations.

\section{Results}\label{sec:results}


\begin{figure}[p]
    \centering
    \includegraphics[width=\linewidth]{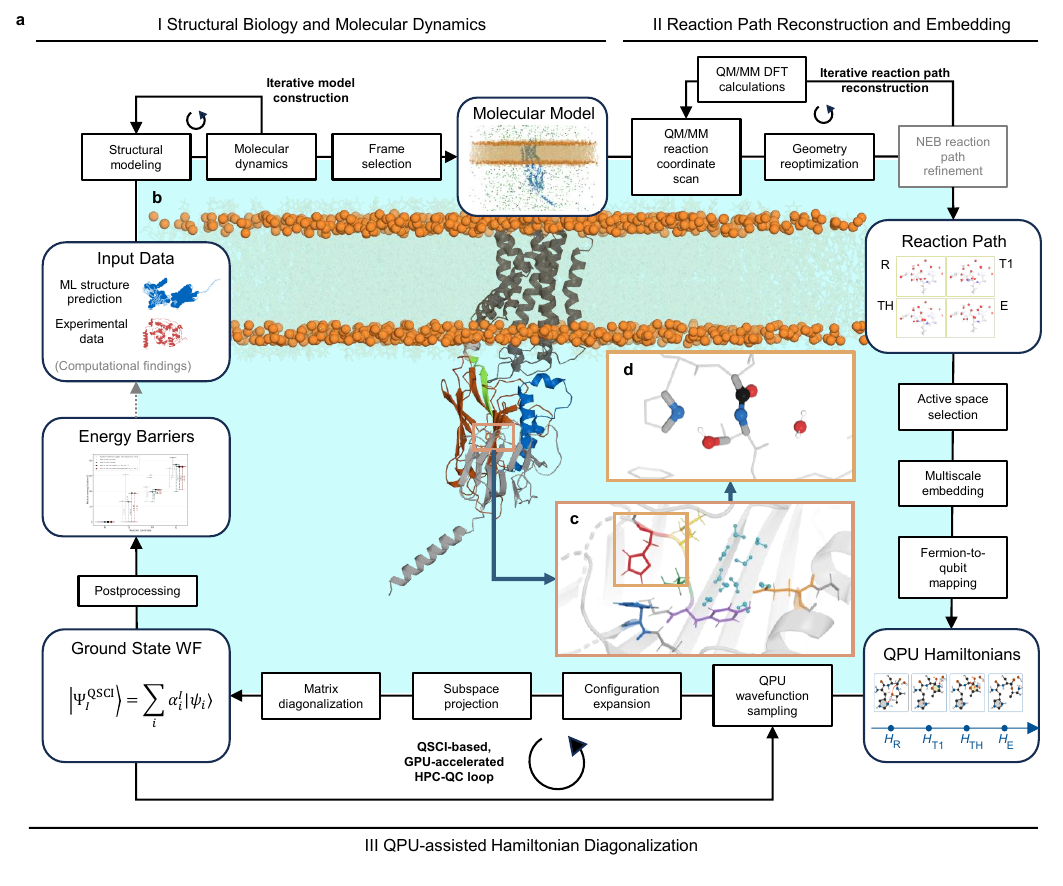}
    \caption{\textbf{WF-in-DFT-in-MD workflow for reaction-barrier calculations in a biomolecular environment, applied to aGPCR autoproteolysis.}
    \textbf{a} Three-stage workflow from structural modeling and molecular dynamics to reaction-path reconstruction and embedding, followed by QPU-assisted Hamiltonian diagonalization. In stage I, the model is built iteratively from computational and experimental structures, embedded in a realistic environment and assessed by MD. A configuration favorable for the reaction of interest is sampled from a final MD trajectory and provides the input to stage II. The reaction pathway is reconstructed at the QM/MM level through an iterative protocol of constrained reaction-coordinate scans. Representative structures are reoptimized. For correlated energy evaluation, the QM region is partitioned into an active space expected to contain most of the relevant electron correlation and a DFT environment polarized by MM point charges. The resulting embedded active-space Hamiltonians are mapped to qubit Hamiltonians and passed to stage III. Each Hamiltonian is treated with an iterative QSCI protocol in which QPU sampling is adapted to previous measurements. The resulting HPC--QC loop approximates the Hamiltonian eigenvalues, whose differences give the corresponding electronic reaction-energy barriers.
    \textbf{b} Full composite ADGRG1 model embedded in an endoplasmic-reticulum-like membrane and solvated, as passed to stage II. The receptor 7TM region spans the orange phospholipid bilayer. GAIN subdomains A and B are shown in blue and dark orange, respectively, and the Stachel $\beta$-strand in green. Water and ions are indicated by the blue background box.
    \textbf{c} Catalytic interface of the GAIN domain, where autoproteolysis occurs and which defines the 163-atom QM region of stages II and III. Residues taken in full are H381 (red), L382 (yellow), T383 (green), F348 (blue), E213 (orange) and Y384 (purple). The ten corridor waters are cyan, the flanking atoms retained for the QM/MM boundary grey and the surrounding protein a translucent cartoon.
    \textbf{d} Atoms on which the active-space selection is centered: H381 N$_{\delta 1}$, the L382 carbonyl C and O, the T383 amide N, O$_{\gamma 1}$ and H$_{\gamma 1}$, and the catalytic water, colored black, blue, red and white by element. The remaining QM region is gray. \label{fig:pipeline}}
\end{figure}

\subsection{Reconstruction of the Autoproteolysis Pathway}

Our WF-in-DFT-in-MD workflow enables the evaluation of reaction energetics along proposed reaction pathways while bridging the gap between realistic biomolecular systems and the size limitations of quantum computing and other correlated electronic structure methods. Figure \ref{fig:pipeline} illustrates both the workflow and its application to ADGRG1 autoproteolysis. The workflow (Figure \ref{fig:pipeline}a) is modular by design, allowing individual components to be reused and adapted without modifying the overall framework.

To investigate autoproteolysis, a composite model of the uncleaved ADGRG1 receptor was embedded within an explicit membrane environment and solvated (Figure \ref{fig:pipeline}b). Molecular dynamics (MD) simulations were then used to sample conformational fluctuations in the native-like environment. The resulting structures were passed to a QM/MM reaction-path reconstruction procedure, where the reactive region of the receptor (Figure \ref{fig:pipeline}c) was treated at the DFT level. Each geometry along the reconstructed pathway was subsequently converted into an embedded active-space Hamiltonian, in which a correlated wavefunction (WF) description of the reactive center (Figure \ref{fig:pipeline}d), leveraging quantum computational resources, is embedded within a DFT region and a molecular-mechanics environment.

The workflow is built around the exchange of well-defined data objects, including molecular geometries, embedded Hamiltonians, quantum circuits and quantum measurement data. Consequently, sampling approaches, embedding schemes, and quantum algorithms can be modified independently without restructuring the pipeline, while allowing each computational stage to run on the hardware best suited to the task. This modular design ensures that advances in either the classical or quantum components of the workflow, such as improved embedding methodologies, more efficient solvers, or new quantum algorithms, can be readily incorporated as they emerge.  Furthermore, the GPU-accelerated subspace diagonalization procedure employed in this work exhibits near-ideal strong-scaling behavior, maintaining high parallel efficiency when scaled to 1200 NVIDIA H100 GPUs. Additional implementation and performance details are provided in Supplementary Note \ref{note:postprocessing}.

\begin{figure}[t]
    \centering
    \includegraphics[width=\linewidth]{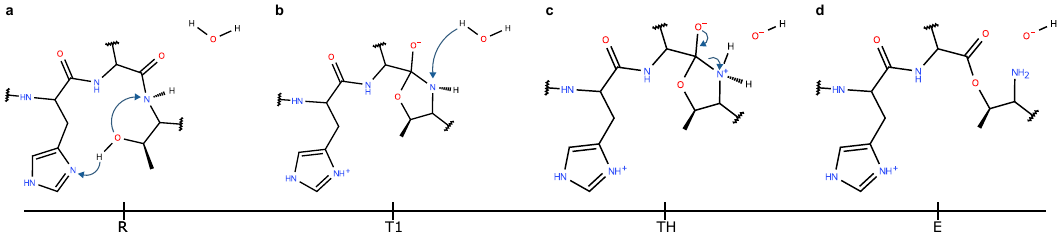}
    \caption{\textbf{The N--O acyl shift reaction pathway considered in this work}, based on the charged pathway of Seufert et al. \cite{seufert2025}, with blue arrows indicating atomic motion relative to the preceding structure. \textbf{a} Initial reactant structure R. \textbf{b} Hydroxy-oxazolidine intermediate T1, after concerted proton transfer and nucleophilic attack. \textbf{c} Protonated intermediate TH, after water-mediated amide protonation. \textbf{d} Ester intermediate E, after ring opening.}
    \label{fig:autoproteolysis_chemd}
\end{figure}

The workflow was applied to the N--O acyl shift mechanism proposed by Seufert \emph{et al.} \cite{seufert2025} for ADGRL1 and transferred to human ADGRG1. We identify H381, L382, and T383 as the residues forming the GPCR proteolysis site (GPS), with cleavage occurring at the L382--T383 peptide bond. The proposed mechanism proceeds through a sequence of reactant, intermediate, and product-like structures, denoted R, T1, TH, and E, respectively (Figure \ref{fig:autoproteolysis_chemd}). 

Our structural model combines an experimentally determined receptor structure with a computationally predicted model to generate a full-length, uncleaved ADGRG1 receptor embedded in a membrane-solvent environment representative of the endoplasmic reticulum. The receptor remained structurally stable throughout all MD simulations and repeatedly sampled conformations favorable for the proton-transfer event that initiates the autoproteolysis reaction. In all trajectories, periods characterized by short donor-acceptor distances were followed by transitions to longer-distance configurations. A representative frame with a proton-acceptor distance of \SI{1.81}{\angstrom} was selected as the starting structure for subsequent QM/MM reaction-path reconstruction.

Constrained QM/MM scans generated the R--T1--TH--E sequence using a two-dimensional scan for the coupled proton-transfer and nucleophilic-attack step, followed by two one-dimensional scans describing amide protonation and ring opening. The resulting geometries exhibit the expected local structural features of the N--O acyl-shift mechanism (Figure \ref{fig:autoproteolysis_chemd}). Consistent with their interpretation as stable reaction endpoints, the R and E structures remain stable upon unconstrained QM/MM reoptimization, whereas T1 and TH are only maintained as constrained intermediates along the reconstructed pathway.

Motivated by the water network observed near the cleavage site, we additionally explored the second stage of autoproteolysis by placing the hydroxide generated during the N--O shift onto the L382 acyl carbon and modeling the reaction as a base-mediated ester hydrolysis. Unconstrained QM/MM optimization of the manually constructed trial structure yielded a product-like geometry consistent with bond cleavage, although the optimization did not satisfy the force convergence criterion. In the resulting structure, the ester bond was elongated and effectively broken (C--O$^{\gamma1}$ distance of \SI{2.09}{\angstrom}). Additionally, the T383 HG1 proton, initially transferred to H381 during the N--O shift, was observed to transit back to T383 OG1. This behavior is consistent with H381 acting as a general acid during the cleavage process. We note that all reoptimizations reached the numerical lower limit of the steepest-descent step size before satisfying the target force criterion, leaving residual forces in the final structures.

\subsection{Impact of Correlation} \label{sec:barriers}

\begin{figure}[t]
    \centering
    \begin{subfigure}[t]{0.24\linewidth}
        \centering
        \includegraphics[width=\linewidth,trim={4.5cm 7cm 4.3cm 5cm},clip]{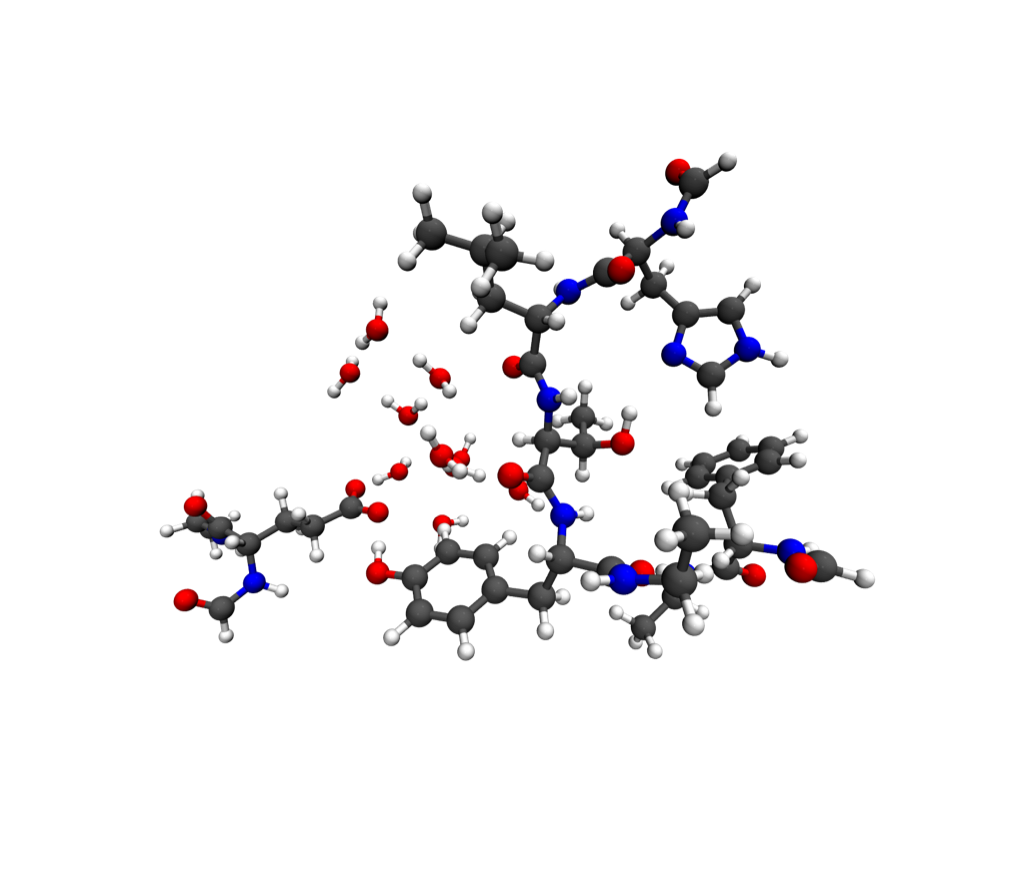}
        \caption{R structure}
        \label{fig:grid:a}
    \end{subfigure}
    \begin{subfigure}[t]{0.24\linewidth}
        \centering
        \includegraphics[width=\linewidth,trim={4.5cm 7cm 4.3cm 5cm},clip]{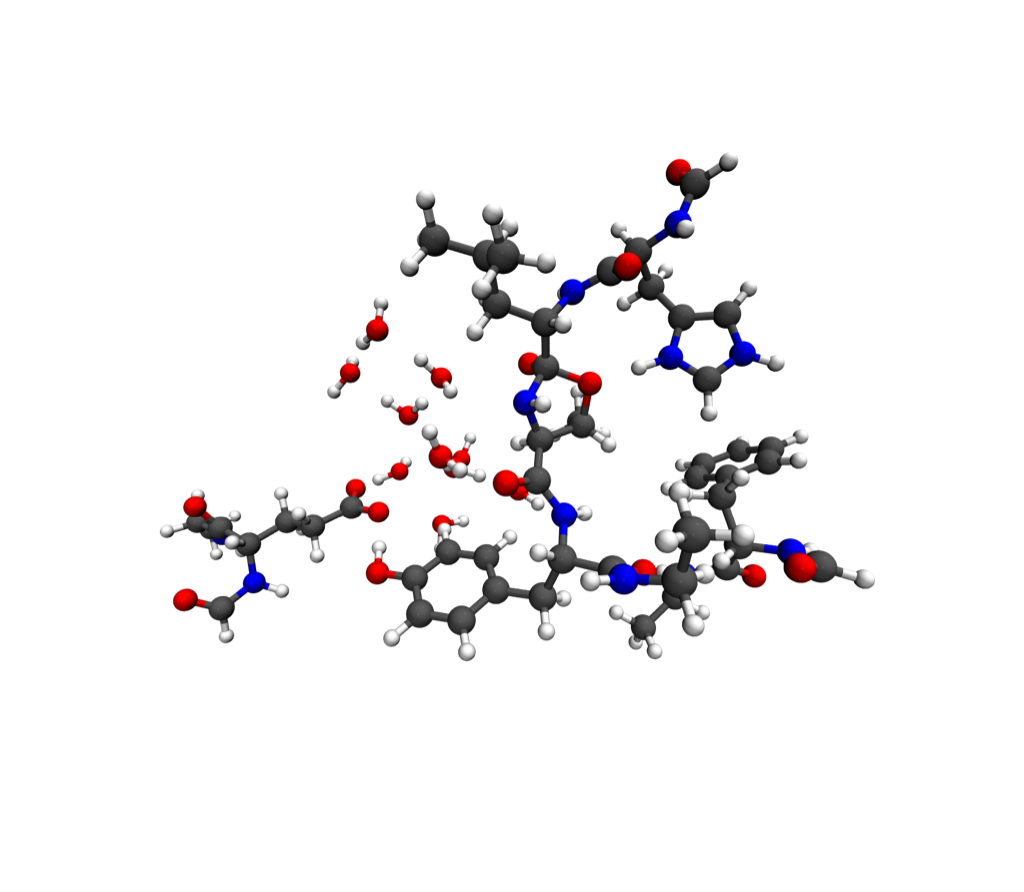}
        \caption{T1 structure}
        \label{fig:grid:b}
    \end{subfigure}
    \begin{subfigure}[t]{0.24\linewidth}
        \centering
        \includegraphics[width=\linewidth,trim={4.5cm 7cm 4.3cm 5cm},clip]{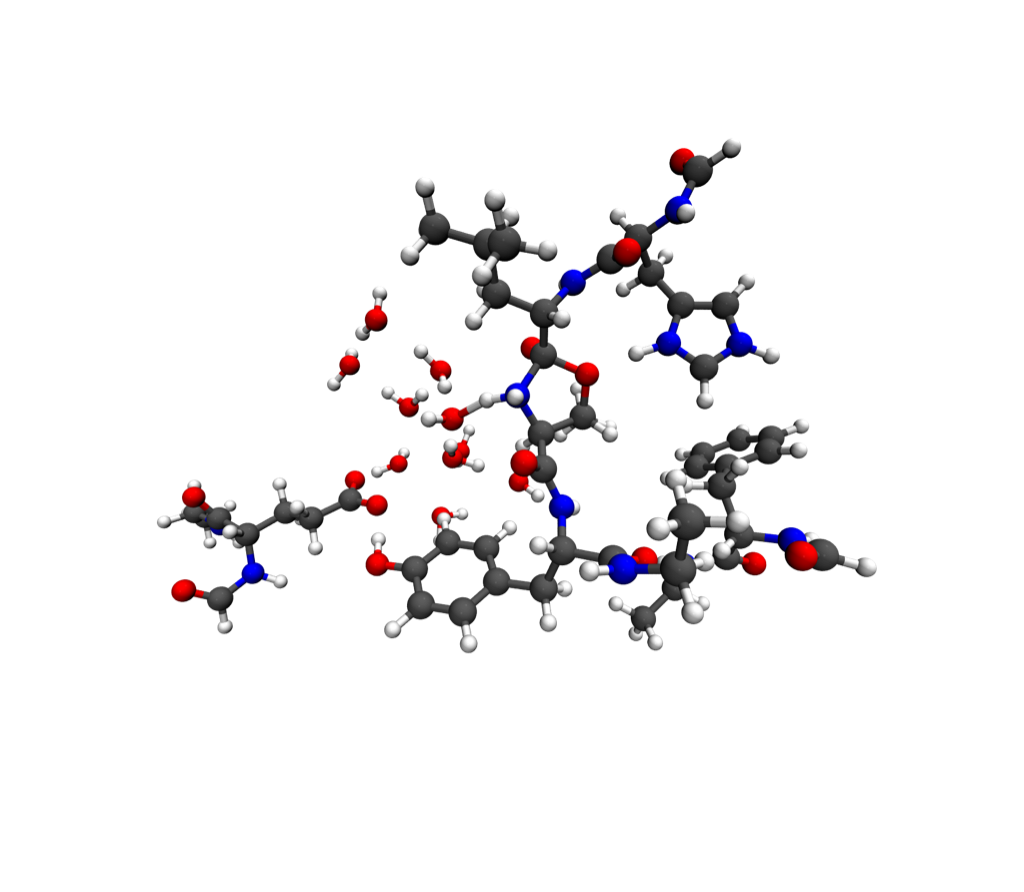}
        \caption{TH structure}
        \label{fig:grid:c}
    \end{subfigure}
    \begin{subfigure}[t]{0.24\linewidth}
        \centering
        \includegraphics[width=\linewidth,trim={4.5cm 7cm 4.3cm 5cm},clip]{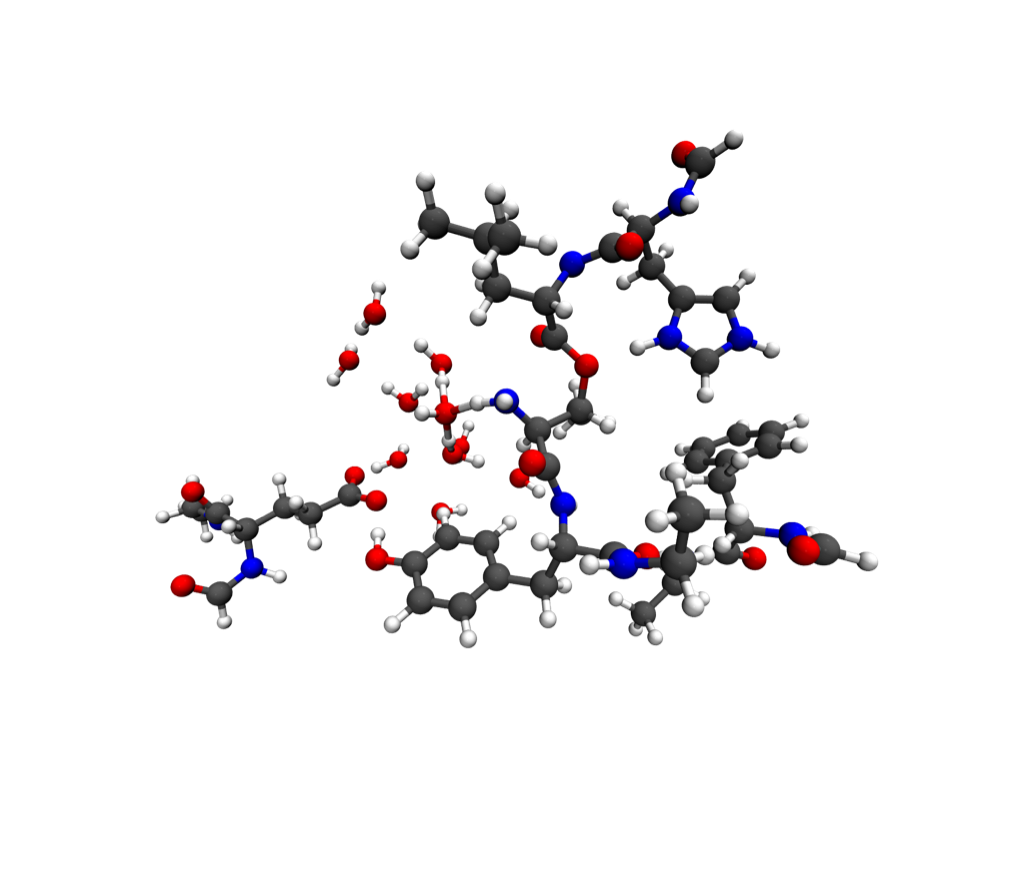}
        \caption{E structure}
        \label{fig:grid:d}
    \end{subfigure}
      
    \includegraphics[width=\textwidth]{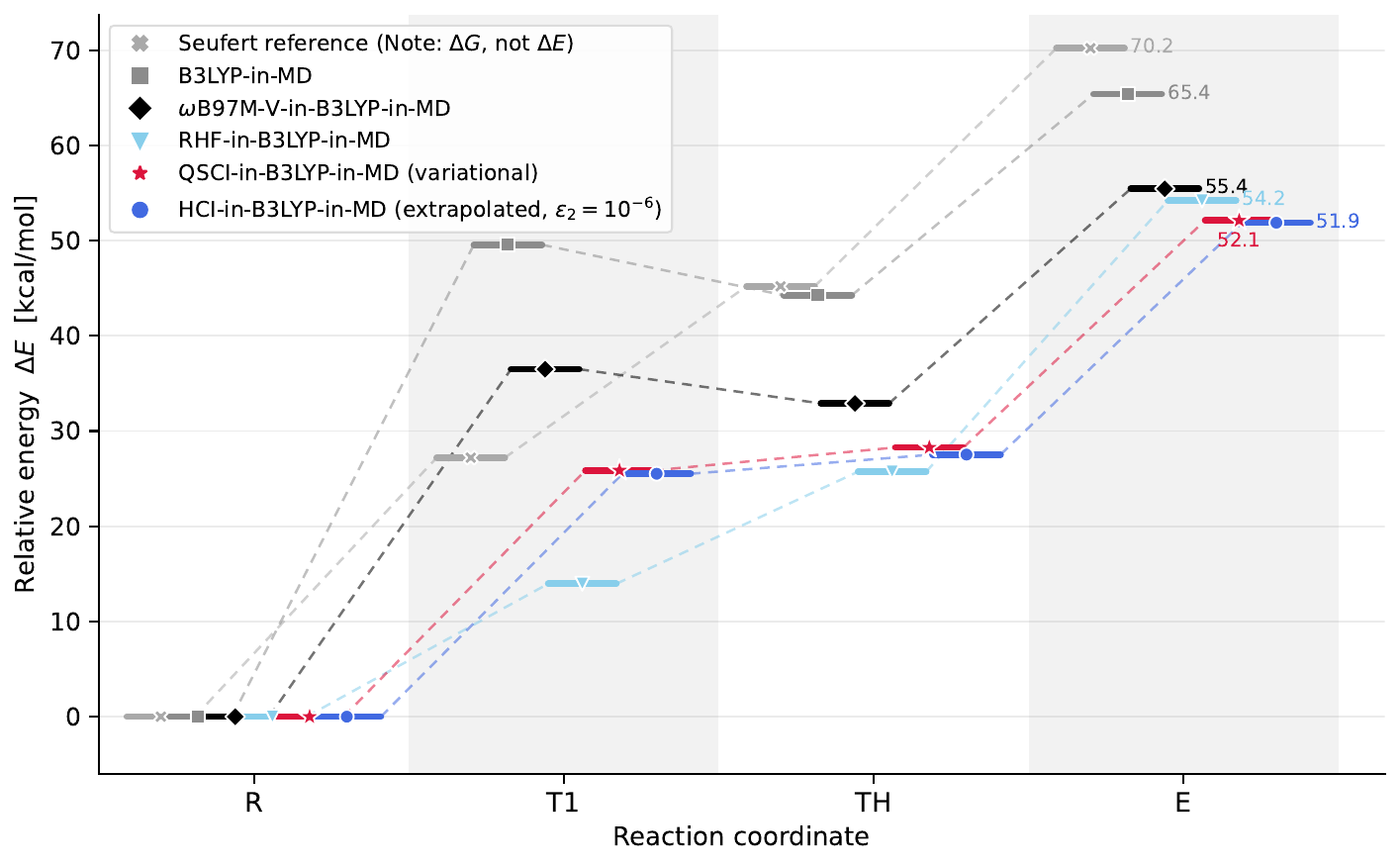}
    
    \caption{\textbf{WF-in-DFT-in-MD electronic barriers.} Full WF-in-DFT-in-MD profiles for the R, T1, TH and E structures shown in \textbf{(a)} -- \textbf{(d)}, comparing the relative energies predicted by DFT, HF, QSCI and HCI benchmarks (the variational energy shown is computed at $\varepsilon_1=5\times10^{-5}$, while the extrapolated results are computed with semistochastic perturbations at $\varepsilon_2=10^{-6}$). The Seufert et al.\ ADGRL1 reference is also shown for context, although we stress those values are Gibbs free energies ($\Delta G$), while we report electronic energies.}
    \label{fig:pooled_qsci_with_hci}
\end{figure}

The reaction profile presented in Figure \ref{fig:pooled_qsci_with_hci} highlights the impact of electron correlation on the predicted energetics of ADGRG1 autoproteolysis. While DFT remains the dominant electronic-structure approach in biomolecular QM/MM simulations owing to its favorable computational cost \cite{tzeliou2022,senn2009}, its treatment of electron correlation is approximate in practice. DFT is formally exact; however, the exact exchange-correlation functional is unknown and practical calculations therefore rely on approximate functionals. Consequently, reaction energetics can be sensitive to the choice of functional, motivating the use of correlated WF methods whenever computationally feasible.

We note that all WF-in-DFT-in-MD energies reported here are evaluated as single-point calculations on geometries obtained from the DFT-based QM/MM pathway reconstruction. The resulting energy profiles therefore reflect differences in the electronic structure treatment along a common reaction pathway rather than fully relaxed reaction profiles at each level of theory. Reoptimizing only the embedded WF region would neglect the structural response of the substantially larger protein, solvent, membrane, and molecular-mechanics environment from which the reaction pathway was derived. The present approach therefore preserves consistency with the underlying multiscale model while enabling a direct comparison of electronic structure methods. Incorporating geometry optimization within the embedding framework remains an important direction for future development of the modular workflow.

Comparison of the different electronic-structure treatments reveals both qualitative and quantitative differences. At the RHF-in-B3LYP-in-MD level, the relative ordering of structures is R, T1, TH, and E in increasing energy. This ordering is consistent with the pathway proposed by Seufert \textit{et al.} \cite{seufert2025}. In contrast, the $\omega$B97M-V-in-B3LYP-in-MD calculations predict a reversal in the energetic ordering of the T1 and TH intermediates and yield a substantially larger barrier for formation of T1 at \SI{36.5}{kcal.mol^{-1}}, versus \SI{14.0}{kcal.mol^{-1}} for RHF-in-B3LYP-in-MD and \SI{27.2}{kcal.mol^{-1}} from the Seufert reference. Such differences have dramatic kinetic implications. From the Eyring equation and approximating the activation free energy by the computed barrier height, we find that reaction rates depend exponentially on the barrier height. Consequently, even small changes in the predicted barrier can determine whether a reaction is feasible on biologically relevant timescales. Further details are provided in Supplementary Note \ref{note:barrier_notes}.


Correlated WF-in-DFT-in-MD treatment reduces the overall electronic barrier relative to the $\omega$B97M-V-in-B3LYP-in-MD result, from \SI{55.4}{kcal.mol^{-1}} to \SI{51.9}{kcal.mol^{-1}}, and recovers the same qualitative ordering of intermediates as the RHF calculations and the pathway reported by Seufert \textit{et al.} (caveated that they report $\Delta G$, not electronic energies). The largest deviations between the DFT and correlated treatments occur for the R$\rightarrow$T1 transition, indicating that electron correlation plays a particularly important role during the initial proton-transfer and nucleophilic-attack step. This sensitivity is not unexpected, since bond-breaking and bond-forming processes are typically challenging for DFT, where certain correlation effects are not properly accounted for. Taken together, these results demonstrate that the predicted energetics of autoproteolysis are highly sensitive to the treatment of electron correlation and suggest that correlated wavefunction approaches may be required for a quantitatively reliable description of this class of biomolecular reactions.


\subsection{Embedded QSCI on the QPU}
\label{sec:qsci_results}

\begin{figure}[t]
  \centering
  \includegraphics[width=\textwidth]{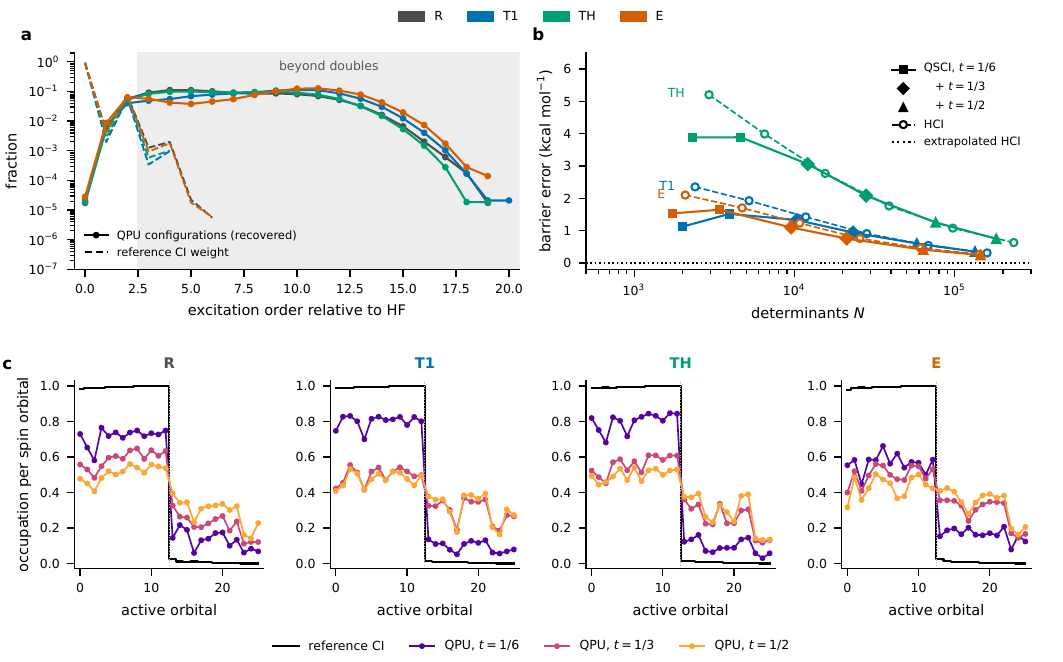}
  \caption{\textbf{Quantum-selected configuration interaction on the QPU.}
  \textbf{(a)}~Distribution of the recovered QPU configurations over excitation order relative to HF (solid) and weight of the reference wavefunction at each order (dashed).
  \textbf{(b)} Error of the barriers relative to R with respect to extrapolated HCI, for occupancy-biased iterative QSCI (filled symbols; squares $t=1/6$, diamonds $t=1/3$, triangles $t=1/2$) and for HCI (open circles). In the smallest spaces the QSCI barriers lie below those of HCI through error cancellation across structures: the QSCI energy of R lies further above HCI of the same size than those of T1, TH and E, which lowers every barrier measured from R. All energies are variational, excluding the extrapolated HCI against which errors are taken; this involved linear regression on the second-order perturbation corrections beyond variational HCI subspaces.
  \textbf{(c)} Measured qubit populations, averaged over spin, at the three evolution times, compared with the occupations of the reference ground-state wavefunction (black). The vertical line separates orbitals occupied and unoccupied in HF.}
  \label{fig:qsci}
\end{figure}

Each embedded CAS(26,26) Hamiltonian was treated with the iterative QSCI protocol of Section \ref{sec:iter-qsci} at evolution times $t=1/6$, $1/3$ and $1/2$. Each timestep was permitted $80$ instances of stochastic Hamiltonian sparsification circuit compilation, measured $4,096$ times per circuit for a total of $\approx10^{6}$ shots per structure. After configuration recovery, the measurements span the full excitation hierarchy, up to 20-fold excitations of the reference determinant as shown in Figure~\ref{fig:qsci}a.

The quality of the samples is influenced by circuit depth (Figure~\ref{fig:qsci}c). At $t=1/6$ the measured populations retain the Aufbau-like pattern of the ground state, and 6--34\,\% of shots lie in the target particle-number sector. At the two later times, with deeper circuits, population spreads and only 0.3--3.1\,\% of shots are in-sector. The raw measurements are biased towards electron deficiency: 90--98\,\% of out-of-sector shots carry too few electrons, with mean electron count 18.7--23.2 at $t=1/6$ and 19.0--20.6 at later times, instead of the physical 26. This is consistent with T1 relaxation and readout errors; configuration recovery therefore mostly restores missing electrons.

The QSCI space is grown through the occupancy-biased expansion of the iterative loop in Section~\ref{sec:iter-qsci}, refined on measured data with time steps entering in order of circuit depth: the $t=1/6$ samples at the two loosest selection thresholds $\epsilon_1$, followed by $t=1/3$ and then $t=1/2$ as the threshold tightens. At each threshold the recovered configurations of the current time step are added, and the space is expanded by a heatbath-like procedure that screens via a modified Hamiltonian that is reweighted by the measured qubit populations of that time step; this favours excitations through fractionally occupied orbitals and recovered configurations that the expansion does not select are retained if their CI coefficient exceeds $\epsilon_1$ (Supplementary Note~\ref{note:qsci_accept}). 

In the smallest spaces the shallow-circuit configurations allow QSCI to undercut HCI of the same size for T1 and TH, by 0.07 and 0.72~kcal\,mol$^{-1}$ at $2$--$2.3\times10^{3}$ determinants, while for R and E the reweighting costs 0.8~kcal\,mol$^{-1}$. Beyond \mbox{$\sim\!5\times10^{3}$} determinants QSCI lies slightly above HCI for all structures, by at most 0.18~kcal\,mol$^{-1}$ and by only 0.01--0.02~kcal\,mol$^{-1}$ at the tightest threshold, where the occupancy bias keeps the selection distinct from HCI while selected configurations from the device remain in the final spaces. Resampling the shots changes the energies in the smallest spaces by at most 0.02~kcal\,mol$^{-1}$. The configurations the device can identify are limited by the $2$--$11\times10^{4}$ in-sector shots at $t=1/6$, which resolve weights above $\sim\!10^{-5}$, whereas the determinants HCI adds beyond $\sim\!10^{4}$ determinants carry weights of $\sim\!10^{-8}$; in larger spaces the measured information therefore enters mainly through the occupancies. The barriers in Figure~\ref{fig:qsci}b converge with expansion size together with those of HCI. In the smallest spaces they lie closer to the extrapolated reference than those of HCI, but this reflects error cancellation across structures rather than more compact wavefunctions: the QSCI energy of R lies furthest above HCI there, which lowers every barrier measured from it. At the tightest threshold QSCI gives 25.9, 28.3 and 52.1~kcal\,mol$^{-1}$ for T1, TH and E relative to R, against 25.5, 27.5 and 51.9~kcal\,mol$^{-1}$ for extrapolated HCI and 25.8, 28.2 and 52.1~kcal\,mol$^{-1}$ for variational HCI at the same threshold.

The energy of these structures is carried predominantly by single and double excitations. The reference determinant holds 92--94\,\% of the wavefunction weight, and only 2--4\,\% of the correlation weight lies beyond doubles in the converged HCI wavefunction (Supplementary Note~\ref{note:qsci_highorder}). In Figure \ref{fig:qsci}a we saw the recovered determinants largely occupy high-order excitations, which are broadly inert for the weakly correlated systems in this study. The picture is sharpened further by truncated near-FCI wavefunctions, which bound how compact a selected CI expansion of a given size can be and show that HCI is close to optimal for these systems (Supplementary Note~\ref{note:qsci_accept}). While these results validate the use of quantum resources within this large multiscale simulation workflow, the true benefit will only be seen for strongly correlated systems, in which high-order excited configurations can carry substantial weight. This is where we anticipate quantum sampling can help most.

\section{Discussion}

The present study of the N--O acyl shift underlying aGPCR autoproteolysis represents, to our knowledge, the first post-mean-field investigation of ADGRG1 and the first to examine this reaction within a full-length receptor embedded in an explicit membrane-solvent environment containing over one million atoms. Despite the larger and thus more realistic biological model, our findings are broadly consistent with the QM/MM study of Seufert \textit{et al.}~\cite{seufert2025}, which investigated the proposed N--O acyl-shift mechanism in the isolated GAIN domain of rat ADGRL1. Although the two studies differ in receptor family member, species, and level of environmental complexity, both yield reaction energetics on a similar overall scale and characterize autoproteolysis as a rare event, lending support to the proposed mechanism across distinct aGPCR systems. We note, however, that this agreement emerges primarily at the correlated level, whereas larger deviations are observed for the DFT-in-MD baseline. These differences may reflect the distinct receptors considered, the increased environmental realism of the present model, or the sensitivity of the energetics to the underlying structural models and geometry optimizations.

The reaction pathway examined here is broadly single-reference and dominated by dynamic correlation; consequently, the classical HCI-in-DFT-in-MD benchmark performs exceptionally well, and quantum-classical QSCI-in-DFT-in-MD workflow closely reproduces the corresponding energetics. While this precludes a demonstration of quantum advantage for the present system, producing a challenging benchmark was not the objective of this work. Instead, we sought to develop a general and transferable workflow and apply it to a biologically relevant problem of genuine scientific interest. Importantly, the QSCI procedure nevertheless identifies and incorporates higher-order wavefunction components that become increasingly expensive to recover using conventional methods such as HCI. Although these determinants contribute negligibly to the total energy in this case, due to their small CI amplitudes, their successful recovery provides evidence that the workflow may be effective in describing systems in which higher-order excitations and stronger correlation effects play a more significant role; this will be investigated through future work. The ADGRG1 application therefore serves both as a biologically relevant case study and as a validation of a problem-agnostic framework capable of capturing more challenging correlated electronic structure when required.


More broadly, this work demonstrates how quantum computers can be integrated into practical multiscale biomolecular simulation workflows. The workflow exemplifies the capabilities of heterogeneous computing ecosystems, combining CPUs, GPUs, and QPUs within a unified framework. The strong scaling observed for the classical post-processing of quantum-generated data further suggests that the approach is well positioned to benefit from continued advances in both classical and quantum hardware. As quantum devices and algorithms mature, such heterogeneous workflows may bridge the gap between quantum computing and real-world scientific applications, enabling correlated electronic structure calculations on biomolecular systems beyond the reach of conventional approaches alone.

\section{Methods}
\label{sec:pipeline}

\subsection{Preparation of the ADGRG1 Protein System for MD Simulations}
As no experimental structure exists for full-length human ADGRG1 in its pre-cleavage state, a composite model combining an AlphaFold2 \cite{jumper2021} prediction with an experimentally determined transmembrane domain was constructed. The extracellular braryGAIN--HormR region was taken from the inactive AlphaFold model of the human N-terminal fragment (NTF) obtained from GPCRdb \cite{herrera2024, gpcrdb_adgrg1_inactive}, validated by alignment against the murine ADGRG1 extracellular domain (PDB 5KVM \cite{salzman2016}), and fused at TM1 to an activated human seven-transmembrane (7TM) domain (PDB 7SF8 \cite{barros2022}), with the post-cleavage N-terminal fragment removed. Missing residues in ECL2 and ICL3 were modeled using Modeller \cite{fiser2003} within Chimera \cite{pettersen2004}, and the fusion region was energy-minimized in MOE \cite{moe2024}. Full details of the model construction are given in Supplementary Note~\ref{note:model}. The model was then prepared in CHARMM-GUI \cite{jo2008, lee2016}, reconstructing five disulphide bonds, four in the extracellular region and one within the TM domain (Supplementary Table~\ref{tab:disulphides}). Default protonation states at neutral pH were assigned to all residues except the catalytic HIS381, protonated at N$\varepsilon$. GLU213, whose protonation state has been shown to affect the autoproteolysis pathway in ADGRL1 \cite{seufert2025}, was retained in its default deprotonated state. 

Following an orientation procedure described in Supplementary Note~\ref{note:model_assembly}, the oriented receptor was embedded in a lipid bilayer and solvated using CHARMM-GUI \cite{jo2008, lee2016}. The bilayer emulated human endoplasmic reticulum composition, the compartment in which autoproteolysis is thought to occur \cite{hamann2015}, with equal proportions of monounsaturated and di-unsaturated lipids in a 3:1:1 ratio of phosphatidylcholine, phosphatidylethanolamine and phosphatidylinositol \cite{ernst2024}. The system was solvated with TIP3P water \cite{jorgensen1983} and neutralized with calcium and chloride ions to 150\,mM, giving a final system of 1,173,254 atoms, of which 10,463 belong to the receptor (Supplementary Table~\ref{tab:composition}).

\subsection{Molecular Dynamics Simulations and Geometry Selection}
The resulting protein--membrane--solvent system was energy-minimized, equilibrated and propagated with GROMACS 2024.4 \cite{spoel2005, abraham2015} using the CHARMM36 force field \cite{brooks2009, huang2013}. Production simulations were run in the NPT ensemble at \SI{310.15}{K} and \SI{1}{bar}. The full MD setup is given in Supplementary Note~\ref{note:md_setup}. The starting geometry for the downstream workflow was selected from an initial \SI{1}{ns} trajectory as the frame minimizing the N$\delta$1--H$\gamma$1 distance between HSE381 and THR383, which reports the approach of the histidine acceptor to the threonine hydroxyl proton in the initial proton-transfer step. Three additional MD runs of approximately \SI{147}{ns} were initiated with independent velocity distributions to assess model stability and the recurrence of molecular arrangements favorable for the N--O acyl shift. All trajectories remained stable over the simulated timescale (Supplementary Figure~\ref{fig:md_stability}), with summary statistics given in Supplementary Table~\ref{tab:md_stability} and the full analysis in Supplementary Note~\ref{note:md_stability}).

\subsection{QM/MM Reaction Path Reconstruction}
Starting from the geometry selected from the MD trajectory, an approximate reaction pathway for the N--O acyl shift was reconstructed by imposing the successive bond rearrangements proposed by Seufert et al.~\cite{seufert2025}, in a manner inspired by their workflow. As bond rearrangement cannot be described by a fixed-valence force field, the reaction was simulated within a QM/MM framework in which the reactive atoms are treated by DFT and their surroundings by MM. Energies and gradients were evaluated through the GROMACS--CP2K interface~\cite{gromacs_manual_qmmm,kuhne2020} using PBE with the DZVP-MOLOPT-GTH basis set~\cite{perdew1996,vandevondele2007}, with electrostatic embedding of the CHARMM36 partial charges~\cite{huang2013}.

The pathway comprises three steps, each traced by restraining the corresponding interatomic distances while relaxing all remaining flexible degrees of freedom. H381 first abstracts the hydroxyl proton of T383 while T383 O$_\gamma$ attacks the L382 carbonyl carbon, forming a five-membered hydroxy-oxazolidine ring. A nearby water molecule then protonates the T383 backbone amide nitrogen. Cleavage of the T383~N--L382~C bond finally opens the ring and forms the ester. Coordinates established in one step were retained at their selected values during all subsequent steps, preventing reversal of the preceding rearrangements and preserving a continuous sequence from the reactant R to the ester E. The endpoint structures R and E were then reoptimized without restraints at the B3LYP level, using the auxiliary density matrix method with the cFIT3 auxiliary basis~\cite{becke1993,lee1988,stephens1994,guidon2010}.

The QM region comprised the catalytic H381--L382--T383 GPS triad, F348, E213, Y384 and ten water molecules positioned between E213 and T383 , extended by backbone and C$\beta$ atoms of flanking residues. Boundaries were placed so that each boundary atom carried exactly one bond across the QM/MM interface and no peptide bond was cut. The selection totaled 163 atoms plus nine hydrogen link atoms, with a net charge of -1 arising from deprotonated E213. The QM region was surrounded by a flexible MM shell comprising all complete residues with at least one atom within \SI{5}{\angstrom} of the QM selection, totaling 986 atoms, with all remaining atoms held frozen. Electrostatic embedding was represented through the Gaussian expansion of the electrostatic potential (GEEP) scheme~\cite{laino2005,laino2006}.

\subsection{QM/MM Product State Construction}
Completing the autoproteolysis requires hydrolysis of the ester intermediate. As the E structure retains the hydroxide generated during the N--O acyl shift, this step is modeled as a base-mediated ester hydrolysis with the distance between the hydroxide oxygen and the L382 acyl carbon as the reaction coordinate. Relaxed scanning along this reaction coordinate, however, did not reach the tetrahedral intermediate, as the hydroxide abstracted the T383 proton on approach rather than adding to the carbonyl. We therefore directly constructed a trial tetrahedral geometry, by manually displacing the hydroxide from its position in E (with a O--C distance of \SI{4.34}{\angstrom}) onto the acyl carbon to form the expected structure with an O--C distance of \SI{1.09}{\angstrom} and a nucleophilic approach angle of \SI{100}{\degree}. This displacement is motivated by the solvation of the site: in E the hydroxide accepts two hydrogen bonds from flanking waters (O--H \SIlist{1.46;1.50}{\angstrom}), and four water oxygens lie within \SI{5}{\angstrom} of the acyl carbon. The bridging protons are therefore partially shared, so the oxygen that adds to the carbonyl need not be the one formally carrying the charge in E. Finally, unconstrained reoptimization was applied to the trial structure with the same setup as for the previous QM/MM reaction path reconstruction.

\subsection{Multiscale Embedding} 
The reaction center was simulated via the projection-based embedding (PBE technique, in which a correlated wavefunction description of an active region is embedded in a larger DFT description of the surrounding environment~\cite{manby2012,goodpaster2014,ralli2024} which itself is polarized by a point charge environment. The QM region is identical to that of the previous QM/MM reaction path reconstruction, while point charges are extracted for all atoms belonging to residues with any atom within \SI{15}{\angstrom} of the nearest QM atom. Each of the nine QM/MM boundary bonds was capped with a link hydrogen placed along the broken bond. 

This region defined the charge embedding. The remaining 175 atom system was simulated using PBE.  Molecular orbitals from a global DFT calculation were localized with the Pipek--Mezey scheme \cite{pipek1989} and assigned to the active or environment subsystem by L\"owdin population analysis \cite{loewdin1950} on the atoms of interest. PySCF \cite{sun2018pyscf} and GPU4PySCF \cite{li2024introducting, wu2024enhancing} were utilized as the quantum chemistry drivers. NBed (version 0.1.0) \cite{ralli2024} was used to perform and generate the projection based embedding problems. The global, lower-cost DFT calculation employed the B3LYP functional \cite{becke1993,lee1988,stephens1994} and the def2-SVP basis set \cite{weigend2005}. The embedded region was treated using the $\omega$B97M-V functional \cite{mardirossian2016omegab97m} within the DFT-in-DFT framework \cite{manby2012,goodpaster2014}. Default numerical integration grids from PySCF were used throughout. An embedded HF-in-DFT calculation was also performed, and post-Hartree-Fock calculations were carried out in the canonical molecular orbital basis obtained from the embedded HF-in-DFT calculation \cite{manby2012,goodpaster2014}. The environment orbitals were then projected out of the active calculation using a Huzinaga projector \cite{huzinaga1971} together with a level-shift projector, both entering a modified core Hamiltonian alongside the embedding potential. Applying the two projectors in combination was found to give the most stable results. Owing to the 54-qubit limit of the available IQM quantum hardware, the embedded Hamiltonian was projected onto an active space comprising 52 spin orbitals, corresponding to a CAS(26,26) description. Details of the active-space construction and projection, together with the orbital-selection criteria, are provided in Supplementary Note~\ref{note:qubit_ham}.

\subsection{Fermion-Qubit Encoding} 
The active-space Hamiltonian was mapped to qubits using the Jordan--Wigner encoding~\cite{jordan1928}, in which fermionic mode occupation corresponds directly to computational basis states, so that Slater determinants are represented by bitstrings and the fermionic vacuum by the all-zero state. This correspondence simplifies preparation of the Hartree-Fock reference and makes the measured bitstrings directly interpretable as electronic configurations. The encoding was further optimized for the target Hamiltonian. The encoding was optimized to minimize the number of Pauli operators in the resulting qubit Hamiltonian using TOPP-HATT optimization~\cite{delabastida2026optimisedfermionqubitencodingsquantum}. The encoding formalism and optimization procedure are given in Supplementary Information~\ref{note:encoding}.

\subsection{Iterative QSCI with Hamiltonian Sparsification Refinement} \label{sec:iter-qsci}
The highest accuracy level of the workflow hierarchy in Figure \ref{fig:pipeline} is the HPC--QC loop. This is realized as an iterative QSCI routine leveraging stochastic Hamiltonian sparsification for wavefunction-biased time evolution, with configuration recovery and expansion through the measured orbital occupancy dynamics. The process iterates over discretized time steps, with each time-evolution circuit being constructed through the stochastic Hamiltonian sparsification simulation framework \cite{ouyang2020compilation}, which generalizes the qDRIFT compilation technique \cite{campbell2019} used by previous time-evolved QSCI \mbox{works~\cite{piccinelli2025quantum, weaving2026selected}.}

The sparsification is controlled through an inclusion probability vector $\bm{p}$, where each Hamiltonian term $h_\ell P_\ell$ is sampled with probability $p_\ell>0$ and the in-circuit excitation is applied with rotation angle $\propto \frac{h_\ell}{p_\ell}$. This ensures the sparsified Hamiltonian, while an approximate compilation strategy, reproduces the desired evolution in expectation but provides a flexible term-wise bias. Each choice of $\bm{p}$ defines a different compilation scheme; when $p_\ell \propto |h_\ell|$ we recover qDRIFT, while $p_\ell = 1$ yields randomized first-order Trotterization. However, the sparsification framework facilitates tailored compilation schemes, which we leverage here for our iterative QSCI method.

Let $\mathcal{D}$ be the configuration subspace obtained in the previous iteration of the process, with $\Pi_\mathcal{D}$ the projector onto the subspace and $\ket{\Psi_{\mathcal{D}}}$ the classically-diagonalized CI solution. Noting that $\mathbbm{1}-\Pi_{\mathcal{D}}$ is the projection outside the subspace, meaning $(\mathbbm{1}-\Pi_{\mathcal{D}}) \ket{\Psi_{\mathcal{D}}} = 0$, we define by
\begin{equation}
    m_\ell \coloneqq \left\lvert\left\lvert (\mathbbm{1}-\Pi_{\mathcal{D}}) P_\ell \ket{\Psi_\mathcal{D}} \right\rvert\right\rvert
\end{equation}
the wavefunction mass that is leaked from the subspace by $P_\ell$. The inclusion probabilities are then chosen as $p_\ell \propto |h_\ell|(m_\ell+\delta)^\alpha$ where $\delta > 0$ sets the sampling floor so that all probabilities are nonzero and $\alpha$ controls the level of biasing ($\alpha=0$ corresponds with qDRIFT). This encourages configuration exploration through the biasing of Hamiltonian terms that carry the most wavefunction mass out of the subspace, weighted by their coefficient magnitudes. In this way, we are able to feed wavefunction information back into the circuit construction, which is refined at each time step to target underexplored regions of the full configuration space. The asymptotic depth-scaling of our QSCI-tailored compilation method with simulation error $\varepsilon>0$ to leading order is $\mathcal{O}\Big(\frac{2\Lambda_\alpha \Lambda_{-\alpha} t^2}{\varepsilon}\Big)$, where $\Lambda_\alpha \coloneqq \sum_\ell |h_\ell| (m_\ell + \delta)^\alpha$ satisfies $\Lambda_\alpha \Lambda_{-\alpha} \geq \lambda^2$; further details are provided in the Supplementary Note~\ref{note:stoch-circ}.

\subsection{Hybrid HPC--QC Setup and Software}
\label{sec:hybrid-setup}
Except as noted otherwise, calculations were executed on BEAST, the Bavarian Energy, Architecture and Software Testbed at the Leibniz Supercomputing Centre (LRZ)~\cite{chung2025qbeast}, an integrated HPC--QC environment in which classical resources and quantum processors are connected through the Munich Quantum Software Stack (MQSS)~\cite{burgholzer2026}. 
MQSS provides middleware for job scheduling and resource allocation through a SLURM plugin~\cite{yoo2003,farooqi2026enahpcqcworkflow} that exposes a QPU as a logical resource on each compute node, allowing classical and quantum resources to be requested within a single job allocation, and manages data exchange between classical nodes and the quantum backend. Quantum computing jobs were submitted through CUDA-Q \cite{cudaq}. The complete hardware and software configuration is given in Supplementary Note~\ref{note:hpcqc}. During quantum circuit runs, Pauli Twirling and Readout Error Mitigation were applied. Details on the error mitigation protocols used are given in Supplementary Note~\ref{note:errormitigation}.

Classical molecular dynamics was performed with GROMACS 2026.0 in mixed
precision with CUDA acceleration. The QM/MM calculations for reaction path reconstruction used a separate double-precision build of GROMACS 2026.0 linked against CP2K 2026.1.

\bmhead{Data availability}
Molecular structures and other data supporting the findings of this study have been deposited in the Zenodo repository at \url{https://doi.org/10.5281/zenodo.21417282}~\cite{orlov2026data}.

\bmhead{Code availability}
The source code used to implement the computational workflow presented in this study is available at \url{https://github.com/UCL-CCS/GPCR}~\cite{gpcr2026github}.

\backmatter

\bmhead{Acknowledgements}
We gratefully acknowledge the use of the quantum system Euro-Q-Exa, co-funded by the EuroHPC JU, BMFTR (grant 13N16690), and the Bavarian State Ministry of Science and the Arts, operated by the Leibniz Supercomputing Centre (LRZ) in Garching, Germany, for providing the computational resources for this work. The research was performed as  part of the Munich Quantum Valley, which is supported by the Bavarian state government with funds from the Hightech Agenda Bayern Plus. This work is also supported by the German Federal Ministry of Research, Technology and Space (BMFTR) with the grant 101194491 (QEX). Computational work has been executed on systems in the test environment BEAST (BAVARIAN Energy Architecture \& Software Testbed) at the Leibniz Supercomputing Centre. The authors would like to thank Dr. Mario Hernández Vera, Thomas M Bickley, Angus Mingare, Martina Nibbi, Niri Govind and Yuri Alexeev for valuable discussions.


\bmhead{Author contributions}
A.O. and P.V.C. conceptualized the project. R.M., A.O. and A.R. performed the structural modeling. A.O., A.R., T.W., M.W.d.B., P.V.S., Z.C. and R.M. developed the software. T.W. designed and implemented the iterative QSCI methodology with wavefunction mass leakage biasing. Z.C. implemented the multi-GPU eigsh solver. P.V.S. implemented error mitigation/reduction techniques and performed the quantum computing experiments. A.O., A.R., T.W., Z.C., M.C. performed the computational work. A.T.N., C.B.M., F.S., M.C. and L.I. supervised aspects of the work; P.V.C. provided overall coordination of the project. A.O., A.R and T.W. prepared the manuscript with input from M.W.D.B., A.T.N., R.M., M.C., Z.C., P.V.S. and P.V.C.

\bmhead{Corresponding authors}
Correspondence to Peter V. Coveney.

\bmhead{Ethics declaration}
The authors declare no competing interests.

\bmhead{Funding}
P.V.C. is grateful for funding from the UK Engineering and Physical Sciences Research Council under the following projects “UK Consortium on Mesoscale Engineering Sciences (UKCOMES)” (Grant No. EP/R029598/1).



\bibliography{sn-bibliography}

\end{document}


\title[Quantum Computing for aGPCR Autoproteolysis]{%
\centering
Supplementary Information\par
\vspace{0.7em}
{\large\bfseries
Multiscale Quantum-Classical Simulation of ADGRG1 Autoproteolysis in a Million-Atom Membrane Environment\par}
}

\input{authors}

\maketitle

\section{Preparation of the ADGRG1 Protein System}
\label{note:model}

\subsection{Construction of the Composite Model}
\label{note:model_construction}

The extracellular domain of murine ADGRG1 (PDB 5KVM) was aligned with the active and inactive AlphaFold models \cite{gpcrdb_adgrg1_active, gpcrdb_adgrg1_inactive, jumper2021, varadi2021} of the human NTF of ADGRG1. The murine structure aligned most closely with the GAIN--HormR domain of the inactive model, which was therefore isolated and used as the extracellular component for our model. Among the loop models generated for ECL2 and ICL3, the structure with the most favourable zDOPE score ($-0.65$) in which both loops pointed away from the membrane surface was selected, avoiding spatial conflict during subsequent membrane embedding. The two regions were joined by aligning TM1 and fusing the chains at the position of maximal backbone overlap, between SER404 of the AlphaFold model and LEU405 of the modelled 7SF8. The five residues on either side of the fusion site were energy-minimised to relieve strain introduced by the junction.

\subsection{Membrane Embedding and System Assembly} 
\label{note:model_assembly} 
The four extracellular disulphide bonds and one TM disulphide bond were identified by comparison of the AlphaFold and 5KVM GAIN domains together with literature data (Supplementary Table~\ref{tab:disulphides}). The receptor was oriented by aligning the vector defined by ILE540 and LEU61 with the membrane normal, followed by translation of \SI{15}{\angstrom} along the $z$-axis to position the transmembrane helices within the bilayer. The composition of the resulting solvated system is given in Supplementary Table~\ref{tab:composition}. 

\begin{table}[h]
\centering
\begin{tabular}{lll}
\hline Region & Residue 1 & Residue 2 \\\hline Extracellular & CYS35 & CYS91 \\Extracellular & CYS121 & CYS177 \\ Extracellular & CYS346 & CYS377 \\Extracellular & CYS366 & CYS379 \\ Transmembrane & CYS469 & CYS556 \\ \hline 
\end{tabular}
\caption{Disulphide bonds reconstructed in the composite ADGRG1 model.}
\label{tab:disulphides}
\end{table}

\begin{table}[h]
\centering
\begin{tabular}{lr}
\hline Component & Atoms \\ \hline Protein & 10,463 \\ Membrane & 131,730 \\ Water & 1,029,069 \\ Ions & 1,992 \\ \hline Total & 1,173,254 \\ \hline 
\end{tabular}
\caption{Composition of the simulated ADGRG1 system.}
\label{tab:composition}
\end{table}

\section{Details of the Molecular Dynamics Simulations}
\label{note:md}

\subsection{Equilibration and Simulation Setup}
\label{note:md_setup}

Simulation inputs were generated with CHARMM-GUI \cite{jo2008, lee2016} and the default settings were retained throughout for constraints, long-range electrostatics and non-bonded cutoffs. Bonds involving hydrogen were constrained with LINCS \cite{hess1997}, allowing a \SI{2}{fs} timestep. Long-range electrostatics were treated by particle-mesh Ewald \cite{darden1993} with a \SI{1.2}{nm} real-space cutoff, and van der Waals interactions with a force-switched cutoff between \SIrange{1.0}{1.2}{nm}. Temperature was maintained by velocity rescaling and pressure by the C-rescale barostat \cite{bernetti2020}, with semiisotropic coupling applied to the membrane system so that the lateral membrane-plane dimensions are coupled while the membrane normal fluctuates independently.

Energy minimisation was performed with the steepest-descent algorithm under position restraints of \SI{4000}{kJ.mol^{-1}.nm^{-2}} on the protein backbone, \SI{2000}{kJ.mol^{-1}.nm^{-2}} on side chains, \SI{1000}{kJ.mol^{-1}.nm^{-2}} on lipids and \SI{1000}{kJ.mol^{-1}.rad^{-2}} on dihedral angles. Equilibration proceeded in six stages. The first two were performed in the NVT ensemble with a \SI{1}{fs} timestep for \SI{125}{ps} each, and the remaining four in the NPT ensemble with a \SI{2}{fs} timestep for \SI{500}{ps} each. Across the protocol the position restraints were progressively reduced for the protein backbone ($4000 \to \SI{50}{kJ.mol^{-1}.nm^{-2}}$), side chains ($2000 \to \SI{0}{kJ.mol^{-1}.nm^{-2}}$) and lipids ($1000 \to \SI{0}{kJ.mol^{-1}.nm^{-2}}$).

\subsection{Stability of the Uncleaved ADGRG1 Ensemble}
\label{note:md_stability}

A summary of the statistics used to assess the stability of the system is provided in Table \ref{tab:md_stability}. All four simulations maintain a mean temperature of \SI{310.2}{K} with a standard deviation of approximately \SI{0.33}{K} and no systematic drift. The instantaneous pressure fluctuates around the \SI{1}{bar} target with standard deviations of approximately \SI{26.7}{bar}, as expected for a finite NPT system under semiisotropic barostat coupling. The potential-energy traces of the three stability trajectories are nearly superimposable, fluctuating around $-1.547\times10^{7}$\,kJ\,mol$^{-1}$ without systematic drift. The backbone RMSD rises over the first \SIrange{20}{40}{ns} before stabilising at plateau values between \SIrange{1.0}{1.9}{nm}, while the \SI{1}{ns} geometry-selection trajectory is too short to reach a plateau.

\begin{figure}[t]
    \centering
    \includegraphics[width=\textwidth]{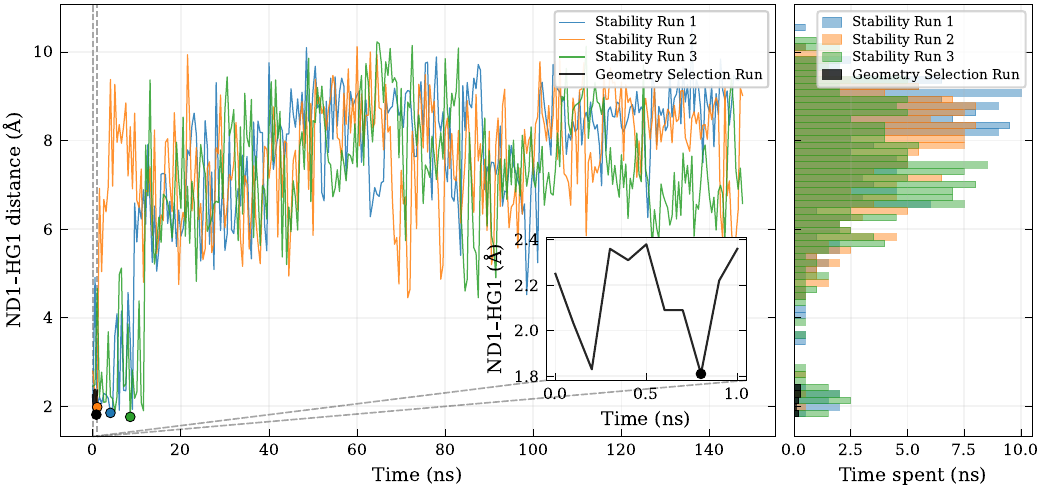}
    \caption{\textbf{MD sampling of the catalytic geometry.} Main, ND1--HG1 (H381--T383) distance across three independent $\sim$\SI{147}{ns} stability runs; filled circles mark the per-run minima. Inset, \SI{1}{ns} geometry-selection run; the $t=\SI{800}{ps}$ frame (\SI{1.81}{\angstrom}) seeds stage~II. Right, cumulative time spent in each distance bin.}
    \label{fig:md_stability}
\end{figure}

\begin{table}[htbp]
    \centering
    \small
    \caption{Summary statistics for the MD stability observables, computed over
    the full trajectory length of each run.}
    \label{tab:md_stability}
    \begin{tabular}{lrrrr}
        \toprule
        & Geometry run & Stability run 1 & Stability run 2 & Stability run 3 \\
        \midrule
        Duration (ns)         & 1.0    & 146.5  & 147.5  & 147.5  \\[0.3em]
        RMSD mean (nm)        & 0.356  & 1.320  & 1.040  & 1.244  \\
        RMSD std (nm)         & 0.152  & 0.441  & 0.172  & 0.211  \\
        RMSD max (nm)         & 0.567  & 1.900  & 1.368  & 1.644  \\[0.3em]
        $E_\text{pot}$ mean (kJ/mol) & $-1.546\times10^7$ & $-1.547\times10^7$
                                     & $-1.547\times10^7$ & $-1.547\times10^7$ \\
        $E_\text{pot}$ std (kJ/mol)  & 4995  & 4690  & 4725  & 4692  \\[0.3em]
        $T$ mean (K)          & 310.2  & 310.2  & 310.2  & 310.2  \\
        $T$ std (K)           & 0.321  & 0.328  & 0.330  & 0.328  \\[0.3em]
        $P$ mean (bar)        & $+0.42$ & $+1.13$ & $+0.89$ & $+0.84$ \\
        $P$ std (bar)         & 26.3   & 26.7   & 26.7   & 26.8   \\
        \bottomrule
    \end{tabular}
\end{table}

\section{Details on the QM/MM Reaction Path Reconstruction}
\label{note:qmmm}

\subsection{QM/MM Partitioning}
\label{note:qmmm_partition}
The QM region selection followed the mechanistic model proposed by Seufert et al.~\cite{seufert2025} for the related ADGRL1 system, adapted to ADGRG1. It comprised the catalytic H381--L382--T383 GPS triad, F348, E213, Y384 and ten water molecules positioned between E213 and T383 , extended by backbone and C$\beta$ atoms of the flanking residues N380, F385, V347, W349, L212 and S214, giving \num{163} atoms in total at a net charge of $-1$ from the deprotonated E213 carboxylate, and singlet multiplicity. The resulting nine cuts between QM and MM region fall on C-–C bonds, except at S214, where the truncation after C$\beta$ places the cut on the C$\beta$–$O\gamma$ bond. Each cut was capped with a link hydrogen placed along the broken bond using the IMOMM scheme. The ten water molecules were selected from a corridor between E213 and T383, requiring each molecule to lie within \SI{9}{\angstrom} of both reference atoms and retaining the ten closest to the corridor midpoint. The flexible MM shell comprised all complete residues with at least one atom within \SI{5}{\angstrom} of the QM selection, totalling 986 atoms, with all remaining atoms held frozen. Electrostatic embedding was represented through the Gaussian expansion of the electrostatic potential (GEEP) scheme~\cite{laino2005,laino2006}.

\subsection{Restrained Scan Protocol}
\label{note:qmmm_scans}
\begin{figure}[t]
    \centering
    \includegraphics[width=\textwidth]{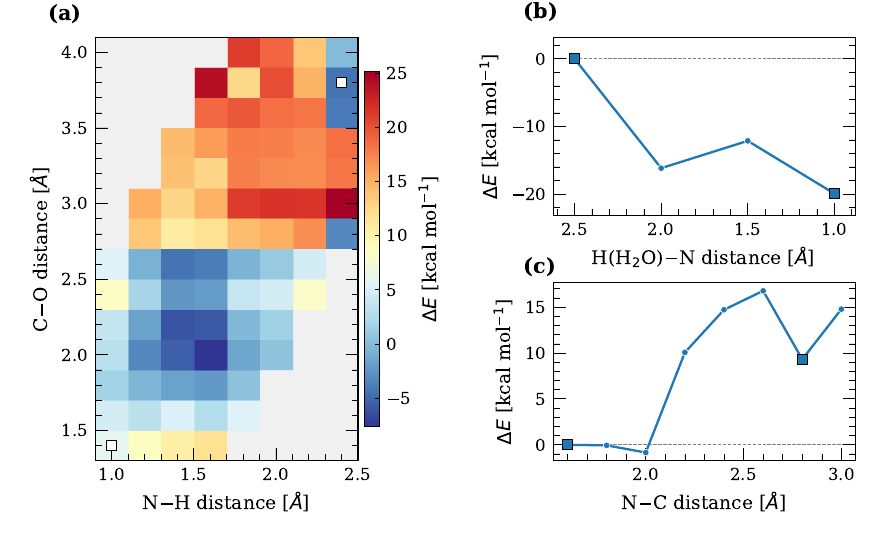}
    \caption{\textbf{QM/MM reconstruction of the N--O acyl-shift path.} \textbf{a,} 2D PBE QM/MM PES for the coupled proton-transfer/nucleophilic-attack step, relative to R (marker in the upper right corner), towards T1 (marker in lower left corner).\textbf{b,c,} 1D PBE scans for amide protonation and ring opening; squares mark the assigned T1, TH and TH, E states respectively.}
    
    \label{fig:qmmm_pes_composite}
\end{figure}

Each reaction step was traced by restraining the corresponding interatomic distances with harmonic umbrella potentials of force constant $k = \SI{250000}{\kilo\joule\per\mole\per\nano\meter\squared}$, while all remaining unfrozen degrees of freedom were relaxed by steepest-descent minimisation at the PBE/DZVP level. The coupled proton transfer and nucleophilic attack were described by two reaction coordinates scanned simultaneously: the H381~N$_{\delta 1}$--T383~H$_{\gamma 1}$ distance, tracking proton transfer, and the L382~C--T383~O$_{\gamma 1}$ distance, tracking nucleophilic attack. These were varied on a two-dimensional grid spanning \SIrange{1.0}{2.4}{\angstrom} and \SIrange{1.4}{4.0}{\angstrom} respectively, in \SI{0.2}{\angstrom} increments and minimized to a force target of \SI{500}{\kilo\joule\per\mole\per\nano\meter}, from which a diagonal-band filter retained the 80 of 112 points lying within the expected reaction corridor. A breadth-first wavefront procedure initialised each point from an adjacent already optimised neighbour, propagating from the point at (\SI{1.8}{\angstrom}, \SI{3.8}{\angstrom}), which was itself initialised from the classical reactant geometry. The resulting potential energy surface is given in Figure \ref{fig:qmmm_pes_composite}a. R was assigned at (\SI{2.4}{\angstrom}, \SI{3.8}{\angstrom}), the lowest-energy point of the reactant region of the surface. T1 was assigned at (\SI{1.0}{\angstrom}, \SI{1.4}{\angstrom}), at the end of the scanned range, where the structural features expected after the first step of the N--O acyl shift are present. The T1 grid point was then reoptimised at the same level with both coordinates restrained at their assigned values of \SI{1.0}{\angstrom} and \SI{1.4}{\angstrom}, tightening the force target to \SI{10}{\kilo\joule\per\mole\per\nano\meter}. 

The two subsequent one-dimensional scans were performed at a force target of \SI{10}{\kilo\joule\per\mole\per\nano\meter} as well. Amide protonation was traced from this reoptimized geometry along a sequential chain that decreased the distance between a hydrogen of the catalytic water and the T383 amide nitrogen from \SIrange{2.5}{1.0}{\angstrom} in \SI{0.5}{\angstrom} increments, converting the catalytic water to a hydroxide, with the two preceding reaction coordinates additionally restrained near the values adopted in the reoptimised T1 geometry, at \SI{1.05}{\angstrom} and \SI{1.45}{\angstrom}, to prevent the five-membered ring from reverting. TH was assigned at \SI{1.0}{\angstrom} and the scan's energetics are given in Figure \ref{fig:qmmm_pes_composite}b. Ring opening was then traced from this geometry by increasing the T383~N--L382~C distance from \SIrange{1.6}{3.0}{\angstrom} in \SI{0.2}{\angstrom} increments, with no other coordinate restrained, and E was assigned at \SI{2.8}{\angstrom}, where the profile, visualized in Figure \ref{fig:qmmm_pes_composite}c, exhibits a local minimum.

\subsection{Convergence of the Refinement}
\label{note:qmmm_convergence}
The B3LYP refinements targeted a maximum-force threshold of 10\,kJ\,mol$^{-1}$\,nm$^{-1}$. Because optimisation of the complete QM region and flexible MM shell involves several thousand nuclear coordinates, all four calculations reached the numerical lower limit of the steepest-descent step size before satisfying this threshold, leaving residual maximum forces of 458, 1046, $1.11 \times 10^{4}$ and 1664\,kJ\,mol$^{-1}$\,nm$^{-1}$ for R, T1, TH and E, respectively.




\section{Qubit Hamiltonian Setup}
\label{note:qubit_ham}

In this section we provide the details on how the qubit Hamiltonian was constructed for the GPCR system. This part of the workflow takes in results from the QM/MM workflow.

\subsection{Projection-Based Embedding} 
Given the size of the GPCR system, a fully quantum-mechanical treatment of the entire structure is computationally prohibitive. Consequently, a multiscale approach is required. In the present work, we are primarily interested in accurately describing the subsystem in which hydrolysis occurs. This region contains the bond-breaking and bond-forming events associated with the reaction and therefore requires a more sophisticated electronic structure treatment than is typically accessible with classical molecular dynamics simulations.

To achieve this, we employed projection-based embedding (PBE) theory to embed a wavefunction (WF) calculation within a density functional theory (DFT) environment \cite{manby2012,goodpaster2014,ralli2024}. In this framework, the chemically active region is treated using a correlated WF method, while the surrounding protein environment is described using DFT. This partitioning enables a substantially higher level of accuracy to be applied where it is most needed, while maintaining a feasible computational cost for the full system. 

The principal advantage of the PBE approach is that it combines the efficiency of DFT for the extended environment with the accuracy of correlated WF methods for the reactive region. This is particularly important for chemical processes involving bond breaking and bond formation, where electron correlation effects can become significant. As a single-reference method, DFT can struggle to accurately describe such situations, especially when the electronic structure deviates from a single-determinant picture. In contrast, correlated WF methods are specifically designed to capture these effects and are therefore better suited for describing the electronic changes that occur along the reaction coordinate. 

To perform a PBE calculation an \emph{active} and \emph{environment} region for the system must be selected each treated at WF and DFT levels of theory respectively. The procedure to do this is composed of four parts, localization,  selection, projection and energy correction. In the first step, full system DFT is run to approximate the electron density. Electrons are then localized to the atoms of either active or environment region. 

We use the Pipek-Mezey orbital localization \cite{pipek1989} approach followed by a L\"{o}wdin population analysis \cite{loewdin1950}:

\begin{equation}
    w_i = \sum_{\mu \in A} \left[ (S^{1/2} C)_{\mu i} \right]^2  \text{ where } 0 \leq w_i \leq 1,
\end{equation}
where $A$ is an atom index, $S_{\mu \nu} = \bra{\chi_{\mu}} \chi_{\nu}\rangle$ is the atomic orbital overlap matrix and $C$ is the orbital coefficient matrix for the localized MOs. To determine which orbitals are active, we also utalize the root-mean-square (RMS) spatial extent of each molecular orbital, defined as the RMS radius about the orbital centroid:

\begin{equation} \label{eq:RMS_orb}
\sigma_{i}
=
\sqrt{
\left\langle \phi_i \middle| r^2 \middle| \phi_i \right\rangle
-
\left|
\left\langle \phi_i \middle| {r} \middle| \phi_i \right\rangle
\right|^2
},
\end{equation}

\noindent where

\begin{equation}
r^2 = x^2 + y^2 + z^2.
\end{equation}

Note that the molecular orbital is expanded in the atomic orbital basis as:

\begin{equation}
\phi_i( r)
=
\sum_\mu C_{\mu i}\chi_\mu( r).
\end{equation}

The expectation values are evaluated from the atomic-orbital integral matrices:

\begin{equation}
    \begin{aligned}
R^{(\alpha)}_{\mu\nu}
&=
\langle\chi_\mu|r_\alpha|\chi_\nu\rangle,
\qquad \alpha=x,y,z,\\
Q_{\mu\nu}
&=
\langle\chi_\mu|r^2|\chi_\nu\rangle.
\end{aligned}
\end{equation}
In PySCF \cite{sun2018pyscf}  $R^{(\alpha)}_{\mu\nu}$ is obtained from: ``int1e\_r" and $Q_{\mu\nu}$``int1e\_r2" integral methods. This allows the construction of:

\begin{equation}
\begin{aligned}
\langle r_\alpha\rangle_i
&=
 C_i^\dagger
R^{(\alpha)}
 C_i,
\qquad \alpha\in \{x,y,z\} \\
\langle r^2\rangle_i
&=
 C_i^\dagger
Q
 C_i,
\end{aligned}
\end{equation}
where

\begin{equation}
\left|
\left\langle
\phi_i
\middle|
 r
\middle|
\phi_i
\right\rangle
\right|^2
=
\sum_{\alpha=x,y,z}
\langle r_\alpha\rangle_i^2 = \langle r_x\rangle_i^2 + \langle r_y\rangle_i^2 + \langle r_z\rangle_i^2.
\end{equation}

Large values of $\sigma_{i}$ (equation \ref{eq:RMS_orb}) indicate an orbital is more spread out in space and we use a threshold of $3$ Bohr to reject diffuse (Rydberg-like) orbitals. The active orbitals are then selected to be MOs that have a spread lower than $3$ Bohr and that have have a high Lowdin population on the atoms of interest. 

Once this choice is made, we have:

\begin{equation}
    \begin{aligned}
        act ~\in~ \mathbb{Z}, \qquad&\text{active MO indices} \\
        env ~\in~ \mathbb{Z},\qquad&\text{environment MO indices} \\
       occ ~\in~ \mathbb{Z},\qquad&\text{set of occupied indices for DFT/HF solution} \\
       vir ~\in~ \mathbb{Z},\qquad&\text{set of virtual indices for DFT/HF solution}
    \end{aligned}
\end{equation}

This defines the active and environment systems. The density matrix from the global DFT calculation can be split into the active and environment as $\gamma_{full} = \gamma_{act} + \gamma_{env}$ using the occupied active and occupied environment MO indices. 

The Huzinaga projector operator can then be defined:

\begin{equation}
    P_{huz} = C_{env\cap occ} C_{env\cap occ}^\dagger S
\end{equation}
where $C_{env\cap occ}$ are the \emph{occupied} columns of $C$ for the environment. As written this projector is idempotent, but is not symmetric. We use it in a symmetrized form as follows:

\begin{equation}
    O_{huz} =  -\left(F P_{huz} + P_{huz}^{\dagger} F\right),
\end{equation}
where $F$ is the AO Fock matrix. Note there is no factor of $0.5$ that is sometimes included in the literature. This is because we defined $P_{huz}$ in terms of $C_{env\cap occ}$ rather than the density matrix $\gamma_{env}$ which for restricted calculations includes an additional factor of two (as the 1-RDM is spin-free in restricted calculations and so includes both spin-up and spin-down contribution).

We can also write the $\mu$ projector for WF-in-DFT embedding as follows:

\begin{equation}
    P_{\mu} =  \mu \big( S C_{env\cap occ} C_{env\cap occ}^\dagger S \big).
\end{equation}

The action of this operator is much easier to understand in the MO basis (rather than the current AO basis). In particular, by transforming to the MO basis we see that this operator becomes:

\begin{equation}
    P_{\mu}^{MO} =  \mu I_{env\cap occ}.
\end{equation}
We observe this is an identity on the rows and columns of the environment orbitals (assuming an orthogonal MO basis). This shifts the energy of the environment orbitals by $\mu$ and if large enough will make them energetically unfavorable to fill in the active calculation (thus keeping the active and environment systems orthogonal). This becomes exact as $\mu \rightarrow \infty$.

Both these operators are used to write a modified core Hamiltonian as:

\begin{equation} \label{eq:hcore_emb}
    h_{emb} =  h_{1e} + O_{huz} + P_{\mu} + v_{emb}.
\end{equation}
where $h_{1e}$ is the standard one-electron core Hamiltonian and $v_{emb} = g[\gamma_{act} + \gamma_{env}] - g[\gamma_{act}]$ where $g$  contains the two electron terms (Coulomb and exchange for Hartree-Fock or Coulomb and exchange-correlation (XC) for DFT).

Note, we use both the Huzinaga and $\mu$ shift method together as we found this gives the most stable results.  The overall energy of the system is then given as:

\begin{subequations} \label{eq:E_embedding}
    \begin{equation}
        E_{\text{WF-in-DFT}} = \bra{\psi_{act}}H_{emb}\ket{\psi_{act}} + E_{DFT}[\gamma_{env}] + E_{cross}[\gamma_{act}, \gamma_{env}] - Tr\big[ \gamma_{act} (h_{emb}-h_{1e}) \big],
    \end{equation}
    \begin{equation}
        E_{\text{DFT-in-DFT}} = \mathcal{E}[\Tilde{\gamma}_{act}]+ E_{DFT}[\gamma_{env}] + E_{cross}[\gamma_{act}, \gamma_{env}] - Tr\big[ \gamma_{act} (h_{emb}-h_{1e}) \big],
    \end{equation}
\end{subequations}
where $\psi_{act}$ and $\Tilde{\gamma}_{act}$ are the solutions for the active system with the modified core Hamiltonian  (equation \ref{eq:hcore_emb}). For the DFT calculation, a ``better" (more expensive) functional over the global DFT calculation can be used hence why we use the $\mathcal{E}$ symbol rather than the $E$. The final term in both equations gives the first order correction for embedding, this is needed as the embedding terms are included in $H_{emb}$ and  $\mathcal{E}$ and so must be subtracted to avoid including it twice. 

In this work, the protein environment was represented using a charge-embedding scheme for the quantum mechanical calculations. The charges were obtained form the QM/MM calculations. The charge embedding was incorporated into both the mean-field and embedded calculations. Subsequently, wavefunction-based quantum (QSCI) and classical (DFT/HCI) approaches were employed to solve the embedded problem ($\mathcal{E}[\Tilde{\gamma}_{act}]$ or $H_{\mathrm{emb}}$). The total energies reported in the main text were each obtained from equation  \ref{eq:E_embedding}.

\subsection{Active Space Selection} 

We were limited to $52$ qubits on a IQM quantum device and thus had to project our $H_{emb}$ into an $52$-spin orbital active space. To achieve this we do the following high level steps:

\begin{enumerate}
    \item Rank the environment-free virtual MOs by their weight in a minimal valence AO space on the active atoms, keeping the leading ones as a screening pool.
    \item We run MP2 within [occupied $+$ pool] and take its natural orbitals, most-occupied virtuals, most-depleted occupied, to define the CASCI subspace.
\end{enumerate}

In detail, for the first step, we projected our virtual orbitals of the active atom indices onto the ``minao" basis in PySCF. This can be written as:

\begin{subequations}
    \begin{equation}
        P_{minao} = S_{cross} S_{minao}^{-1} S_{cross}^{\dagger}
    \end{equation}
    \begin{equation}
        [P_{minao}]_{ij} = \sum_{a,b}^{M} \bra{\chi_{i}} \phi_{a}\rangle (S_{minao}^{-1})_{ab} \bra{\phi_{b}} \chi_{j} \rangle,
    \end{equation}
\end{subequations}

\noindent where:
\begin{equation}
    [S_{cross}]_{\nu i} = \bra{\chi_{\nu}^{mol}} \phi_{i}^{minao} \rangle,
\end{equation}

\begin{equation}
    [S_{minao}]_{ij} = \bra{\phi_{i}^{minao}} \phi_{j}^{minao} \rangle.
\end{equation}
$P_{minao}$ acts as a projector onto the minimal valence AO space. However, we want to only look at the  MOs on active atoms and so we write this as:

\begin{equation}
        [P_{minao}^{\text{act-atom}}]_{ij} = \sum_{a,b\in \text{act-atom}} \bra{\chi_{i}} \phi_{a}\rangle (S_{minao}^{-1})_{ab} \bra{\phi_{b}} \chi_{j} \rangle
\end{equation}
where $a,b$ are restricted to be the AO indices of the active atoms.

We use this projector (onto the valence orbitals of the active atoms) to project the valence orbitals as:

\begin{equation}
        \Omega = C_{vir}^{\dagger} P_{minao}^{\text{act-atom}} C_{vir}.
\end{equation}

We then perform as SVD on $\Omega$

\begin{equation}
        U \Sigma V^{\dagger} = SVD(\Omega),
\end{equation}
and transform $C_{vir}$ as:

\begin{equation}
    \Tilde{C}_{vir} = C_{vir} U.
\end{equation}
The singular values $\Sigma$ are the projection weights of the rotated virtual orbitals onto the retained MINAO subspace. Sorting them in descending order ranks the virtual orbitals by how strongly they resemble the selected minao-derived valence space. This is used to select a virtual space that does not focus on the diffuse functions and instead focuses on the virtual orbitals of the active atoms.

In summary, within the environment-free virtual block we diagonalise a projector onto the minimal valence AO space of the active atoms, giving rotated virtuals $\Tilde{C}_{vir}$ ordered by decreasing fragment valence weight.

To define an active space there is a combinatorial choice of which orbitals to include. To construct a ``good" active space one needs to include the orbitals that are necessary to describe the physics of the system of interest, for example, orbitals that are involved in a bond breaking / forming step being studied rather than orbitals far away from the site of interest.

We take the leading $n_{pool}$ MOs of  $\Tilde{C}_{vir}$ and semicanonicalise them, since MP2 denominators are orbital-energy differences only when the Fock matrix is diagonal in the correlated block. An MP2 calculation is then performed over [all embedded occupied MOs + pool], with all the other MOs frozen (in PySCF) so the cost is set by $n_{pool}$ rather than the basis for the MP2 calculation. Note we do not allow any occupied environment orbitals into the active space (instead they are always frozen by the reference state). The occupied and virtual blocks of the resulting 1-RDM are diagonalised separately: virtuals ordered by decreasing occupation, occupied by increasing. The CAS is then the $n\_cas\_vir$ most-occupied virtual natural orbitals, plus, if a reduced occupied space is requested, the $n\_cas\_occ$ occupied orbitals that MP2 has depleted furthest below 2, the remainder becoming CASCI core. We targeted a CASCI of size $(26,26)$. The resulting active space can be found at \cite{orlov2026data}, including the one- and two-body Hamiltonian integrals needed to construct the second-quantized molecular Hamiltonian.

\subsection{Details on the Fermion-Qubit Encoding}
\label{note:encoding}
Digital quantum simulation of fermionic Hamiltonians requires an encoding from the anti-symmetric Fock space of fermionic operators to the Hilbert space of qubit operators, $\mathcal{E}:\mathcal{F}_{m,M}^- \to \mathcal{H}_{N}$. The most straightforward of these is the Jordan--Wigner encoding~\cite{jordan1928}, in which fermionic mode occupation is associated with the computational basis states of qubits. This encoding simplifies the preparation of Hartree--Fock input states, as computational basis states represent Slater determinants and the all-zero state encodes the fermionic vacuum. While the Jordan--Wigner encoding is local in mode occupation, correctly accounting for fermionic exchange antisymmetry requires non-local operations spanning a number of qubits that grows linearly with system size. The Jordan--Wigner encoding can be numerically optimised by exploiting features of the target Hamiltonian~\cite{delabastida2026optimisedfermionqubitencodingsquantum}, in this case the electronic structure Hamiltonian under the Born-Oppenheimer approximation,
\begin{equation}
\label{eq:esh}
H_f = \sum_{ij}c_{ij} a^{\dagger}_i a_j + \sum_{ijkl}c_{ijkl}a^{\dagger}_j a^{\dagger}_i a_k a_l ,
\end{equation}
where $c_{ij}$ and $c_{ijkl}$ are the
one- and two-electron integrals expressed in an orthonormal molecular orbital (MO) basis. A given encoding need only describe the algebra of fermionic operators formed by terms with non-zero coefficients. Optimisation therefore proceeds by minimising the number of Pauli operators required to encode non-vanishing terms appearing in the encoded qubit Hamiltonian. We employ TOPP-HATT optimisation \cite{topphatt} to obtain a near-optimal Jordan--Wigner encoding, which is applied to the input Hamiltonian and reference states, while Z-basis measurements from the QPU are decoded to Fock states using the inverse encoding~\cite{delabastida2026optimisedfermionqubitencodingsquantum}.

\section{Background and Implementation Details of Iterative Quantum-Selected Configuration Interaction}
\label{note:qsci}

The central goal of quantum chemistry is to solve the time-independent many-body Schr\"odinger equation $H\ket{\Psi}=E \ket{\Psi}$, where $H$ is the electronic structure Hamiltonian of Equation \eqref{eq:esh}, $E\in\mathbb{R}$ is the electronic energy and $\ket{\Psi} \in (\mathbb{C}^{2})^{\otimes N}$ its wavefunction. The complexity of this problem increases rapidly with system size, meaning the exact solution $\ket{\Psi}$ is only possible for the smallest of systems. Instead, we must turn to approximate methods, for which a vast body of techniques has been developed. A powerful and commonly adopted approach arises from expanding the wavefunction in terms of powers of the Hamiltonian, applied to some reference ``guess" state $\ket{\psi}$. Explicitly, this involves constructing the space
\begin{equation}
    \mathrm{span}\big( \{\ket{\psi}, H\ket{\psi}, H^2 \ket{\psi}, \dots, H^{K}\ket{\psi}\} \big),
\end{equation}
where an ansatz/trial wavefunction takes the form $\ket{\Psi(\bm{\theta})} = \sum_{k=0}^{K} \theta_k H^k \ket{\psi}$. This works since, expanding the reference in the eigenbasis of $H$, $\ket{\psi} = \sum_\ell \alpha_\ell \ket{\Psi_\ell}$, application of a degree $K$ polynomial $f$ in terms of $H$ yields $f(H)\ket{\psi} = \sum_\ell \alpha_\ell f(E_\ell) \ket{\Psi_\ell}$ and through careful selection of $f$, defined by the parameters $\bm{\theta}$, we can attempt to isolate particular eigenvalues/states. Such approaches are called \textit{Krylov} methods, a direct example being the Lanczos algorithm \cite{lanczos1950iteration} and, more loosely, Davidson's algorithm \cite{DAVIDSON197587}, which augments the subspace with preconditioned residual vectors rather than pure Krylov vectors.

Access to quantum computers offers a variation on the standard Krylov set-up, namely through access to Hamiltonian exponentials $e^{iHt}$, rather than powers of the Hamiltonian. The effective space in \textit{quantum} Krylov \cite{parrish2019filter, yoshioka2025krylov} is instead
\begin{equation}
    \mathrm{span}\big( \{\ket{\psi}, e^{iHt_1}\ket{\psi}, e^{iHt_2} \ket{\psi}, \dots, e^{iHt_K}\ket{\psi}\} \big),
\end{equation}
where each $t_k$ can be viewed as an evolution time. Taking a fixed step $\tau$, we can set $t_k = k\tau$ and then $e^{iHt_k} = (e^{iH\tau})^k$, i.e. evolution by $k$ steps via the time propagator $e^{iH\tau}$. On a quantum computer, one can therefore prepare an initial state $\ket{\psi}$ and subsequently apply powers of the propagator $e^{iH\tau}$, with the caveat that we are only permitted to draw finite samples from the evolved state $e^{iHk\tau}\ket{\psi}$, not full knowledge of the wavefunction. Expanding $\ket{\psi}$ as we did before and writing each eigenstate in the computational basis $\ket{\Psi_\ell} = \sum_{m} \beta_{\ell m} \ket{m}$, in the quantum setting we get $e^{iHk\tau}\ket{\psi} = \sum_\ell \alpha_\ell e^{i E_\ell k\tau} \ket{\Psi_\ell} = \sum_{\ell, m} \alpha_\ell \beta_{\ell m} e^{i E_\ell k\tau} \ket{m}$ and thus we measure $\ket{m}$ with probability $p(m) = |\sum_\ell\alpha_\ell \beta_{\ell m} e^{i E_\ell k\tau}|^2$. In Sample-based Krylov Quantum Diagonalization (SKQD)~\cite{Yu2025}, one samples a collection of measurements $\mathcal{D} = \{\ket{m} \}_m$ taken across several time steps and project the Hamiltonian into the subspace spanned by $\mathcal{D}$, explicitly $H \mapsto PHP = \sum_{mn} H_{mn} \ket{m}\bra{n}$ where $P = \sum_m \ket{m}\bra{m}$ is a projector and $H_{mn} = \braket{m|H|n}$ are Hamiltonian matrix elements. The subspace matrix $\bm{H}$ defined by entries $H_{mn}$ is of size $|\mathcal{D}|\times|\mathcal{D}|$, so as long as the number of basis elements is not prohibitively large, we can obtain an approximate eigenstate using reduced (classical) computational resources as compared with solving the full problem.

The quality of the resulting approximation is dependent on the samples $\mathcal{D}$, so if the sampling is performed on a noisy quantum device then hardware errors can degrade the quality of the subspace. While SKQD offers certain performance guarantees \cite{Yu2025, piccinelli2025quantum}, there are many alternative sample-based quantum diagonalization (SQD), or quantum-selected configuration interaction (QSCI), approaches that are differentiated through how the sampling circuit is designed. The most extensively investigated circuit construction for molecular electronic structure is that based on the Unitary Cluster Jastrow (UCJ) \cite{matsuzawa2020jastrow} and local (LUCJ) \cite{motta2023bridging} restrictions to fixed hardware layouts, facilitating calculations using up to 77 qubits \cite{robledo2025, kaliakin2025accurate, barison2025quantum, duriez2025computing, barroca2025surface, nogaki2025symmetry, liepuoniute2024quantum, kaliakin2025implicit, bazayeva2025quantum, sriluckshmy2025quantum, shajan2025toward, smith2025quantum, danilov2025enhancing, shirakawa2025closed, mcfarthing2026noise, park2026cluster, li2026protein, shajan2026molecular, anurag2026towards}. The UCJ parameters are obtained via a factorization of the $T_2$ amplitudes in coupled cluster theory, while LUCJ imposes locality constraints by zeroing interactions between non-neighbouring orbitals on the target device graph. 

However, without performing subsequent optimization of the variational parameters in the LUCJ ansatz, it has been questioned whether the accessible space is truly non-classical. While there has been some effort to formalize the hardness of LUCJ circuit sampling \cite{hafid2025hardness}, other work has found practical instances in the literature to be efficiently simulable by tensor networks \cite{rudolph2025simulating}. The challenge in optimizing variational parameters in sample-based methods is the indirect relationship between the samples and corresponding energies, which is not a smooth function of the circuit parameters and renders gradients ill-defined; one work attempted to search the parameter space with populations of walkers governed by a differential evolution scheme \cite{shirakawa2025closed}. Besides LUCJ, other works have explored different variational circuit constructions \cite{yoffe2024qubit, nutzel2025solving, bauer2025efficient, ohgoe2025quantum, shirai2025enhancing, pellow2025hivqe}, MPS-to-circuit mappings \cite{bickley2025}, generative models \cite{kemmoku2026generative} and neural networks \cite{chen2026neural}. While many of these approaches have produced successful hardware demonstrations, the non-classicality of each method is difficult to assess. The more provably non-classical approaches are those based on stochastic time evolution \cite{sugisaki2024hamiltonian, mikkelsen2024quantum, piccinelli2025quantum, oumarou2025molecular, graves2026resource, weaving2026selected}, which naturally align with the framework of SKQD. 

A commonly used stochastic compilation technique is qDRIFT \cite{campbell2019}. The present work adopts the generalised stochastic Hamiltonian sparsification framework \cite{ouyang2020compilation} as detailed in Section \ref{note:stoch-circ}, of which qDRIFT is a special case. This stochastic time evolution approach to QSCI is moreover embedded within broader classical theories, namely a hierarchy consisting of an active space approximation, DFT and an MM environment at the highest level. This workflow is designed to leverage the benefits of quantum-centric supercomputing \cite{robledo2025, alexeev2024quantum}, utilizing co-located HPC and QPU resources as detailed in Section \ref{sec:hybrid-setup} and with GPU acceleration for critical computational workloads, described in Section \ref{note:postprocessing}. We also propose an iterative approach, which refines the stochastic Hamiltonian sparsification bias with information from the classically-diagonalized wavefunction at each iteration.

\subsection{Stochastic Circuit Construction}\label{note:stoch-circ}

Given $H=\sum_{\ell=1}^{L} h_\ell P_\ell$, a sampling budget $N_{\mathrm{sample}}$ and inclusion probabilities \mbox{$p_\ell\in[0,1]$} with mean sparsity \mbox{$\mu=\sum_\ell p_\ell$}, each of $R=\lceil \frac{N_{\mathrm{sample}}}{\mu} \rceil$ repetitions independently retains term $\ell$ with probability $p_\ell$, reweighted by $1/p_\ell$ to ensure $H$ is recovered in expectation. Explicitly, per slice we sample a sparsifed Hamiltonian 
\begin{equation}
\tilde{H}_r=\sum_{k=1}^{K \approx \mu}\frac{h_{\ell_k} \mu}{p_{\ell_k} N_{\mathrm{sample}}}P_{\ell_k}
\end{equation}
so that \mbox{$H= \mathbb{E}(\sum_{r=1}^R \tilde{H}_r)$} and moreover, within each slice, we apply the sampled terms in forward or reverse order with equal probability, i.e. $\prod_{k=1}^{K \approx \mu} \exp\big(-it{\frac{h_{\ell_k} \mu}{p_{\ell_k} N_{\mathrm{sample}}}P_{\ell_k}}\big)$ or $\prod_{k=K \approx \mu}^{1} \exp\big(-it{\frac{h_{\ell_k} \mu}{p_{\ell_k} N_{\mathrm{sample}}}P_{\ell_k}}\big)$. This stochastic reversal makes the average a symmetric (second-order) product formula \cite{ouyang2020compilation}. 

Taking $p_\ell = |h_\ell| / \lambda$ where $\lambda=\sum_{\ell} |h_\ell|$ is the Hamiltonian coefficient $\ell_1$-norm recovers qDRIFT, noting that we sample on average $\mu = 1$ term per slice over $R=M$ repetitions. On the other end of the spectrum, fixing $p_\ell=1$ for all $\ell$ corresponds with randomised first-order Trotter, in which $\mu=L$. Note that it can occur that there is some leftover budget and we append a supplementary slice to account for this. Alternative choices of the inclusion probability vector $\bm{p}$ can yield tailored compilation schemes that bias the selection of certain terms within the Hamiltonian simulation. Our approach uses information from the classically-diagonalised wavefunction solution to encourage the probing of under-explored configuration spaces.

Let $\mathcal{D} = \{\ket{m}\}_m$ be the configuration subspace at a particular iteration of our QSCI workflow and define by $\Pi_{\mathcal{D}} = \sum_m \ket{m}\bra{m}$ the projector onto the subspace. We solve the projected Hamiltonian 
\begin{equation}
    H_\mathcal{D} = \Pi_\mathcal{D} H \Pi_\mathcal{D} = \sum_{m,n} H_{mn} \ket{m} \bra{n}
\end{equation}
to obtain a wavefunction $\ket{\Psi_\mathcal{D}} = \sum_{m} \alpha_m\ket{m}$ with energy $E_\mathcal{D}$. A criticism of many QSCI implementations is a lack of feedback loop from this wavefunction back into circuit construction; we use this information to inform an update to the inclusion probability vector.

To update each individual probability $p_\ell$, we consider how much wavefunction ``mass'' is leaked out of the subspace by the corresponding Hamiltonian term $P_\ell$, defined by the quantity
\begin{equation}\label{eq:wfn_leakage}
    m_\ell \coloneqq \left\lvert\left\lvert \big(\mathbbm{1} - \Pi_\mathcal{D}\big) P_\ell \ket{\Psi_{\mathcal{D}}} \right\rvert\right\rvert.
\end{equation}
In this expression, $P_\ell \ket{\Psi_{\mathcal{D}}}$ contains the configurations coupled to the variational space via the Hamiltonian term $P_\ell$, and $\mathbbm{1} - \Pi_\mathcal{D}$ is the projector outside the configuration subspace, i.e. $(\mathbbm{1} - \Pi_\mathcal{D})\ket{\Psi_\mathcal{D}} = 0$. We then choose 
\begin{equation}\label{eq:inclusion_prob}
    p_\ell \propto |h_\ell|(m_\ell+\delta)^\alpha
\end{equation}
where $\delta > 0$ sets the sampling floor so that all probabilities are nonzero and $\alpha$ controls the level of biasing ($\alpha=0$ corresponds with qDRIFT).

To further encourage the exploration of new configuration spaces, we also trial several dominant determinants in the wavefunction as the reference state for circuit initialisation, alongside several instances of Hamiltonian sparsification per time step. More concretely, we allow a circuit budget of $N_{\mathrm{circuit}}$ which is allocated among the top $N_{\mathrm{ref}}$ determinants $\ket{\Phi_\mathrm{ref}}$ in $\ket{\Psi_\mathcal{D}}$ proportional to their wavefunction coefficients $|\alpha_m|^2$.

\subsubsection{Asymptotic Depth Scaling}\label{note:asymptotics}

Stochastic circuit compilation techniques were developed to improve on the asymptotic scaling of deterministic product formulae such as Trotterization. To simulate the Hamiltonian time evolution operator $e^{-iHt}$ to error $\varepsilon>0$ (in diamond-norm), the first-order Trotter expansion requires a circuit depth of $\mathcal{O}(\frac{L^3 \Lambda^2 t^2}{2\varepsilon})$ \cite{suzuki1976generalized}, recalling that $L$ is the number of Hamiltonian terms and $\Lambda=\max_\ell |h_\ell|$ is the Hamiltonian coefficient with greatest magnitude. A shift from deterministic to stochastic approaches decouples us from $L$, scaling instead with the coefficient $\ell_1$-norm $\lambda = \sum_\ell |h_\ell|$. This was first formalised with qDRIFT, permitting a depth of $\mathcal{O}(\frac{2\lambda^2 t^2}{\varepsilon})$ \cite{campbell2019}. Although, note that $\lambda \leq L \Lambda$, with equality when all coefficients are equal, so the decoupling from the number of Hamiltonian terms is at worst implicit. We now derive the scaling of the stochastic sparsfication by wavefunction leakage introduced in Section \ref{note:stoch-circ}. Referring to the original compilation by stochastic Hamiltonian sparsification work \cite[Therom 1]{ouyang2020compilation}, the leading order error term given gate depth $G$ is
\begin{equation}\label{eq:sparsto_leading}
    \varepsilon = \frac{2t^2\mu}{G}\|\bm{u}\|, \;\; \text{where}\;\; u_\ell = \bigg( \frac{1}{p_\ell} - 1 \bigg) |h_\ell|^2.
\end{equation}
Defining
\begin{equation}
    \Lambda_\alpha \coloneqq \sum_\ell |h_\ell| (m_\ell + \delta)^\alpha
\end{equation}
where $m_\ell$ is the wavefunction leakage as per Equation \eqref{eq:wfn_leakage}, we choose the inclusion probability $p_\ell = \frac{|h_\ell| (m_\ell + \delta)^\alpha}{\Lambda_\alpha}$ as in Equation \eqref{eq:inclusion_prob}. Due to the normalising factor $\Lambda_\alpha$ we have $\mu = \sum_\ell p_\ell = 1$, the same as qDRIFT. Substituting into Equation \eqref{eq:sparsto_leading} we obtain
\begin{equation}
    u_\ell = \Lambda_\alpha |h_\ell| (m_\ell+\delta)^{-\alpha} - |h_\ell|^2
\end{equation}
and therefore
\begin{equation}
    \|\bm{u}\|_1 = \Lambda_\alpha \Lambda_{-\alpha} - \|\bm{h}\|_2^2
\end{equation}
where $\|\bm{h} \|_2$ is the coefficient $\ell_2$-norm. Finally, the circuit depth to achieve error $\varepsilon>0$ to leading order is
\begin{equation}
    \mathcal{O} \bigg(\frac{2 \Lambda_\alpha \Lambda_{-\alpha} t^2}{\varepsilon}\bigg),
\end{equation}
which lands in the qDRIFT scaling regime. Taking a ratio of this scaling to qDRIFT we get $\frac{\Lambda_\alpha \Lambda_{-\alpha}}{\lambda^2}\geq 1$, obtained through application of the Cauchy-Schwarz inequality on $\sqrt{|h_\ell|} (m_\ell + \delta)^{\pm\alpha/2}$. These results are summarised in Table \ref{tab:gate-depth-comparison}.

\begin{table*}[b]
  \centering
  \begin{tabular}{lc}
    \hline
    Protocol & Circuit depth \\
    \hline
    1st-order Trotter
      & $\displaystyle\frac{L^{3}\Lambda^{2}t^{2}}{2\varepsilon}$
       \\[1.2em]
    qDRIFT
      & $\displaystyle\frac{2\lambda^{2}t^{2}}{\varepsilon}$
       \\[1.2em]
    Sparsification by wavefunction leakage
      & $\displaystyle\frac{2\Lambda_{\alpha}\Lambda_{-\alpha}\,t^{2}}{\varepsilon}$
       \\
    \hline
  \end{tabular}
  \caption{Leading-order gate depth $G$ to simulate $e^{-iHt}$ to diamond-norm error $\varepsilon>0$,
  for $H=\sum_{\ell=1}^{L}h_\ell P_\ell$ with
  $\lambda=\sum_\ell | h_\ell |$.
  First-order Trotter uses one gadget per term per slice ($G=Lr$).
  qDRIFT and wavefunction mass/leakage-biased sparsification are stochastic and independent of $L$ (noting $\lambda \leq L\Lambda$ where $\Lambda = \max_{\ell} |h_\ell|$).
  The inclusion probability for sparsification is $p_\ell\propto |h_\ell|(g_\ell+\varepsilon_g)^\alpha$ at
  $\mu=\sum_\ell p_\ell=1$, with
  $\Lambda_{\pm\alpha}=\sum_\ell |h_\ell|(g_\ell+\varepsilon_g)^{\pm\alpha}$.
  The $\alpha=0$ (coefficient-only) case recovers the qDRIFT row,
  since $\Lambda_0=\lambda$.}
  \label{tab:gate-depth-comparison}
\end{table*}

\subsection{Recovery and Expansion}\label{note:recovery}

At each time step of the iterative QSCI loop, Section \ref{note:stoch-circ} describes the approach to circuit construction, producing $N_\mathrm{circuit}$ circuits each approximating $e^{-iHt} \ket{\Phi_\mathrm{ref}}$ where $\ket{\Phi_\mathrm{ref}}$ is one of $N_\mathrm{ref}$ dominant configurations in the wavefunction obtained at the previous step. The time evolution operator $e^{-iHt}$ is approximated through $N_\mathrm{sample}$ in-circuit excitations, selected via the stochastic sparsification method given above and biased to include terms that encourage excitations outside the current variational space. We then perform $N_\mathrm{shot}$ measurements on the QPU, for a total of $N_\mathrm{circuit} \times N_\mathrm{shot}$ shots per QSCI iteration.

This produces a collection of quantum measurements $\{\mathcal{M}_n\}_{n=1}^{N_\mathrm{circuit}}$ where each $\mathcal{M}_n$ consists of $N$-qubit binary strings $\bm{b} \in \mathbb{Z}_2^{\times N}$. Wherever $\bm{b}$ corresponds with a valid electronic configuration, i.e. it consists of the correct number of $\alpha/\beta$ electrons, we append such elements to $\mathcal{D}$. However, it is most likely the vast majority of measurements will not produce valid configurations, owing to hardware noise corrupting the desired particle sector. In such cases, we defer to a configuration recovery protocol to improve the retention rate of configurations sampled from the QPU -- a post-processing step that other works also employ in one form or another \cite{robledo2025}. In our approach, we estimate the orbital occupancies from our collections of quantum measurements, specifically 
\begin{equation}\label{eq:occ}
o_q = \frac{1}{2}\bra{\Phi_\mathrm{ref}} e^{iHt} (1 - Z_q) e^{-iHt} \ket{\Phi_\mathrm{ref}} \approx \frac{\left\lvert\{\bm{b}\in\mathcal{M}_n: b_q=1\}\right\rvert}{N_{\mathrm{shot}}}
\end{equation}
evaluated for each qubit $q\in\{1, \dots, N\}$. Any measurement $\bm{b}$ that deviates from a valid configuration by at most a pre-determined Hamming distance $W$ is then pushed into the correct particle sector by flipping the bits favoured by their measured orbital occupancies.

This capped-Hamming recovery improves the retention of configurations in the subspace, but the subspace growth is still restricted by the number of measurements we can extract from a quantum device. This differs in methodology to conventional SCI methods such as heatbath CI, which involve rapid expansion via a screening protocol and explores large spaces of candidate configurations. To bridge these methodologies, in previous work we investigated the use of an SCI-like expansion stage biased by the measured orbital occupancies of Equation \eqref{eq:occ} \cite{weaving2026selected}. This allows us to inject classically-inaccessible information into the configuration screening process while facilitating broader exploration than just what we sample directly from the QPU. Whereas the previous work leveraged this idea in a non-iterative way, here we adopt the qubit population dynamics expansion step within the iterative QSCI loop -- iterating between QPU sampling, capped-Hamming occupancy-biased recovery, expansion and subsequent configuration interaction to yield an updated wavefunction. This information feeds back into the inclusion probability vector described in Section \ref{note:stoch-circ}, in turn refining the circuit construction and the process repeats.

\providecommand{\kcal}{kcal\,mol$^{-1}$}

\section{QSCI QPU Results}

This section characterises the quantum measurements for the four structures R, T1, TH and E, describes how they enter the configuration space, and compares the resulting energies with HCI. QSCI and HCI are compared on a common ladder of selection thresholds $\epsilon_1$ and at equal determinant count $N$. The benchmark energy for each structure is HCI extrapolated with semistochastic perturbation theory at $\epsilon_2=10^{-6}\,E_\mathrm{h}$ (Supplementary Note~\ref{note:extrapolated_hci}), giving reference barriers relative to R of 25.51 (T1), 27.53 (TH) and 51.88~\kcal{} (E). Classical CI calculations used Dice~\cite{holmes2016heat,sharma2017semistochastic}.

\subsection{Measurements and configuration recovery}\label{note:qsci_sampling}

Each structure was treated with the iterative QSCI protocol of Supplementary Note~\ref{note:qsci} at evolution times $t=1/6$, $1/3$ and $1/2$. The
first time step starts from the reference determinant, while the second and third distribute the circuit budget over the three dominant determinants of the preceding CI solution (HF and two double excitations). Each time step comprises 80 circuit instances of 4096 shots with circuit randomization and local readout-error mitigation (Supplementary Note~\ref{note:errormitigation}), giving $\approx 10^{6}$ shots and $7.5$--$8.2\times10^{5}$ distinct bitstrings per structure.

The fraction of shots in the target particle-number sector falls steeply with evolution time. It is 13.3, 34.4, 28.5 and 5.7\,\% at $t=1/6$ for R, T1, TH and E, but only 0.3--3.1\,\% at $t=1/3$ and $1/2$. Out-of-sector shots are overwhelmingly electron-deficient, where it was found that at the three time steps, 90.2--97.5, 96.2--97.2 and 96.4--98.2\,\% of them carry fewer than 26 electrons, with mean electron counts of 18.7--23.2, 19.8--20.6 and 19.0--20.2. This is consistent with T1 relaxation, where qubits decohere towards $\ket{0}$. The measured populations follow the same trend (main-text Figure~\ref{fig:qsci}c); at $t=1/6$ they retain the HF-like character seen in the ground state occupancies, while at later time steps the distribution flattens. Noting that a fully depolarized state yields a uniform distribution, the shallow first-step circuits therefore provide the most reliable samples. QSCI introduces the time steps in order of circuit depth as its expansion tightens (Supplementary Note~\ref{note:qsci_accept}).

\begin{figure}[t]
  \centering
  \includegraphics[width=0.75\textwidth]{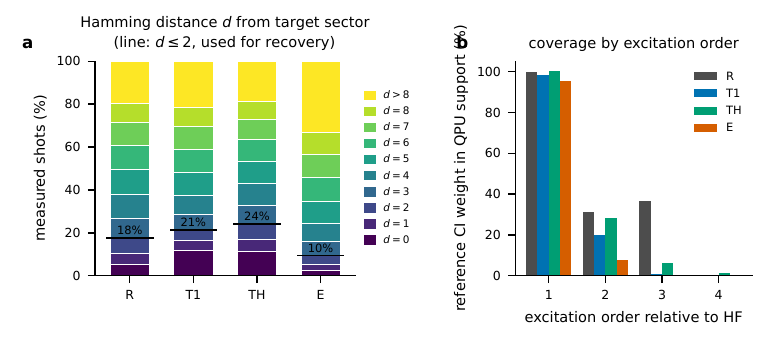}
  \caption{\textbf{QPU measurements and the recovered configuration space.}
  \textbf{a},~Distribution of measured shots over the Hamming distance $d$ from the target particle-number sector; horizontal lines mark the shots with $d\leq2$ used for recovery.
  \textbf{b},~Fraction of the reference-wavefunction weight at each excitation order carried by determinants in the recovered QPU support.}
  \label{sfig:sampling}
\end{figure}

Bitstrings within Hamming distance $d=|\Delta N_\alpha|+|\Delta N_\beta|\leq2$ of the target sector (9.5--24.1\,\% of shots; Supplementary Figure~\ref{sfig:sampling}a) are recovered into it by flipping bits within each spin block~\cite{robledo2025}, separately for each circuit reference. For an excess of $|\Delta N_\sigma|$ electrons, the occupied spin orbitals with the smallest occupancy $o_q$ are emptied. For a deficit, the unoccupied spin orbitals with the largest $o_q$ are filled, where $o_q$ are the occupancies of Equation~\eqref{eq:occ} for that reference, rescaled to span $[0,1]$. Pooled over all time steps, the recovered QPU support contains $3.6$--$5.7\times10^{4}$ determinants and holds 95--100\,\% of the single-excitation weight of the reference wavefunction and 8--31\,\% of its double-excitation weight (Supplementary Figure~\ref{sfig:sampling}b). Per time step, the recovered pools contain $1.0$--$1.7\times10^{4}$ ($t=1/6$), $1.4$--$2.2\times10^{4}$ ($t=1/3$) and $1.2$--$2.2\times10^{4}$ ($t=1/2$) determinants.

\subsection{Occupancy-biased iterative QSCI expansion}\label{note:qsci_accept}\label{note:qsci_occupancy}

The classical post-processing of the measurements follows the iterative QSCI loop of Supplementary Note~\ref{note:recovery}. At each time step, we run two expansion rounds on a threshold ladder that is successively refined along the time schedule. In each round the recovered determinants are appended to the configuration space and then expanded through a heatbath-like selection via a modified Hamiltonian whose one- and two-electron integrals are reweighted by the measured qubit populations of that time step. Each spatial orbital receives the activity factor
\begin{equation}\label{eq:occ_bias}
f_i = \max\!\left[0.2,\; \frac{o_i(1-o_i) + 10^{-4}}{\max_j \left(o_j(1-o_j) + 10^{-4}\right)}\right], \qquad w_i(\lambda) = 1-\lambda\,(1-f_i),
\end{equation}
where $o_i$ is the spin-averaged occupancy of Equation~\eqref{eq:occ} and $\lambda$ sets the strength of the biasing ($\lambda=0$ recovers unbiased heatbath selection). Integrals are scaled by the mean factor of their orbital indices, so that selection favours excitations through orbitals the device finds fractionally occupied. The current space, the biased selection and all recovered configurations are diagonalised together in the exact Hamiltonian.


The selected protocol yields the compactness profile of Supplementary Figure~\ref{sfig:crossover}. In the smallest spaces the result depends on the structure: T1 and TH lie 0.07 and 0.72~\kcal{} below HCI of the same size at $2.0$--$2.3\times10^{3}$ determinants and cross over to above HCI at $3.9$ and $4.6\times10^{3}$ determinants, whereas R and E start 0.77 and 0.76~\kcal{} above HCI. With the $t=1/3$ samples ($\epsilon_1=2^{-11}$ and $2^{-12}\,E_\mathrm{h}$, $0.6$--$2.8\times10^{4}$ determinants) all four structures lie slightly above HCI, by 0.04--0.18~\kcal{}, narrowing to 0.009--0.016~\kcal{} at the tightest threshold. The configurations retained from the device account for 11--37 determinants of the final spaces, and the occupancy reweighting determines which classical excitations are added around them, so that the selection remains distinct from HCI at convergence. Ten multinomial resamplings of the shots of every circuit, propagated through both the recovery and the population-reweighted integrals, change the energies at the two loosest thresholds by at most 0.020~\kcal{} (standard deviation).

A configuration can only be identified from the samples if its weight exceeds roughly the inverse number of in-sector shots, $w_\mathrm{min}\approx1/N_\mathrm{ins}$. At $t=1/6$, $N_\mathrm{ins}=1.9$--$11.4\times10^{4}$, so $w_\mathrm{min}=0.9$--$5.3\times10^{-5}$, and only 164--705 reference determinants per structure are heavier than this. The determinants that heat-bath selection adds at successive thresholds have median weights of $3\times10^{-7}$ ($\epsilon_1=2^{-10}\,E_\mathrm{h}$), $7\times10^{-8}$ ($2^{-11}$), $2\times10^{-8}$ ($2^{-12}$) and $4\times10^{-9}$ ($2^{-13}$), and no determinant added beyond $0.7$--$3.9\times10^{4}$ determinants exceeds $w_\mathrm{min}$. The quantum samples can therefore only identify important configurations in the smaller spaces; at larger $N$ the measured information enters through the orbital occupancies. Extending the configuration-level information to $N\sim10^{5}$ would require $\sim\!10^{3}$ times more in-sector shots, which could come from more shots, lower noise or shallower circuits.

\begin{figure}[t]
  \centering
  \includegraphics[width=\textwidth]{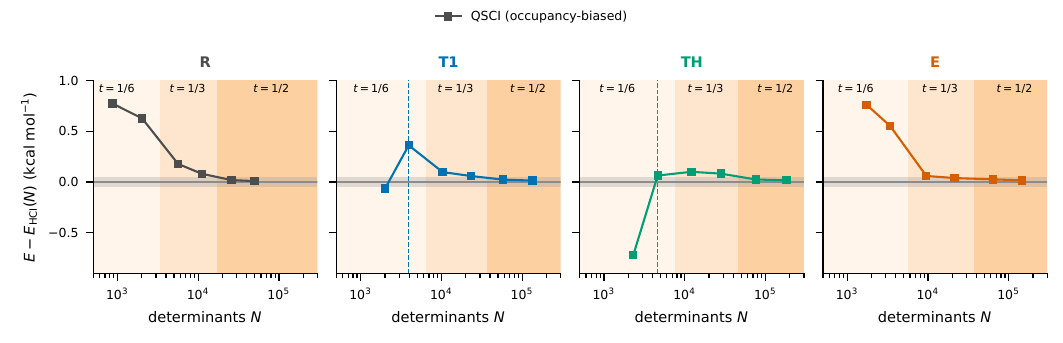}
  \caption{\textbf{Compactness of occupancy-biased QSCI relative to HCI.} Energy relative to HCI of the same size for the selected occupancy-biased iterative QSCI protocol. Shading marks the time step supplying recovered configurations and occupancies; dashed vertical lines mark the crossover to above HCI. The grey band is the point-to-point scatter of the HCI frontier.}
  \label{sfig:crossover}
\end{figure}


\subsection{Correlation character and high-order configurations}\label{note:qsci_highorder}

\begin{figure}[t]
  \centering
  \includegraphics[width=\textwidth]{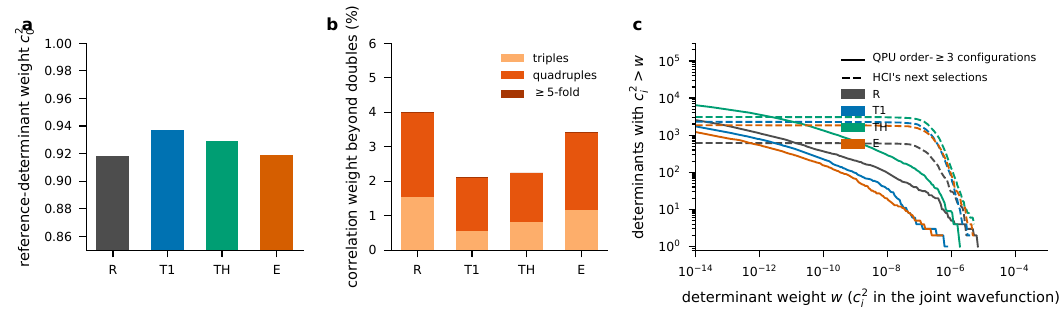}
  \caption{\textbf{Correlation character and weights of high-order configurations.}
  \textbf{a},~Weight of the reference determinant in the reference wavefunction.
  \textbf{b},~Share of the correlation weight $1-c_0^2$ carried by triple, quadruple and higher excitations.
  \textbf{c},~Number of determinants with weight above $w$ in the joint variational wavefunction of $B$, HCI's next selections $P$ (dashed) and the recovered QPU determinants of order $\geq3$ (solid).}
  \label{sfig:correlation}
\end{figure}

The four structures are single-reference and dynamically correlated. The reference determinant carries 92--94\,\% of the reference weight, and only 2.1--4.0\,\% of the correlation weight $1-c_0^2$ lies in triple or higher excitations, mostly quadruples, i.e.\ products of doubles (Supplementary Figure~\ref{sfig:correlation}a,b).

Between 93 and 95\,\% of the recovered QPU determinants are triple or higher excitations, up to 20-fold. To assess them, we start from an HCI space $B$ ($\epsilon_1=2\times10^{-3}\,E_\mathrm{h}$; $1.2$--$2.9\times10^{3}$ determinants) and form two sets. $P$ contains the determinants that HCI would add next, i.e.\ those connected to $B$ with $|H_{ij}c_j|>8\times10^{-4}\,E_\mathrm{h}$, while the second set contains the recovered QPU determinants of excitation order $\geq3$. None of the $3.3$--$5.4\times10^{4}$ high-order QPU determinants is in $P$, so the device reaches configurations that a weighted transition energy selection does not visit at this stage. Diagonalising $B\cup P$ together with all high-order QPU determinants gives every determinant a weight $c_i^2$ in one variational wavefunction (Supplementary Figure~\ref{sfig:correlation}c, Supplementary Table~\ref{stab:value}). A tail of 3--49 high-order QPU determinants per structure reaches the median weight of HCI's own selections ($\sim2\times10^{-7}$), while the large majority carry negligible weight. Together they lower the energy of $B$ by 0.004--0.085~\kcal{}, compared with 0.24--1.39~\kcal{} for the $0.6$--$3.1\times10^{3}$ determinants of $P$.

\begin{figure}[h]
  \centering
  \includegraphics[width=\textwidth]{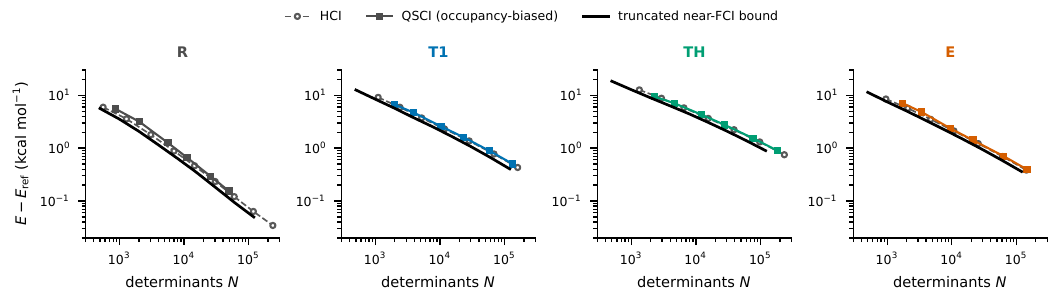}
  \caption{\textbf{Compactness.} Variational energy error versus number of determinants for HCI, occupancy-biased iterative QSCI and the truncated near-FCI bound, for each structure.}
  \label{sfig:compactness}
\end{figure}

The implication of this is that the R, T1, TH and E systems studied in this work are not natural candidates for demonstrating the benefit of quantum sampling in SCI, since the wavefunction is overwhelmingly single-reference and the bulk of its weight lies in low-order excitations. As such, we conclude that HCI is near-optimal in terms of wavefunction compactness. For fixed $N$, the lowest variational energy of any selected CI is approximately bounded by the truncated near-FCI wavefunction: the $N$ determinants of largest $|c_i|$ in a tightly converged HCI wavefunction, re-diagonalised. HCI lies within 0.17--0.88 and 0.05--0.53~\kcal{} of this bound at $N=8\times10^{3}$ and $3\times10^{4}$, respectively (Supplementary Figure~\ref{sfig:compactness}), leaving little room for any selection strategy in these single-reference structures. Occupancy-biased QSCI approaches HCI from above as the expansion grows and remains slightly less compact at convergence, consistent with Supplementary Note~\ref{note:qsci_accept}. Pruning a larger space to its largest coefficients approaches the bound regardless of the method and applies identically to HCI, so for fairness we do not use it in QSCI.

\begin{table}[t]
\centering
\caption{\textbf{Weights of high-order QPU determinants in a joint variational wavefunction.} Energy lowerings are relative to the variational energy of $B$.}
\label{stab:value}
\small
\begin{tabular}{lcccc}
\toprule
 & R & T1 & TH & E \\
\midrule
HCI space $B$ (determinants)                        & 1245 & 2337 & 2858 & 2026 \\
HCI's next selections $P$ (determinants)            & 616 & 2298 & 3101 & 1856 \\
QPU determinants of order $\geq3$                   & 53\,717 & 45\,297 & 50\,780 & 33\,000 \\
\quad of which in $P$                               & 0 & 0 & 0 & 0 \\
\quad of which above median weight of $P$           & 24 & 3 & 49 & 3 \\
energy lowering by $P$ (\kcal)                      & 0.244 & 0.951 & 1.388 & 0.821 \\
energy lowering by QPU order-$\geq3$ (\kcal)        & 0.050 & 0.006 & 0.085 & 0.004 \\
\bottomrule
\end{tabular}
\end{table}

\subsection{Reaction barriers}\label{note:qsci_barrier}

\begin{table}[b]
\centering
\caption{\textbf{Barriers relative to R (\kcal) along the threshold ladder.} $N$ is the size of the QSCI space of the respective structure. All values are variational except the extrapolated reference.}
\label{stab:barrier}
\setlength{\tabcolsep}{3pt}
\begin{tabular}{@{}l rcc rcc rcc@{}}\toprule
 & \multicolumn{3}{c}{T1} & \multicolumn{3}{c}{TH} & \multicolumn{3}{c}{E} \\
\cmidrule(lr){2-4}\cmidrule(lr){5-7}\cmidrule(lr){8-10}
$\epsilon_1$ ($E_\mathrm{h}$) & $N$ & HCI & QSCI & $N$ & HCI & QSCI & $N$ & HCI & QSCI \\
\midrule
$1.95\times 10^{-3}$ & 1997 & 27.87 & 26.63 & 2304 & 32.73 & 31.41 & 1739 & 53.98 & 53.41 \\
$9.77\times 10^{-4}$ & 3930 & 27.44 & 27.03 & 4625 & 31.52 & 31.41 & 3410 & 53.58 & 53.52 \\
$4.88\times 10^{-4}$ & 10336 & 26.94 & 26.84 & 12098 & 30.30 & 30.59 & 9539 & 53.11 & 52.97 \\
$2.44\times 10^{-4}$ & 23284 & 26.42 & 26.45 & 28005 & 29.29 & 29.61 & 21204 & 52.63 & 52.63 \\
$1.22\times 10^{-4}$ & 58051 & 26.06 & 26.11 & 76169 & 28.60 & 28.78 & 63747 & 52.34 & 52.29 \\
$6.10\times 10^{-5}$ & 133992 & 25.82 & 25.86 & 181864 & 28.15 & 28.27 & 145332 & 52.14 & 52.12 \\
\midrule
HF & & \multicolumn{2}{c}{13.96} & & \multicolumn{2}{c}{25.73} & & \multicolumn{2}{c}{54.21} \\
extrapolated HCI (reference) & & \multicolumn{2}{c}{25.51} & & \multicolumn{2}{c}{27.53} & & \multicolumn{2}{c}{51.88} \\
\bottomrule
\end{tabular}
\end{table}

Supplementary Table~\ref{stab:barrier} lists the barriers relative to R along the common threshold ladder, whose convergence is shown in main-text Figure~\ref{fig:qsci}b. At the loosest threshold the QSCI barriers lie 1.23 (T1), 1.32 (TH) and 0.57~\kcal{} (E) closer to the extrapolated reference than those of HCI. Barriers from truncated near-FCI wavefunctions of the same sizes as the QSCI spaces of each structure and of R are not variational bounds, since truncation errors partly cancel in a difference, and the QSCI barriers even lie 0.59, 0.53 and 0.05~\kcal{} below them. This reflects error cancellation across structures rather than more compact wavefunctions: every QSCI energy lies above the truncated near-FCI energy of the same size, but at this threshold the excess is largest for R (1.57~\kcal{}, against 0.98, 1.03 and 1.52~\kcal{} for T1, TH and E), which lowers every barrier measured from R. As the spaces grow, the QSCI barriers of T1 and TH cross to slightly above those of HCI, while those of E remain marginally below them. At the tightest threshold QSCI gives 25.86 (T1), 28.27 (TH) and 52.12~\kcal{} (E), 0.35, 0.75 and 0.24~\kcal{} above the extrapolated reference, compared with 0.31, 0.63 and 0.26~\kcal{} for HCI. This residual is the truncation error of the variational expansion.

\section{Rates of reaction}\label{note:barrier_notes}

The energies of molecular systems determine their thermodynamic stability, while energy differences associated with reaction barriers govern reaction rates and kinetics. In particular, barrier heights determine whether a reaction can occur on a practically relevant timescale. Reaction rates can be estimated using the Eyring equation:

\begin{equation}
k = \frac{k_{\mathrm B}T}{h} e^{\left(-\frac{\Delta G^{\ddagger}}{RT}\right)},
\end{equation}
where $k$ is the rate constant, $T$ is the temperature, $\Delta G^{\ddagger}$ is the entropy of activation, $h$ is the Planck constant, $k_{\mathrm B}$ is the Boltzmann constant and $R$ is the molar gas constant. 
Approximating the activation free energy by the computed barrier height, a barrier of \SI{50}{kcal.mol^{-1}} corresponds to an estimated half-life of approximately $5.21 \times 10^{23}$ seconds under standard conditions, whereas a barrier of \SI{25}{kcal.mol^{-1}} yields an estimated half-life of approximately $2.41 \times 10^{5}$ seconds, as shown in Table \ref{tab:rxt_rate}. Here we assume first-order kinetics, defined as:
\begin{equation} \label{eq:first_order_rate}
[R]_t = [R]_0 e^{-kt},
\end{equation}
with a half-life $t_{1/2}$ defined by

\begin{equation}
\begin{aligned}
\frac{[R]_{0}}{2} &= [R]_0 e^{-kt_{1/2}} \\&\implies \\
\ln{ \bigg( \frac{[R]_{0}}{2 [R]_{0}} \bigg)} &= -k t_{1/2} \\
&\implies \\
t_{1/2} &= \frac{\ln(2)}{k}.
\end{aligned}
\end{equation}
The half-life $t_{1/2}$ represents the time required for the reactant concentration to decrease to half of its initial value $[R]_{t/2} = \frac{[R]_{0}}{2}$. Although the barrier estimates in this work neglect entropic and zero-point energy contributions, and should therefore be interpreted only qualitatively, they clearly illustrate the extreme sensitivity of reaction rates to activation barriers. A reduction in the barrier height from \SI{50}{kcal.mol^{-1}} to \SI{25}{kcal.mol^{-1}} decreases the estimated half-life by approximately seventeen orders of magnitude, highlighting how even modest errors in predicted barrier heights can lead to dramatic differences in the inferred reaction kinetics and, consequently, in conclusions regarding whether a reaction is biologically relevant on observable timescales.

\begin{table}[h]
\centering
\begin{tabular}{ccc}
\hline
$\Delta G^{\ddagger}$ (kcal/mol) & $k$ ($s^{-1}$) & $t_{1/2}$ ($s$) \\ \hline
15 & $62.0$  & $11.2\times 10^{-3}$       \\
25 & $2.88 \times 10^{-6}$  & $2.41 \times 10^{5}$       \\
50 & $1.33\times 10^{-24}$  & $5.21 \times 10^{23}$       \\
\hline
\end{tabular}
\caption{Approximate rate of reactions for different $\Delta G^{\ddagger}$ under ambient conditions assuming a first-order rate law (equation \ref{eq:first_order_rate}) and $T=298K$. \label{tab:rxt_rate}}
\end{table}

\section{Extrapolated Heatbath Confinguration Interaction Benchmark Energies}\label{note:extrapolated_hci}

To assess the accuracy of the iterative QSCI implementation detailed in Sections \ref{note:stoch-circ} and \ref{note:recovery}, we perform semistochastic heatbath CI (SHCI) \cite{sharma2017semistochastic} calculations as an energy benchmark against which to compare. SHCI consists of two stages: computation of a variational energy $E_{\rm var}$ within a configuration subspace $\mathcal{D}$, followed by a second-order perturbation correction $E_{\rm pt2}$ that captures correlations outside the subspace. The variational space is built iteratively; if $\mathcal{D}$ is the current space and $\ket{\Psi}=\sum_n \alpha_n \ket{n}$ is the CI solution in $\mathcal{D}$, we append any configuration $\ket{m} \not\in\mathcal{D}$ that satisfies the screening criterion $ |H_{mn} \alpha_n | > \varepsilon_1$ for some element $\ket{n} \in \mathcal{D}$. $\varepsilon_1 > 0$ is the screening threshold and controls the accuracy of the variational energy, noting that as $\varepsilon_1 \rightarrow 0$ the FCI energy is eventually recovered.

The semistochastic designation of SHCI enters in the perturbative stage, since the full Epstein-Nesbet expansion \cite{song2020multi}
\begin{equation}
    E_{\rm enpt2} = -\sum_{\ket{m} \not\in \mathcal{D}} \frac{\left\lvert \sum_{n}\alpha_n H_{mn} \right\rvert^2}{H_{mm} - E_{\rm var}}
\end{equation}
is prohibitively memory intensive for large configuration subspaces. To circumvent this, SHCI samples elements from the expansion to form an unbiased estimator of the perturbation correction. A second parameter $\varepsilon_2 > 0$ defines a subset of terms in the expansion that are always evaluated, using the same criterion as the variational inclusion $|H_{mn}\alpha_n|>\varepsilon_2$. The combination of this deterministic portion of the perturbation expansion, supplemented with sampled corrections in the tail of the distribution, is where the semistochastic name comes from.

Following this procedure, we obtain a corrected energy $E_{\rm SHCI} = E_{\rm var} + E_{\rm pt2}$, noting that as the configuration subspace expands, we will have $E_{\rm var} \rightarrow E_0$ and thus $E_{\rm pt2} \rightarrow 0$. Moreover, up to second order the error with respect to the true ground state energy is $\left\lvert E_0 - E_{\rm SHCI} \right\rvert  = \mathcal{O}(E_{\rm pt2}^2)$. Therefore, as the correction term approaches zero the relationship between $E_{\rm var}$ and $E_{\rm pt2}$ is approximately linear and motivates an extrapolation scheme. For varying small $\varepsilon_1>0$, we evaluate $E_{\rm var}^{(\varepsilon_1)}$, $E_{\rm pt2}^{(\varepsilon_1)}$ and perform linear regression to approximate $E_{\rm var}$ as $E_{\rm pt2}$ is extrapolated to zero. This can be viewed below in Figure \ref{fig:extrapolated_hci} for the R, T1, TH and E structures studied in this work, with exact values in Table \ref{tab:extrapolated_hci}.

\begin{figure}[t]
  \centering
  \begin{subfigure}[t]{0.48\textwidth}
    \centering
    \includegraphics[width=0.9\textwidth]{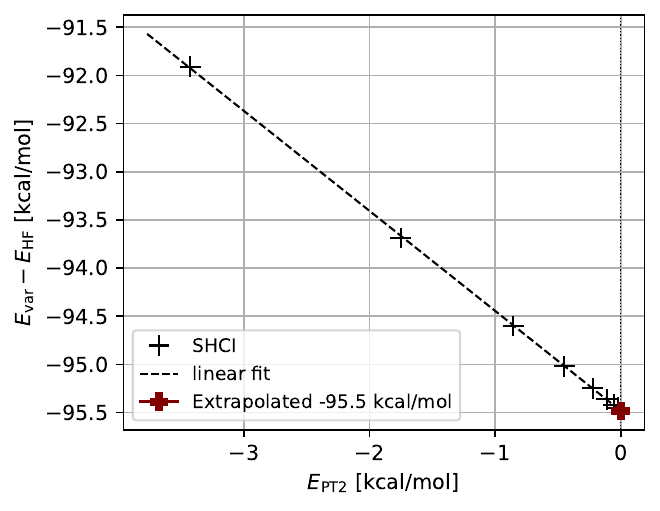}
    \caption{R system}
    \label{fig:grid:aa}
  \end{subfigure}
  \hfill
  \begin{subfigure}[t]{0.48\textwidth}
    \centering
    \includegraphics[width=0.9\textwidth]{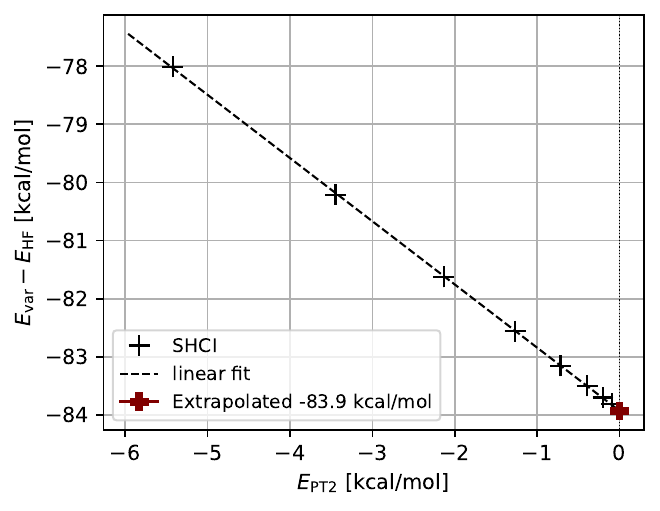}
    \caption{T1 system}
    \label{fig:grid:bb}
  \end{subfigure}
  \vspace{0.6em}
  \begin{subfigure}[t]{0.48\textwidth}
    \centering
    \includegraphics[width=0.9\textwidth]{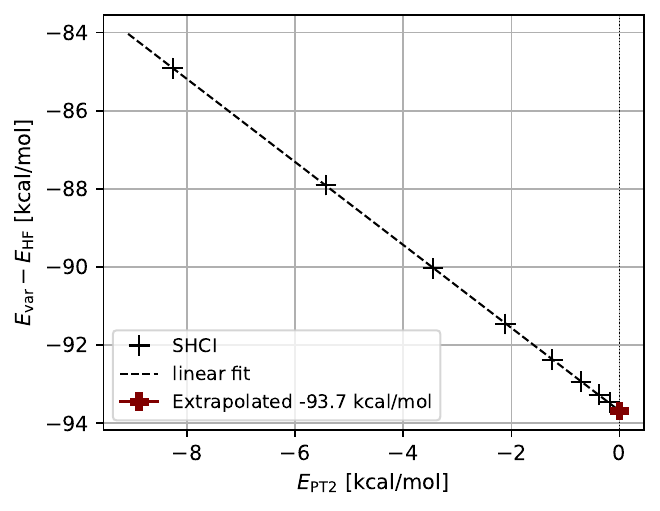}
    \caption{TH system}
    \label{fig:grid:cc}
  \end{subfigure}
  \hfill
  \begin{subfigure}[t]{0.48\textwidth}
    \centering
    \includegraphics[width=0.9\textwidth]{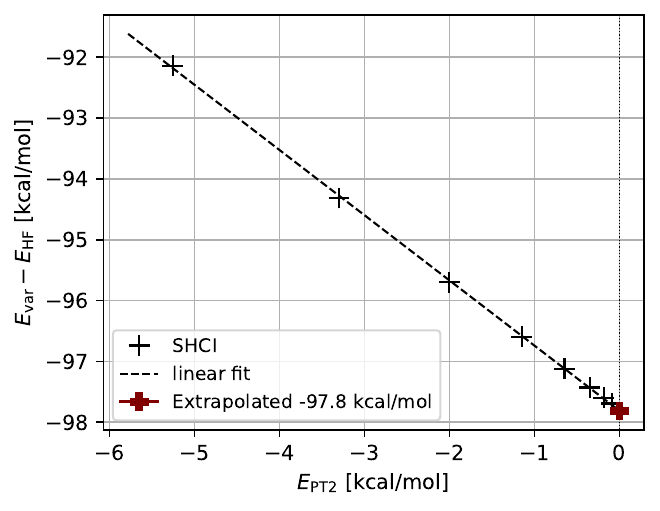}
    \caption{E system}
    \label{fig:grid:dd}
  \end{subfigure}
  \caption{Extrapolated SHCI correlation energies against perturbation correction with screening parameter $1.5 \times 10^{-5} \leq \varepsilon_1 \leq 2.0 \times 10^{-3}$ and $\varepsilon_2 = 10^{-6}$ for the \textbf{(a)} R \textbf{(b)} T1 \textbf{(c)} TH \textbf{(d)} E systems. See Table \ref{tab:extrapolated_hci} for exact values.}
  \label{fig:extrapolated_hci}
\end{figure}

\begin{table}[t]
  \centering
  \begin{subtable}[t]{0.48\textwidth}
    \centering
    \begin{tabular}{S[table-format=1.1e-1] S[table-format=6.0] S[table-format=-2.2] S[table-format=-1.2]}
      \toprule
      {$\epsilon_1$} & {$N_{\mathrm{det}}$} & {$E_{\mathrm{corr}}$} & {$E_{\mathrm{pt2}}$} \\
      \midrule
        2.0e-03 &    1284 & -91.912 & -3.428 \\
       9.8e-04 &    3058 & -93.687 & -1.751 \\
       4.9e-04 &    6984 & -94.602 & -0.855 \\
       2.4e-04 &   14519 & -95.020 & -0.451 \\
       1.2e-04 &   30300 & -95.250 & -0.224 \\
       6.1e-05 &   59727 & -95.362 & -0.112 \\
       3.1e-05 &  118124 & -95.420 & -0.055 \\
       1.5e-05 &  236674 & -95.449 & -0.026 \\
       {--}    &    {--} & \textbf{-95.483} & \textbf{0.000} \\
    
      \bottomrule
    \end{tabular}
    \caption{R, $E_{\rm HF}=-2691448.375$ kcal/mol}
    \label{tab:hci:grid:R}
  \end{subtable}
  \hfill
  \begin{subtable}[t]{0.48\textwidth}
    \centering
    \begin{tabular}{S[table-format=1.1e-1] S[table-format=6.0] S[table-format=-2.2] S[table-format=-1.2]}
      \toprule
      {$\epsilon_1$} & {$N_{\mathrm{det}}$} & {$E_{\mathrm{corr}}$} & {$E_{\mathrm{pt2}}$} \\
      \midrule
       2.0e-03 &    2401 & -78.008 & -5.422 \\
       9.8e-04 &    5213 & -80.209 & -3.447 \\
       4.9e-04 &   11774 & -81.624 & -2.135 \\
       2.4e-04 &   28300 & -82.562 & -1.268 \\
       1.2e-04 &   68512 & -83.156 & -0.717 \\
       6.1e-05 &  159353 & -83.500 & -0.393 \\
       3.1e-05 &  358800 & -83.699 & -0.202 \\
       1.5e-05 &  778159 & -83.806 & -0.097 \\
       {--}    &    {--} & \textbf{-83.930} & \textbf{0.000} \\

      \bottomrule
    \end{tabular}
    \caption{T1, $E_{\rm HF}=-2691434.413$ kcal/mol}
    \label{tab:hci:grid:T1}
  \end{subtable}

  \vspace{1em}

  \begin{subtable}[t]{0.48\textwidth}
    \centering
    \begin{tabular}{S[table-format=1.1e-1] S[table-format=7.0] S[table-format=-2.2] S[table-format=-1.2]}
      \toprule
      {$\epsilon_1$} & {$N_{\mathrm{det}}$} & {$E_{\mathrm{corr}}$} & {$E_{\mathrm{pt2}}$} \\
      \midrule
        2.0e-03 &    2932 & -84.909 & -8.265 \\
       9.8e-04 &    6519 & -87.901 & -5.431 \\
       4.9e-04 &   15564 & -90.036 & -3.452 \\
       2.4e-04 &   38937 & -91.460 & -2.111 \\
       1.2e-04 &   97355 & -92.378 & -1.246 \\
       6.1e-05 &  233914 & -92.937 & -0.712 \\
       3.1e-05 &  546643 & -93.273 & -0.381 \\
       1.5e-05 & 1000000 & -93.466 & -0.185 \\
       {--}    &    {--} & \textbf{-93.687} & \textbf{0.000} \\

      \bottomrule
    \end{tabular}
    \caption{TH, $E_{\rm HF}=-2691422.645$ kcal/mol}
    \label{tab:hci:grid:TH}
  \end{subtable}
  \hfill
  \begin{subtable}[t]{0.48\textwidth}
    \centering
    \begin{tabular}{S[table-format=1.1e-1] S[table-format=6.0] S[table-format=-2.2] S[table-format=-1.2]}
      \toprule
      {$\epsilon_1$} & {$N_{\mathrm{det}}$} & {$E_{\mathrm{corr}}$} & {$E_{\mathrm{pt2}}$} \\
      \midrule
      2.0e-03 &    2085 & -92.141 & -5.255 \\
       9.8e-04 &    4720 & -94.315 & -3.300 \\
       4.9e-04 &   10783 & -95.699 & -2.003 \\
       2.4e-04 &   25656 & -96.594 & -1.149 \\
       1.2e-04 &   61406 & -97.120 & -0.648 \\
       6.1e-05 &  142732 & -97.429 & -0.350 \\
       3.1e-05 &  322198 & -97.597 & -0.186 \\
       1.5e-05 &  709518 & -97.691 & -0.092 \\
       {--}    &    {--} & \textbf{-97.813} & \textbf{0.000} \\

      \bottomrule
    \end{tabular}
    \caption{E, $E_{\rm HF}=-2691394.167$ kcal/mol}
    \label{tab:hci:grid:E}
  \end{subtable}

  \caption{SHCI $\epsilon_1$ scans with $\varepsilon_2 = 10^{-6}$, values reported in kcal/mol. Correlation energy given as $E_{\mathrm{corr}}=E_{\mathrm{SHCI}}-E_{\mathrm{HF}}$. These tables correspond with the extrapolation plots in Figure \ref{fig:extrapolated_hci}.}
  \label{tab:extrapolated_hci}
\end{table}

\section{Details on the GPU-Accelerated Post-Processing } \label{note:postprocessing}

    The iterative QSCI routine comprised repeated Hamiltonian diagonalization steps based on the bitstrings measured. These were decoded as electronic configurations, filtered to the target particle-number and spin sector, and accumulated into the selected space. The Hamiltonian was then diagonalised within this space using \texttt{distributed\_eigsh}, a matrix-free distributed thick-restart Lanczos eigensolver. The solver removes the need to construct or store the selected-space Hamiltonian in compressed sparse form. For the projected Hamiltonian \begin{equation} H_{\mathcal{S}} = P_{\mathcal{S}} H P_{\mathcal{S}},\end{equation} the selected configurations and Pauli operators are encoded as compact bit masks, and a custom CUDA kernel evaluates each matrix--vector product 
\begin{equation} 
\mathbf{y} = H_{\mathcal{S}}\mathbf{x}
\end{equation}
on demand using XOR transformations, population-count phase evaluation and GPU-resident hash-table lookups. Across $G$ GPUs, the rows of $H_{\mathcal{S}}$ and the Lanczos vectors are partitioned into contiguous blocks, reducing the dominant Lanczos-storage requirement from $O(rD)$ to $O(rD/G)$, where $D$ is the selected-space dimension and $r$ is the classical Lanczos dimension. NCCL \texttt{AllGather} operations provide each rank with the complete input vector required to evaluate its independently owned output rows, while \texttt{AllReduce} operations synchronise the global inner products, norms, reorthogonalisation coefficients and convergence criterion. This design replaces the memory-intensive materialisation of all nonzero Hamiltonian entries with repeated, highly parallel operator application, enabling selected-space eigensolutions beyond the memory capacity of a single accelerator while preserving numerical equivalence to explicit sparse-matrix diagonalization.

\begin{figure}[t]
    \centering
    \includegraphics[width=\textwidth]{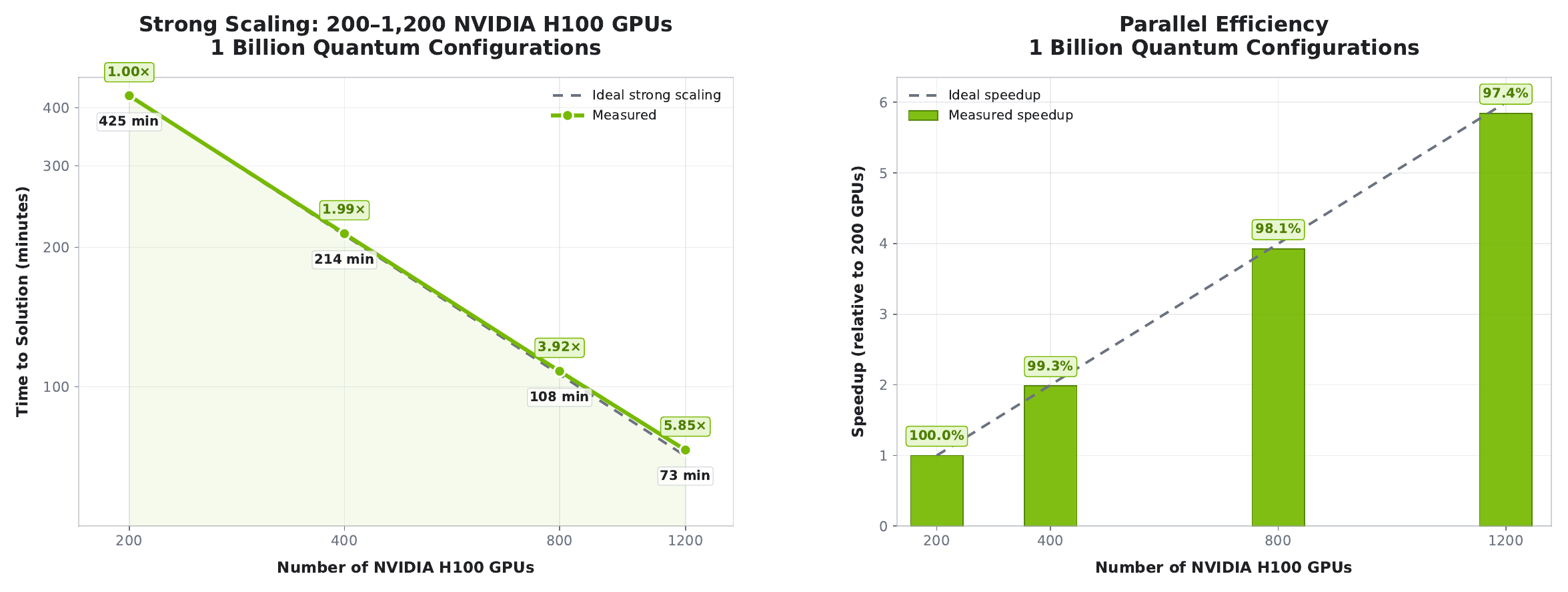}    \caption{\textbf{Strong scaling of the GPU-accelerated QSCI post-processing.} \textbf{a}, Time to solution for a selected space of one
    billion quantum configurations, from 200 to 1200 NVIDIA H100 GPUs on the
    NVIDIA Eos cluster, compared with ideal strong scaling. \textbf{b}, Speedup
    relative to 200 GPUs, with parallel efficiencies annotated.}
    \label{fig:scaling}
\end{figure}

To characterise how the GPU-accelerated subspace diagonalization scales, the configuration interaction was executed on a selected space of one billion quantum configurations across increasing GPU counts on the NVIDIA Eos cluster (Figure \ref{fig:scaling}). Time to solution falls from \SI{425}{min} on 200 NVIDIA H100 GPUs to \SI{73}{min} on 1200, tracking the ideal strong-scaling curve across the full range (Figure \ref{fig:scaling}a). The corresponding speedups relative to 200 GPUs are $1.99\times$ at 400, $3.92\times$ at 800 and $5.85\times$ at 1200 (Figure \ref{fig:scaling}b), giving parallel efficiencies of 99.3\%, 98.1\% and 97.4\%. The matrix-free formulation therefore retains near-ideal efficiency at a sixfold increase in device count, with the modest departure at 1200 GPUs consistent with the growing share of collective communication as the per-device workload shrinks. This behaviour determines the size of selected space that the workflow can treat: because the dominant Lanczos storage requirement is distributed across devices, the accessible subspace dimension grows with the available GPU count rather than with the memory of a single accelerator.

\section{Details on the Hybrid HPC--QC Setup}
\label{note:hpcqc}
At the system level, integration is realized through MQSS~\cite{burgholzer2026}, which provides middleware components for job scheduling and resource allocation via a SLURM integration plugin~\cite{yoo2003,farooqi2026enahpcqcworkflow} in which a QPU is defined as a logical resource on each compute node (\texttt{--gres=qpu:1}). SLURM serves as the sole resource manager for the hybrid workflow. On the classical side, it performs its conventional role: allocating compute nodes, CPU cores, and GPUs; launching the distributed-memory execution, so that the degree of parallelism of the post-processing stage is set entirely by the job specification (number of nodes, tasks per node, and GPUs per node). Communication between GPUs within and across nodes is handled by MPI over InfiniBand~\cite{macarthur2017mpiinfiniband}. 

\begin{figure}[t]
    \centering
    \includegraphics[width=\textwidth]{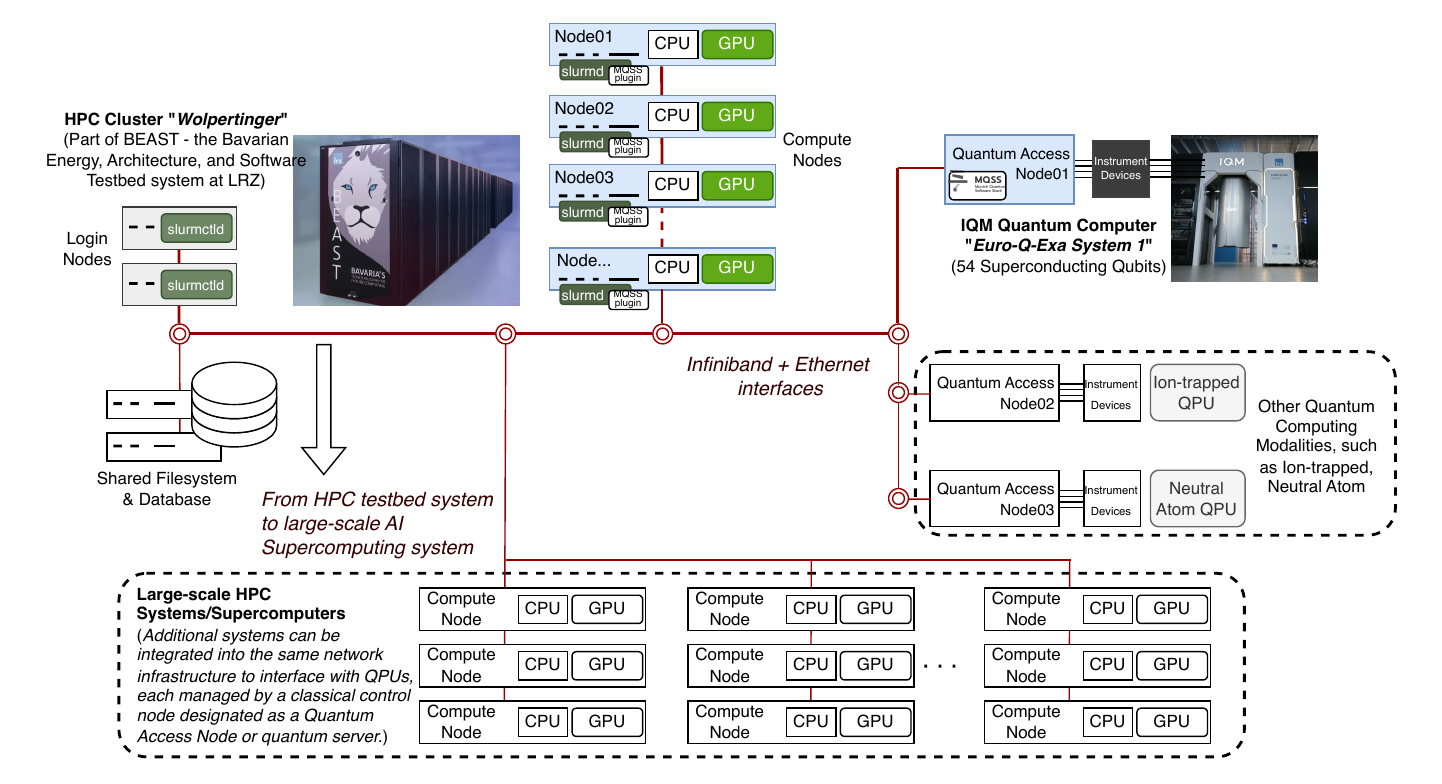}    \caption{\textbf{High-level overview of the hybrid HPC-quantum network architecture.} The shared InfiniBand and Ethernet interconnect enables HPC systems (Wolpertinger within BEAST - the Bavarian Energy, Architecture, and Software Testbed system at LRZ) to interface with quantum processors (Euro-Q-Exa System 1) via dedicated classical host instances, designated as Quantum Access Nodes (QANs). SLURM serves as the HPC workload manager, with the controller daemon (\texttt{slurmctld}) residing on the login nodes and compute daemons (\texttt{slurmd}) on the compute nodes. While CPUs and GPUs are managed as native hardware, the QPU is abstracted as a logical resource, allowing users to request exclusive, coordinated access in a single job script. Utilizing the shared network fabric, compute nodes can connect and offload quantum tasks seamlessly to the quantum computers through QAN.}
    \label{fig:hpcqc-system}
\end{figure}

On the quantum side, the MQSS plugin registers the QPU as a co-processor attached to the compute nodes, in the same way an accelerator is exposed to the scheduler. Physically, the IQM quantum device is tethered to a dedicated classical host, which we name the Quantum Access Node (QAN). QAN can also be considered a classical quantum server that runs the vendor-specific drivers and control software necessary to operate the hardware. As illustrated in Figure~\ref{fig:hpcqc-system}, this QAN is integrated into the same network infrastructure demonstrated in this work as the Wolpertinger cluster within BEAST. The architecture utilizes dual network interfaces: Ethernet for external connectivity and InfiniBand for high-throughput inter-node communication (e.g., MPI applications). Consequently, submitting tasks to the QPU triggers a background offloading process. The MQSS runtime handles this offloading by transmitting quantum circuits and associated execution data directly from the classical compute nodes to the Quantum Access Node over the shared InfiniBand fabric. Requesting \texttt{--gres=qpu:1} therefore reserves coordinated access to the device for the lifetime of the job, while the circuit submission and measurement result retrieval are handled by the MQSS at runtime. The practical consequence is that both classical and quantum resources are requested in a single job script. The QPU is co-scheduled with the CPU--GPU resources that prepare its input and consume its output.

The workflow is organised into five steps, each executed on the appropriate hardware resource: (1) environment setup for the full biomolecular system and selection of the representative geometry for each reaction-path state, (2) projection-based embedding and Hamiltonian construction, (3) classical benchmarks providing reference correlated-wavefunction calculations, (4) qDRIFT circuit compilation and submission to the quantum processor, and (5) GPU post-processing for subspace diagonalization to extract the final correlated energies. Steps (1)--(3) are executed on the CPU--GPU nodes, step (4) is offloaded to the QPU, and step (5) requires GPU-accelerated computing and can be scaled to thousands of GPUs. Each Wolpertinger node features two Intel Xeon Platinum 8360Y CPU sockets, each with 36 physical cores, and one NVIDIA A100 GPU with 80\,GB of memory. The nodes are directly connected to the IQM Emerald QPU, a 54-qubit device designated Euro-Q-Exa (EQE System 1). The classical computation runs on Wolpertinger nodes while the qDRIFT circuits are submitted to the QPU; the resulting bitstring counts are written to the shared filesystem, and a post-processing run produces correlated energies.

Distributed post-processing was carried out on up to 16 nodes of the LRZ AI cluster operated within the BayernKI consortium~\cite{bayernki2025}, each equipped with four NVIDIA H100 GPUs with 94\,GB of memory, for a total of 64 GPUs. 
This scaling is expressed through the SLURM allocation: the same post-processing code is launched across a requested number of nodes and GPUs per node, with one MPI rank bound to each GPU, so that moving from a single node to 16 nodes (64 GPUs) requires a change of the resource request.
Deployment on distributed-memory machines accelerates the configuration expansion and subspace diagonalization steps, which are computationally intensive owing to the large number of configurations sampled on the quantum processor. Both systems are co-located and connected via InfiniBand to minimize communication overhead. To characterize scaling behavior further, the post-processing computation was extracted and scaled separately to 1200 NVIDIA H100 GPUs on the NVIDIA Eos cluster~\cite{boyle2024nvidia}.

\section{Details on Error Mitigation} 
\label{note:errormitigation}

Circuit execution for the GPCR reaction-path system was performed on the 54-qubit IQM superconducting quantum processor hosted on-premises at the Leibniz Supercomputing Centre (LRZ), using the 52 best-calibrated qubits for the required 26-orbital active space. Three complementary error-reduction techniques, implemented via IQM's \texttt{error-reduction} \cite{iqm-error-reduction-tools} tool library and accessed through IQM's Pulla framework, were benchmarked individually and in combination against unmitigated (raw) circuit outputs.

The first method is readout error mitigation (REM), applied to single-qubit occupation probabilities that bias the population-guided configuration recovery step of the iterative QSCI expansion. Under Jordan-Wigner mapping, the occupation probability of qubit $i$ is a single-qubit $Z$-type $n_i = (1-\langle Z_i\rangle)/2$. Rather than characterizing the full joint measurement-outcome distribution -- computationally impractical at this qubit count -- each qubit's readout assignment probabilities, $P(\mathrm{measured}=i \mid \mathrm{prepared}=j)$ for $i,j \in \{0,1\}$, were characterized independently by preparing and measuring $|0\rangle$ and $|1\rangle$ on every qubit simultaneously, and the resulting single-qubit assignment probabilities were inverted to correct each qubit's own measured occupation probability entering $n_i$. This scales as $\mathcal{O}(n_\mathrm{qubits})$, independent of any correlations between qubits, in contrast to joint mitigation schemes whose cost grows exponentially with the number of qubits.

The second is dynamical decoupling (DD), which inserts pulse sequences into otherwise idle qubit periods during circuit execution, suppressing the effects of dephasing. We use the toolbox's default strategy, which selects among three gate-level sequences according to each qubit's idle duration: a center-aligned $XX$ type for short idle windows, a left-aligned $YXYX$ type for medium windows, and a left-aligned eight-pulse $XYXYYXYX$ type for the longest windows. Each sequence is logically equivalent to the identity when executed on a noiseless backend, so that only the physical noise accumulated during the idle window is affected. DD, therefore, incurs no additional overhead beyond the gates themselves.

The third is circuit randomization (CR), which randomizes each circuit by adding a $\pi$ rotation to the polar angle of pseudo-randomly selected existing single-qubit $R(\theta,\phi)$ gates, with the resulting virtual-$Z$ phase corrections propagated forward through the remaining gates so that every randomized circuit stays logically equivalent to the original. Because no additional two-qubit gates are introduced, the circuit depth and gate count are essentially unchanged. Measurement outcomes were averaged over $N=8$ independently randomized instances per circuit. Randomization tends to turn coherent gate errors into a stochastic noise channel, whose accumulated effect on the sampled quantities is typically smaller than that of the coherent errors. This channel can still carry a systematic bias, so averaging over randomizations reduces the statistical spread, but is not expected to recover the ideal output exactly. We therefore assess the benefit of CR empirically.
Each circuit's $4096$ shots were divided equally among the $N=8$ randomized instances ($512$ shots each), so the total shot budget per circuit matches the unrandomized runs. At each QSCI iteration $80$ circuit instances are sampled, each with $4096$ shots in total.

\begin{figure}[h]
    \centering
    \includegraphics[width=0.85\linewidth]{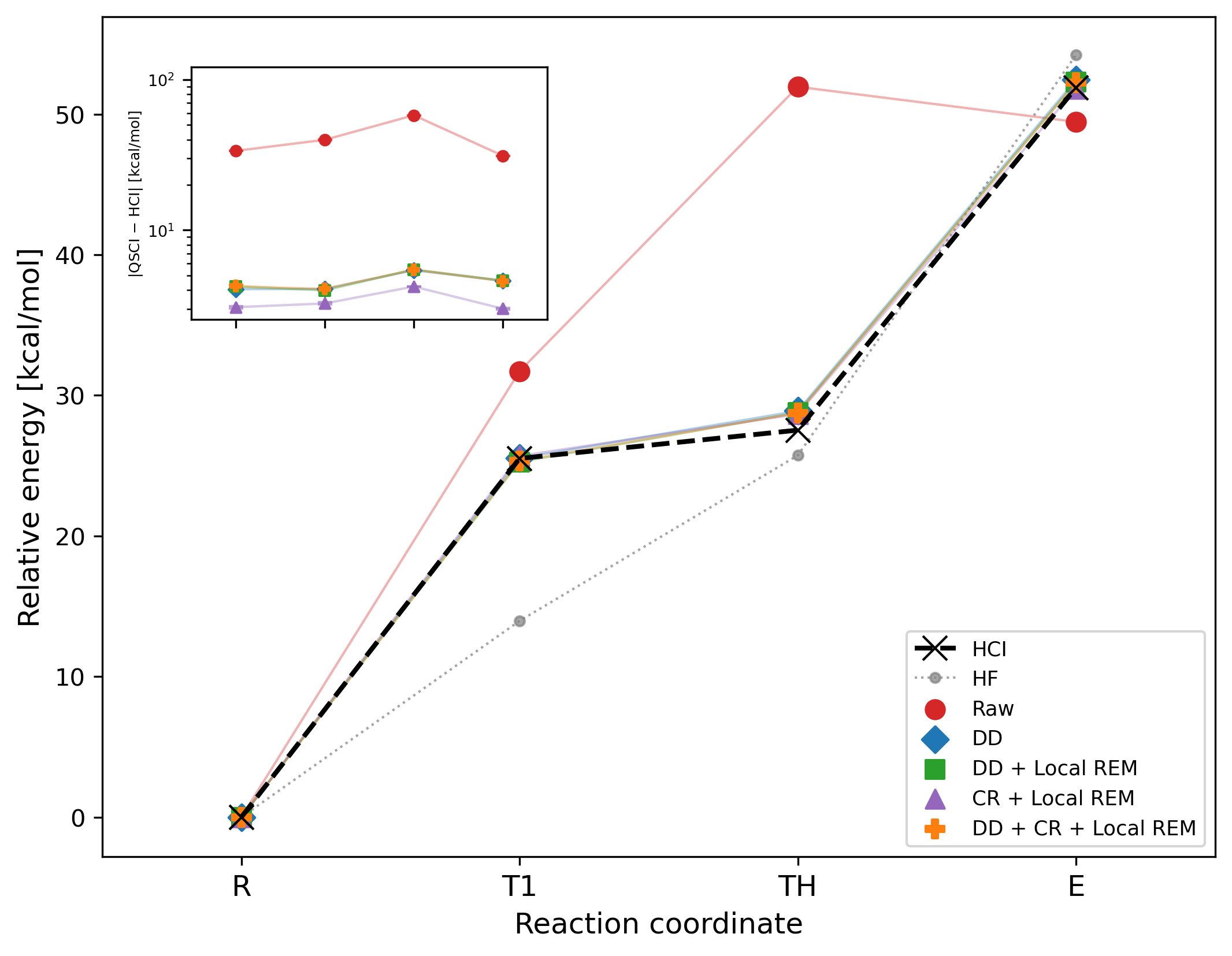}
    \caption{Ground-state energy along the aGPCR R$\to$E reaction coordinate (R, T1, TH, E), relative to R, for the raw circuit and combinations of circuit randomization (CR), dynamical decoupling (DD), and readout error mitigation (REM), against the HCI reference (dashed) and the HF reference (dotted). Inset: absolute error against the HCI reference per structure in log scale. All circuits were run at $\epsilon=0.1$.}
    \label{fig:error_reduction}
\end{figure}

We compare the resulting ground-state energy along the reaction coordinates (R, T1, TH, E) for the raw circuits and combinations of REM, DD, and CR against the HCI reference in Figure \ref{fig:error_reduction}. The main panel shows the energies relative to R, while the inset shows the absolute error $|E - E_{\mathrm{HCI}}|$ in the total energy of each structure. Energies are taken from the final completed QSCI iteration for each structure. The raw circuits show the largest deviation from the HCI reference. DD alone brings a substantial improvement, and adding REM or combining all three techniques together gives comparable but no further gains. For DD+REM and DD+CR+REM, the $R$ point is not converged, as the classical configuration-recovery post-processing did not complete within the available compute time; the value shown is from the last completed iteration. CR combined with REM instead gives the lowest errors, $3.0$-$4.2$, kcal/mol across all four structures, roughly an order of magnitude below the raw circuits.

\clearpage

\bibliography{sn-bibliography}

%% file: authors.tex
\author[1,2,3]{\fnm{Alexander} \sur{Orlov}}
\author[2,4]{\fnm{Alexis} \sur{Ralli}}
\author[2,4]{\fnm{Tim} \sur{Weaving}}
\author[2]{\fnm{Michael} \sur{Williams de la Bastida}}
\author[2,5]{\fnm{Reuben} \sur{Martin}}
\author[6]{\fnm{Zohim} \sur{Chandani}}
\author[3,7]{\fnm{Minh} \sur{Chung}}
\author[8]{\fnm{P. V.} \sur{Sriluckshmy}}
\author[8]{\fnm{Fedor} \sur{Šimkovic IV}}
\author[3]{\fnm{Luigi} \sur{Iapichino}}
\author[9]{\fnm{Peter J.} \sur{Love}}
\author[1]{\fnm{Christian B.} \sur{Mendl}}
\author[2]{\fnm{Andrea} \sur{Townsend-Nicholson}}
\author*[2]{\fnm{Peter V.} \sur{Coveney}}
\email{p.v.coveney@ucl.ac.uk}

\affil[1]{%
\orgname{Technical University of Munich},
\orgaddress{\city{Munich}, \country{Germany}}}

\affil[2]{%
\orgname{University College London},
\orgaddress{\city{London}, \country{UK}}}

\affil[3]{%
\orgname{Leibniz Supercomputing Centre of the Bavarian Academy of Sciences and Humanities},
\orgaddress{\city{Garching}, \country{Germany}}}

\affil[4]{%
\orgname{QMatter, Inc.},
\orgaddress{\city{Newark}, \state{DE}, \country{USA}}}

\affil[5]{%
\orgname{Evotec UK Ltd},
\orgaddress{\city{Abingdon}, \country{UK}}}

\affil[6]{%
\orgname{NVIDIA Corp},
\orgaddress{\city{Santa Clara}, \state{CA}, \country{USA}}}

\affil[7]{%
\orgname{Ludwig-Maximilians-Universität München},
\orgaddress{\city{Munich}, \country{Germany}}}

\affil[8]{%
\orgname{IQM Quantum Computers},
\orgaddress{\city{Munich}, \country{Germany}}}

\affil[9]{%
\orgname{University of Toronto},
\orgaddress{\city{Toronto}, \country{Canada}}}

%% file: main.bbl

\begin{thebibliography}{69}
\ifx \bisbn   \undefined \def \bisbn  #1{ISBN #1}\fi
\ifx \binits  \undefined \def \binits#1{#1}\fi
\ifx \bauthor  \undefined \def \bauthor#1{#1}\fi
\ifx \batitle  \undefined \def \batitle#1{#1}\fi
\ifx \bjtitle  \undefined \def \bjtitle#1{#1}\fi
\ifx \bvolume  \undefined \def \bvolume#1{\textbf{#1}}\fi
\ifx \byear  \undefined \def \byear#1{#1}\fi
\ifx \bissue  \undefined \def \bissue#1{#1}\fi
\ifx \bfpage  \undefined \def \bfpage#1{#1}\fi
\ifx \blpage  \undefined \def \blpage #1{#1}\fi
\ifx \burl  \undefined \def \burl#1{\textsf{#1}}\fi
\ifx \doiurl  \undefined \def \doiurl#1{\url{https://doi.org/#1}}\fi
\ifx \betal  \undefined \def \betal{\textit{et al.}}\fi
\ifx \binstitute  \undefined \def \binstitute#1{#1}\fi
\ifx \binstitutionaled  \undefined \def \binstitutionaled#1{#1}\fi
\ifx \bctitle  \undefined \def \bctitle#1{#1}\fi
\ifx \beditor  \undefined \def \beditor#1{#1}\fi
\ifx \bpublisher  \undefined \def \bpublisher#1{#1}\fi
\ifx \bbtitle  \undefined \def \bbtitle#1{#1}\fi
\ifx \bedition  \undefined \def \bedition#1{#1}\fi
\ifx \bseriesno  \undefined \def \bseriesno#1{#1}\fi
\ifx \blocation  \undefined \def \blocation#1{#1}\fi
\ifx \bsertitle  \undefined \def \bsertitle#1{#1}\fi
\ifx \bsnm \undefined \def \bsnm#1{#1}\fi
\ifx \bsuffix \undefined \def \bsuffix#1{#1}\fi
\ifx \bparticle \undefined \def \bparticle#1{#1}\fi
\ifx \barticle \undefined \def \barticle#1{#1}\fi
\bibcommenthead
\ifx \bconfdate \undefined \def \bconfdate #1{#1}\fi
\ifx \botherref \undefined \def \botherref #1{#1}\fi
\ifx \url \undefined \def \url#1{\textsf{#1}}\fi
\ifx \bchapter \undefined \def \bchapter#1{#1}\fi
\ifx \bbook \undefined \def \bbook#1{#1}\fi
\ifx \bcomment \undefined \def \bcomment#1{#1}\fi
\ifx \oauthor \undefined \def \oauthor#1{#1}\fi
\ifx \citeauthoryear \undefined \def \citeauthoryear#1{#1}\fi
\ifx \endbibitem  \undefined \def \endbibitem {}\fi
\ifx \bconflocation  \undefined \def \bconflocation#1{#1}\fi
\ifx \arxivurl  \undefined \def \arxivurl#1{\textsf{#1}}\fi
\csname PreBibitemsHook\endcsname

\bibitem[\protect\citeauthoryear{S{\'a}nchez~Lorente et~al.}{2025}]{sanchezlorente2025}
\begin{barticle}
\bauthor{\bsnm{S{\'a}nchez~Lorente}, \binits{J.}},
\bauthor{\bsnm{Sokolov}, \binits{A.V.}},
\bauthor{\bsnm{Ferguson}, \binits{G.}},
\bauthor{\bsnm{Schi{\"o}th}, \binits{H.B.}},
\bauthor{\bsnm{Hauser}, \binits{A.S.}},
\bauthor{\bsnm{Gloriam}, \binits{D.E.}}:
\batitle{{GPCR} drug discovery: new agents, targets and indications}.
\bjtitle{Nature Reviews Drug Discovery}
\bvolume{24}(\bissue{6}),
\bfpage{458}--\blpage{479}
(\byear{2025})
\doiurl{10.1038/s41573-025-01139-y}
\end{barticle}
\endbibitem

\bibitem[\protect\citeauthoryear{Hamann et~al.}{2015}]{hamann2015}
\begin{barticle}
\bauthor{\bsnm{Hamann}, \binits{J.}},
\bauthor{\bsnm{Aust}, \binits{G.}},
\bauthor{\bsnm{Ara{\c{c}}}, \binits{D.}},
\bauthor{\bsnm{Engel}, \binits{F.B.}},
\bauthor{\bsnm{Formstone}, \binits{C.}},
\bauthor{\bsnm{Fredriksson}, \binits{R.}},
\bauthor{\bsnm{Hall}, \binits{R.A.}},
\bauthor{\bsnm{Harty}, \binits{B.L.}},
\bauthor{\bsnm{Kirchhoff}, \binits{C.}},
\bauthor{\bsnm{Knapp}, \binits{B.}},
\bauthor{\bsnm{Krishnan}, \binits{A.}},
\bauthor{\bsnm{Liebscher}, \binits{I.}},
\bauthor{\bsnm{Lin}, \binits{H.-H.}},
\bauthor{\bsnm{Martinelli}, \binits{D.C.}},
\bauthor{\bsnm{Monk}, \binits{K.R.}},
\bauthor{\bsnm{Peeters}, \binits{M.C.}},
\bauthor{\bsnm{Piao}, \binits{X.}},
\bauthor{\bsnm{Pr{\"o}mel}, \binits{S.}},
\bauthor{\bsnm{Sch{\"o}neberg}, \binits{T.}},
\bauthor{\bsnm{Schwartz}, \binits{T.W.}},
\bauthor{\bsnm{Singer}, \binits{K.}},
\bauthor{\bsnm{Stacey}, \binits{M.}},
\bauthor{\bsnm{Ushkaryov}, \binits{Y.A.}},
\bauthor{\bsnm{Vallon}, \binits{M.}},
\bauthor{\bsnm{Wolfrum}, \binits{U.}},
\bauthor{\bsnm{Wright}, \binits{M.W.}},
\bauthor{\bsnm{Xu}, \binits{L.}},
\bauthor{\bsnm{Langenhan}, \binits{T.}},
\bauthor{\bsnm{Schi{\"o}th}, \binits{H.B.}}:
\batitle{International union of basic and clinical pharmacology.{XCIV}. adhesion {G} protein-coupled receptors}.
\bjtitle{Pharmacological Reviews}
\bvolume{67}(\bissue{2}),
\bfpage{338}--\blpage{367}
(\byear{2015})
\doiurl{10.1124/pr.114.009647}
\end{barticle}
\endbibitem

\bibitem[\protect\citeauthoryear{Ara{\c{c}} et~al.}{2012}]{arac2012}
\begin{barticle}
\bauthor{\bsnm{Ara{\c{c}}}, \binits{D.}},
\bauthor{\bsnm{Boucard}, \binits{A.A.}},
\bauthor{\bsnm{Bolliger}, \binits{M.F.}},
\bauthor{\bsnm{Nguyen}, \binits{J.}},
\bauthor{\bsnm{Soltis}, \binits{S.M.}},
\bauthor{\bsnm{S{\"u}dhof}, \binits{T.C.}},
\bauthor{\bsnm{Brunger}, \binits{A.T.}}:
\batitle{A novel evolutionarily conserved domain of cell-adhesion {GPCRs} mediates autoproteolysis}.
\bjtitle{The EMBO Journal}
\bvolume{31}(\bissue{6}),
\bfpage{1364}--\blpage{1378}
(\byear{2012})
\doiurl{10.1038/emboj.2012.26}
\end{barticle}
\endbibitem

\bibitem[\protect\citeauthoryear{Yeung et~al.}{2020}]{yeung2020}
\begin{barticle}
\bauthor{\bsnm{Yeung}, \binits{J.}},
\bauthor{\bsnm{Adili}, \binits{R.}},
\bauthor{\bsnm{Stringham}, \binits{E.N.}},
\bauthor{\bsnm{Luo}, \binits{R.}},
\bauthor{\bsnm{Vizurraga}, \binits{A.}},
\bauthor{\bsnm{Rosselli-Murai}, \binits{L.K.}},
\bauthor{\bsnm{Stoveken}, \binits{H.M.}},
\bauthor{\bsnm{Yu}, \binits{M.}},
\bauthor{\bsnm{Piao}, \binits{X.}},
\bauthor{\bsnm{Holinstat}, \binits{M.}},
\bauthor{\bsnm{Tall}, \binits{G.G.}}:
\batitle{{GPR56/ADGRG1} is a platelet collagen-responsive {GPCR} and hemostatic sensor of shear force}.
\bjtitle{Proceedings of the National Academy of Sciences}
\bvolume{117}(\bissue{45}),
\bfpage{28275}--\blpage{28286}
(\byear{2020})
\doiurl{10.1073/pnas.2008921117}
\end{barticle}
\endbibitem

\bibitem[\protect\citeauthoryear{Bahi-Buisson et~al.}{2010}]{bahibuisson2010}
\begin{barticle}
\bauthor{\bsnm{Bahi-Buisson}, \binits{N.}},
\bauthor{\bsnm{Poirier}, \binits{K.}},
\bauthor{\bsnm{Boddaert}, \binits{N.}},
\bauthor{\bsnm{Fallet-Bianco}, \binits{C.}},
\bauthor{\bsnm{Specchio}, \binits{N.}},
\bauthor{\bsnm{Bertini}, \binits{E.}},
\bauthor{\bsnm{Caglayan}, \binits{O.}},
\bauthor{\bsnm{Lascelles}, \binits{K.}},
\bauthor{\bsnm{Elie}, \binits{C.}},
\bauthor{\bsnm{Rambaud}, \binits{J.}},
\bauthor{\bsnm{Baulac}, \binits{M.}},
\bauthor{\bsnm{An}, \binits{I.}},
\bauthor{\bsnm{Dias}, \binits{P.}},
\bauthor{\bsnm{Portes}, \binits{V.}},
\bauthor{\bsnm{Moutard}, \binits{M.L.}},
\bauthor{\bsnm{Soufflet}, \binits{C.}},
\bauthor{\bsnm{El~Maleh}, \binits{M.}},
\bauthor{\bsnm{Beldjord}, \binits{C.}},
\bauthor{\bsnm{Villard}, \binits{L.}},
\bauthor{\bsnm{Chelly}, \binits{J.}}:
\batitle{Gpr56-related bilateral frontoparietal polymicrogyria: further evidence for an overlap with the cobblestone complex}.
\bjtitle{Brain}
\bvolume{133}(\bissue{11}),
\bfpage{3194}--\blpage{3209}
(\byear{2010})
\doiurl{10.1093/brain/awq259}
\end{barticle}
\endbibitem

\bibitem[\protect\citeauthoryear{Zhou et~al.}{2026}]{zhou2026}
\begin{barticle}
\bauthor{\bsnm{Zhou}, \binits{Z.}},
\bauthor{\bsnm{Zhong}, \binits{H.}},
\bauthor{\bsnm{Wang}, \binits{H.}},
\bauthor{\bsnm{Wang}, \binits{S.}},
\bauthor{\bsnm{Zhang}, \binits{N.}},
\bauthor{\bsnm{Wang}, \binits{Z.}},
\bauthor{\bsnm{Huang}, \binits{R.}},
\bauthor{\bsnm{Shen}, \binits{K.}}:
\batitle{Targeting adgrg1 in cancer-associated adipocytes reprograms the immunosuppressive microenvironment and enhances immunotherapy responsiveness in triple-negative breast cancer models}.
\bjtitle{BMC Medicine}
(\byear{2026})
\doiurl{10.1186/s12916-026-05049-7}
\end{barticle}
\endbibitem

\bibitem[\protect\citeauthoryear{Manby et~al.}{2012}]{manby2012}
\begin{barticle}
\bauthor{\bsnm{Manby}, \binits{F.R.}},
\bauthor{\bsnm{Stella}, \binits{M.}},
\bauthor{\bsnm{Goodpaster}, \binits{J.D.}},
\bauthor{\bsnm{Miller~III}, \binits{T.F.}}:
\batitle{A simple, exact density-functional-theory embedding scheme}.
\bjtitle{Journal of Chemical Theory and Computation}
\bvolume{8}(\bissue{8}),
\bfpage{2564}--\blpage{2568}
(\byear{2012})
\doiurl{10.1021/ct300544e}
\end{barticle}
\endbibitem

\bibitem[\protect\citeauthoryear{Goodpaster et~al.}{2014}]{goodpaster2014}
\begin{barticle}
\bauthor{\bsnm{Goodpaster}, \binits{J.D.}},
\bauthor{\bsnm{Barnes}, \binits{T.A.}},
\bauthor{\bsnm{Manby}, \binits{F.R.}},
\bauthor{\bsnm{Miller~III}, \binits{T.F.}}:
\batitle{Accurate and systematically improvable density functional theory embedding for correlated wavefunctions}.
\bjtitle{The Journal of Chemical Physics}
\bvolume{140}(\bissue{18}),
\bfpage{18}--\blpage{507}
(\byear{2014})
\doiurl{10.1063/1.4864040}
\end{barticle}
\endbibitem

\bibitem[\protect\citeauthoryear{Ralli et~al.}{2024}]{ralli2024}
\begin{barticle}
\bauthor{\bsnm{Ralli}, \binits{A.}},
\bauthor{\bsnm{Bastida}, \binits{M.}},
\bauthor{\bsnm{Coveney}, \binits{P.V.}}:
\batitle{Scalable approach to quantum simulation via projection-based embedding}.
\bjtitle{Physical Review A}
\bvolume{109},
\bfpage{022418}
(\byear{2024})
\doiurl{10.1103/PhysRevA.109.022418}
\end{barticle}
\endbibitem

\bibitem[\protect\citeauthoryear{Warshel and Levitt}{1976}]{warshel1976}
\begin{barticle}
\bauthor{\bsnm{Warshel}, \binits{A.}},
\bauthor{\bsnm{Levitt}, \binits{M.}}:
\batitle{Theoretical studies of enzymic reactions: dielectric, electrostatic and steric stabilization of the carbonium ion in the reaction of lysozyme}.
\bjtitle{Journal of Molecular Biology}
\bvolume{103}(\bissue{2}),
\bfpage{227}--\blpage{249}
(\byear{1976})
\doiurl{10.1016/0022-2836(76)90311-9}
\end{barticle}
\endbibitem

\bibitem[\protect\citeauthoryear{Senn and Thiel}{2009}]{senn2009}
\begin{barticle}
\bauthor{\bsnm{Senn}, \binits{H.M.}},
\bauthor{\bsnm{Thiel}, \binits{W.}}:
\batitle{{QM/MM} methods for biomolecular systems}.
\bjtitle{Angewandte Chemie International Edition}
\bvolume{48}(\bissue{7}),
\bfpage{1198}--\blpage{1229}
(\byear{2009})
\doiurl{10.1002/anie.200802019}
\end{barticle}
\endbibitem

\bibitem[\protect\citeauthoryear{Bennie et~al.}{2016}]{bennie2016}
\begin{barticle}
\bauthor{\bsnm{Bennie}, \binits{S.J.}},
\bauthor{\bsnm{Stella}, \binits{M.}},
\bauthor{\bsnm{Miller~III}, \binits{T.F.}},
\bauthor{\bsnm{Manby}, \binits{F.R.}}:
\batitle{A projector-embedding approach for multiscale coupled-cluster calculations applied to citrate synthase}.
\bjtitle{Journal of Chemical Theory and Computation}
\bvolume{12}(\bissue{6}),
\bfpage{2689}--\blpage{2697}
(\byear{2016})
\doiurl{10.1021/acs.jctc.6b00285}
\end{barticle}
\endbibitem

\bibitem[\protect\citeauthoryear{Seufert et~al.}{2025}]{seufert2025}
\begin{barticle}
\bauthor{\bsnm{Seufert}, \binits{F.}},
\bauthor{\bsnm{Chung}, \binits{Y.K.}},
\bauthor{\bsnm{Schick}, \binits{R.}},
\bauthor{\bsnm{Pohl}, \binits{F.}},
\bauthor{\bsnm{Str{\"a}ter}, \binits{N.}},
\bauthor{\bsnm{Langenhan}, \binits{T.}},
\bauthor{\bsnm{Batebi}, \binits{H.}},
\bauthor{\bsnm{Hildebrand}, \binits{P.W.}}:
\batitle{Mechanistic insights into adhesion {GPCR} autoproteolysis by a multiscale computational approach}.
\bjtitle{Research Square (preprint)}
(\byear{2025})
\doiurl{10.21203/rs.3.rs-7121662/v1}
\end{barticle}
\endbibitem

\bibitem[\protect\citeauthoryear{Chung et~al.}{2025}]{chung_self-cleavage_2025}
\begin{barticle}
\bauthor{\bsnm{Chung}, \binits{Y.K.}},
\bauthor{\bsnm{Ihling}, \binits{C.H.}},
\bauthor{\bsnm{Zielke}, \binits{L.}},
\bauthor{\bsnm{Mathiasen}, \binits{S.}},
\bauthor{\bsnm{Sinz}, \binits{A.}},
\bauthor{\bsnm{Langenhan}, \binits{T.}}:
\batitle{Self-cleavage of the {GAIN} domain of adhesion {G} protein-coupled receptors requires multiple domain-extrinsic factors}.
\bjtitle{Nature Communications}
\bvolume{16}(\bissue{1}),
\bfpage{8736}
(\byear{2025})
\doiurl{10.1038/s41467-025-64589-3}
\end{barticle}
\endbibitem

\bibitem[\protect\citeauthoryear{Salzman et~al.}{2016}]{salzman2016}
\begin{barticle}
\bauthor{\bsnm{Salzman}, \binits{G.S.}},
\bauthor{\bsnm{Ackerman}, \binits{S.D.}},
\bauthor{\bsnm{Ding}, \binits{C.}},
\bauthor{\bsnm{Koide}, \binits{A.}},
\bauthor{\bsnm{Leon}, \binits{K.}},
\bauthor{\bsnm{Luo}, \binits{R.}},
\bauthor{\bsnm{Stoveken}, \binits{H.M.}},
\bauthor{\bsnm{Fernandez}, \binits{C.G.}},
\bauthor{\bsnm{Tall}, \binits{G.G.}},
\bauthor{\bsnm{Piao}, \binits{X.}},
\bauthor{\bsnm{Monk}, \binits{K.R.}},
\bauthor{\bsnm{Koide}, \binits{S.}},
\bauthor{\bsnm{Ara{\c{c}}}, \binits{D.}}:
\batitle{Structural basis for regulation of {GPR56/ADGRG1} by its alternatively spliced extracellular domains}.
\bjtitle{Neuron}
\bvolume{91},
\bfpage{1292}--\blpage{1304}
(\byear{2016})
\doiurl{10.1016/j.neuron.2016.08.022}
\end{barticle}
\endbibitem

\bibitem[\protect\citeauthoryear{Aspuru-Guzik et~al.}{2005}]{aspuru2005simulated}
\begin{barticle}
\bauthor{\bsnm{Aspuru-Guzik}, \binits{A.}},
\bauthor{\bsnm{Dutoi}, \binits{A.D.}},
\bauthor{\bsnm{Love}, \binits{P.J.}},
\bauthor{\bsnm{Head-Gordon}, \binits{M.}}:
\batitle{Simulated quantum computation of molecular energies}.
\bjtitle{Science}
\bvolume{309}(\bissue{5741}),
\bfpage{1704}--\blpage{1707}
(\byear{2005})
\doiurl{10.1126/science.1113479}
\end{barticle}
\endbibitem

\bibitem[\protect\citeauthoryear{Bickley et~al.}{2025}]{bickley2025}
\begin{barticle}
\bauthor{\bsnm{Bickley}, \binits{T.M.}},
\bauthor{\bsnm{Mingare}, \binits{A.}},
\bauthor{\bsnm{Weaving}, \binits{T.}},
\bauthor{\bsnm{Bastida}, \binits{M.}},
\bauthor{\bsnm{Wan}, \binits{S.}},
\bauthor{\bsnm{Nibbi}, \binits{M.}},
\bauthor{\bsnm{Seitz}, \binits{P.}},
\bauthor{\bsnm{Ralli}, \binits{A.}},
\bauthor{\bsnm{Love}, \binits{P.J.}},
\bauthor{\bsnm{Chung}, \binits{M.}},
\bauthor{\bsnm{Hernández~Vera}, \binits{M.}},
\bauthor{\bsnm{Schulz}, \binits{L.}},
\bauthor{\bsnm{Coveney}, \binits{P.V.}}:
\batitle{Extending quantum computing through subspace{,} embedding and classical molecular dynamics techniques}.
\bjtitle{Digital Discovery}
\bvolume{4},
\bfpage{3427}--\blpage{3444}
(\byear{2025})
\doiurl{10.1039/D5DD00225G}
\end{barticle}
\endbibitem

\bibitem[\protect\citeauthoryear{Shajan et~al.}{2026}]{shajan2026}
\begin{barticle}
\bauthor{\bsnm{Shajan}, \binits{A.}},
\bauthor{\bsnm{Kaliakin}, \binits{D.}},
\bauthor{\bsnm{Liang}, \binits{F.}},
\bauthor{\bsnm{Pellegrini}, \binits{T.}},
\bauthor{\bsnm{Doga}, \binits{H.}},
\bauthor{\bsnm{Bhowmik}, \binits{S.}},
\bauthor{\bsnm{Das}, \binits{S.}},
\bauthor{\bsnm{Mezzacapo}, \binits{A.}},
\bauthor{\bsnm{Motta}, \binits{M.}},
\bauthor{\bsnm{Merz~Jr}, \binits{K.M.}}:
\batitle{Molecular quantum computations on a protein}.
\bjtitle{Journal of Chemical Theory and Computation}
\bvolume{22}(\bissue{12}),
\bfpage{6041}
(\byear{2026})
\end{barticle}
\endbibitem

\bibitem[\protect\citeauthoryear{Merz~Jr et~al.}{2026}]{merz2026}
\begin{botherref}
\oauthor{\bsnm{Merz~Jr}, \binits{K.M.}},
\oauthor{\bsnm{Shajan}, \binits{A.}},
\oauthor{\bsnm{Kaliakin}, \binits{D.}},
\oauthor{\bsnm{Liang}, \binits{F.}},
\oauthor{\bsnm{Otsuka}, \binits{Y.}},
\oauthor{\bsnm{Shirakawa}, \binits{T.}},
\oauthor{\bsnm{Broers}, \binits{L.}},
\oauthor{\bsnm{Xu}, \binits{H.}},
\oauthor{\bsnm{Tsuji}, \binits{M.}},
\oauthor{\bsnm{Sato}, \binits{M.}}, et al.:
Crossing the 12,000-atom barrier with heterogeneous quantum-classical supercomputing: quantum chemistry of protein-ligand complexes.
arXiv preprint arXiv:2605.01138
(2026)
\end{botherref}
\endbibitem

\bibitem[\protect\citeauthoryear{Kanno et~al.}{2026}]{kanno2026}
\begin{barticle}
\bauthor{\bsnm{Kanno}, \binits{K.}},
\bauthor{\bsnm{Kohda}, \binits{M.}},
\bauthor{\bsnm{Imai}, \binits{R.}},
\bauthor{\bsnm{Koh}, \binits{S.}},
\bauthor{\bsnm{Mitarai}, \binits{K.}},
\bauthor{\bsnm{Mizukami}, \binits{W.}},
\bauthor{\bsnm{Nakagawa}, \binits{Y.O.}}:
\batitle{Quantum-selected configuration interaction: Classical diagonalization of hamiltonians in subspaces selected by quantum computers}.
\bjtitle{Phys. Rev. Res.}
\bvolume{8},
\bfpage{023268}
(\byear{2026})
\doiurl{10.1103/dmn4-snfx}
\end{barticle}
\endbibitem

\bibitem[\protect\citeauthoryear{Robledo-Moreno et~al.}{2025}]{robledo2025}
\begin{barticle}
\bauthor{\bsnm{Robledo-Moreno}, \binits{J.}},
\bauthor{\bsnm{Motta}, \binits{M.}},
\bauthor{\bsnm{Haas}, \binits{H.}},
\bauthor{\bsnm{Javadi-Abhari}, \binits{A.}},
\bauthor{\bsnm{Jurcevic}, \binits{P.}},
\bauthor{\bsnm{Kirby}, \binits{W.}},
\bauthor{\bsnm{Martiel}, \binits{S.}},
\bauthor{\bsnm{Sharma}, \binits{K.}},
\bauthor{\bsnm{Sharma}, \binits{S.}},
\bauthor{\bsnm{Shirakawa}, \binits{T.}},
\bauthor{\bsnm{Sitdikov}, \binits{I.}},
\bauthor{\bsnm{Sun}, \binits{R.-Y.}},
\bauthor{\bsnm{Sung}, \binits{K.J.}},
\bauthor{\bsnm{Takita}, \binits{M.}},
\bauthor{\bsnm{Tran}, \binits{M.C.}},
\bauthor{\bsnm{Yunoki}, \binits{S.}},
\bauthor{\bsnm{Mezzacapo}, \binits{A.}}:
\batitle{Chemistry beyond the scale of exact diagonalization on a quantum-centric supercomputer}.
\bjtitle{Science Advances}
\bvolume{11}(\bissue{25}),
\bfpage{9991}
(\byear{2025})
\doiurl{10.1126/sciadv.adu9991}
\end{barticle}
\endbibitem

\bibitem[\protect\citeauthoryear{Alexeev et~al.}{2024}]{alexeev2024quantum}
\begin{barticle}
\bauthor{\bsnm{Alexeev}, \binits{Y.}},
\bauthor{\bsnm{Amsler}, \binits{M.}},
\bauthor{\bsnm{Barroca}, \binits{M.A.}},
\bauthor{\bsnm{Bassini}, \binits{S.}},
\bauthor{\bsnm{Battelle}, \binits{T.}},
\bauthor{\bsnm{Camps}, \binits{D.}},
\bauthor{\bsnm{Casanova}, \binits{D.}},
\bauthor{\bsnm{Choi}, \binits{Y.J.}},
\bauthor{\bsnm{Chong}, \binits{F.T.}},
\bauthor{\bsnm{Chung}, \binits{C.}}, \betal:
\batitle{Quantum-centric supercomputing for materials science: A perspective on challenges and future directions}.
\bjtitle{Future Generation Computer Systems}
\bvolume{160},
\bfpage{666}--\blpage{710}
(\byear{2024})
\doiurl{10.1016/j.future.2024.04.060}
\end{barticle}
\endbibitem

\bibitem[\protect\citeauthoryear{Weaving et~al.}{2026}]{weaving2026selected}
\begin{barticle}
\bauthor{\bsnm{Weaving}, \binits{T.}},
\bauthor{\bsnm{Mingare}, \binits{A.}},
\bauthor{\bsnm{Ralli}, \binits{A.}},
\bauthor{\bsnm{Coveney}, \binits{P.V.}}:
\batitle{Selected configuration interaction using time-evolved population statistics}.
\bjtitle{Journal of Chemical Theory and Computation}
\bvolume{22}(\bissue{9}),
\bfpage{4315}--\blpage{4328}
(\byear{2026})
\doiurl{10.1021/acs.jctc.5c01994}
\end{barticle}
\endbibitem

\bibitem[\protect\citeauthoryear{Ouyang et~al.}{2020}]{ouyang2020compilation}
\begin{barticle}
\bauthor{\bsnm{Ouyang}, \binits{Y.}},
\bauthor{\bsnm{White}, \binits{D.R.}},
\bauthor{\bsnm{Campbell}, \binits{E.T.}}:
\batitle{Compilation by stochastic hamiltonian sparsification}.
\bjtitle{Quantum}
\bvolume{4},
\bfpage{235}
(\byear{2020})
\doiurl{10.22331/q-2020-02-27-235}
\end{barticle}
\endbibitem

\bibitem[\protect\citeauthoryear{Campbell}{2019}]{campbell2019}
\begin{barticle}
\bauthor{\bsnm{Campbell}, \binits{E.}}:
\batitle{Random compiler for fast {H}amiltonian simulation}.
\bjtitle{Physical Review Letters}
\bvolume{123},
\bfpage{070503}
(\byear{2019})
\doiurl{10.1103/PhysRevLett.123.070503}
\end{barticle}
\endbibitem

\bibitem[\protect\citeauthoryear{Tzeliou et~al.}{2022}]{tzeliou2022}
\begin{botherref}
\oauthor{\bsnm{Tzeliou}, \binits{C.E.}},
\oauthor{\bsnm{Mermigki}, \binits{M.A.}},
\oauthor{\bsnm{Tzeli}, \binits{D.}}:
Review on the qm/mm methodologies and their application to metalloproteins.
Molecules
\textbf{27}(9)
(2022)
\doiurl{10.3390/molecules27092660}
\end{botherref}
\endbibitem

\bibitem[\protect\citeauthoryear{Jumper et~al.}{2021}]{jumper2021}
\begin{barticle}
\bauthor{\bsnm{Jumper}, \binits{J.}},
\bauthor{\bsnm{Evans}, \binits{R.}},
\bauthor{\bsnm{Pritzel}, \binits{A.}},
\bauthor{\bsnm{Green}, \binits{T.}},
\bauthor{\bsnm{Figurnov}, \binits{M.}},
\bauthor{\bsnm{Ronneberger}, \binits{O.}},
\bauthor{\bsnm{Tunyasuvunakool}, \binits{K.}},
\bauthor{\bsnm{Bates}, \binits{R.}},
\bauthor{\bsnm{{\v{Z}}{\'i}dek}, \binits{A.}},
\bauthor{\bsnm{Potapenko}, \binits{A.}},
\bauthor{\bsnm{Bridgland}, \binits{A.}},
\bauthor{\bsnm{Meyer}, \binits{C.}},
\bauthor{\bsnm{Kohl}, \binits{S.A.A.}},
\bauthor{\bsnm{Ballard}, \binits{A.J.}},
\bauthor{\bsnm{Cowie}, \binits{A.}},
\bauthor{\bsnm{Romera-Paredes}, \binits{B.}},
\bauthor{\bsnm{Nikolov}, \binits{S.}},
\bauthor{\bsnm{Jain}, \binits{R.}},
\bauthor{\bsnm{Adler}, \binits{J.}},
\bauthor{\bsnm{Back}, \binits{T.}},
\bauthor{\bsnm{Petersen}, \binits{S.}},
\bauthor{\bsnm{Reiman}, \binits{D.}},
\bauthor{\bsnm{Clancy}, \binits{E.}},
\bauthor{\bsnm{Zielinski}, \binits{M.}},
\bauthor{\bsnm{Steinegger}, \binits{M.}},
\bauthor{\bsnm{Pacholska}, \binits{M.}},
\bauthor{\bsnm{Berghammer}, \binits{T.}},
\bauthor{\bsnm{Bodenstein}, \binits{S.}},
\bauthor{\bsnm{Silver}, \binits{D.}},
\bauthor{\bsnm{Vinyals}, \binits{O.}},
\bauthor{\bsnm{Senior}, \binits{A.W.}},
\bauthor{\bsnm{Kavukcuoglu}, \binits{K.}},
\bauthor{\bsnm{Kohli}, \binits{P.}},
\bauthor{\bsnm{Hassabis}, \binits{D.}}:
\batitle{Highly accurate protein structure prediction with alphafold}.
\bjtitle{Nature}
\bvolume{596}(\bissue{7873}),
\bfpage{583}--\blpage{589}
(\byear{2021})
\doiurl{10.1038/s41586-021-03819-2}
\end{barticle}
\endbibitem

\bibitem[\protect\citeauthoryear{Herrera et~al.}{2024}]{herrera2024}
\begin{barticle}
\bauthor{\bsnm{Herrera}, \binits{L.P.T.}},
\bauthor{\bsnm{Andreassen}, \binits{S.N.}},
\bauthor{\bsnm{Caroli}, \binits{J.}},
\bauthor{\bsnm{Rodríguez-Espigares}, \binits{I.}},
\bauthor{\bsnm{Kermani}, \binits{A.A.}},
\bauthor{\bsnm{Keserű}, \binits{G.M.}},
\bauthor{\bsnm{Kooistra}, \binits{A.J.}},
\bauthor{\bsnm{Pándy-Szekeres}, \binits{G.}},
\bauthor{\bsnm{Gloriam}, \binits{D.E.}}:
\batitle{{GPCRdb} in 2025: adding odorant receptors, data mapper, structure similarity search and models of physiological ligand complexes}.
\bjtitle{Nucleic Acids Research}
\bvolume{53}(\bissue{D1}),
\bfpage{425}--\blpage{435}
(\byear{2024})
\doiurl{10.1093/nar/gkae1065}
\end{barticle}
\endbibitem

\bibitem[\protect\citeauthoryear{{GPCRdb}}{2026}]{gpcrdb_adgrg1_inactive}
\begin{botherref}
\oauthor{\bsnm{{GPCRdb}}}:
{ADGRG1} ({GPR56}) human inactive state homology model.
Accessed: 2026-02-18
(2026).
\url{https://gpcrdb.org/structure/homology_models/agrg1_human_inactive}
\end{botherref}
\endbibitem

\bibitem[\protect\citeauthoryear{Barros-{\'A}lvarez et~al.}{2022}]{barros2022}
\begin{barticle}
\bauthor{\bsnm{Barros-{\'A}lvarez}, \binits{X.}},
\bauthor{\bsnm{Nwokonko}, \binits{R.M.}},
\bauthor{\bsnm{Vizurraga}, \binits{A.}},
\bauthor{\bsnm{Matzov}, \binits{D.}},
\bauthor{\bsnm{He}, \binits{F.}},
\bauthor{\bsnm{Papasergis-Scott}, \binits{M.M.}},
\bauthor{\bsnm{Robertson}, \binits{M.J.}},
\bauthor{\bsnm{Panova}, \binits{O.}},
\bauthor{\bsnm{Yardeni}, \binits{E.H.}},
\bauthor{\bsnm{Seven}, \binits{A.B.}},
\bauthor{\bsnm{Kwarcinski}, \binits{F.E.}},
\bauthor{\bsnm{Su}, \binits{H.}},
\bauthor{\bsnm{Peroto}, \binits{M.C.}},
\bauthor{\bsnm{Meyerowitz}, \binits{J.G.}},
\bauthor{\bsnm{Shalev-Benami}, \binits{M.}},
\bauthor{\bsnm{Tall}, \binits{G.G.}},
\bauthor{\bsnm{Skiniotis}, \binits{G.}}:
\batitle{The tethered peptide activation mechanism of adhesion {GPCRs}}.
\bjtitle{Nature}
\bvolume{604},
\bfpage{757}--\blpage{762}
(\byear{2022})
\doiurl{10.1038/s41586-022-04575-7}
\end{barticle}
\endbibitem

\bibitem[\protect\citeauthoryear{Fiser and Sali}{2003}]{fiser2003}
\begin{barticle}
\bauthor{\bsnm{Fiser}, \binits{A.}},
\bauthor{\bsnm{Sali}, \binits{A.}}:
\batitle{Modeller: Generation and refinement of homology-based protein structure models}.
\bjtitle{Methods in Enzymology}
\bvolume{374},
\bfpage{461}--\blpage{491}
(\byear{2003})
\doiurl{10.1016/S0076-6879(03)74020-8}
\end{barticle}
\endbibitem

\bibitem[\protect\citeauthoryear{Pettersen et~al.}{2004}]{pettersen2004}
\begin{barticle}
\bauthor{\bsnm{Pettersen}, \binits{E.F.}},
\bauthor{\bsnm{Goddard}, \binits{T.D.}},
\bauthor{\bsnm{Huang}, \binits{C.C.}},
\bauthor{\bsnm{Couch}, \binits{G.S.}},
\bauthor{\bsnm{Greenblatt}, \binits{D.M.}},
\bauthor{\bsnm{Meng}, \binits{E.C.}},
\bauthor{\bsnm{Ferrin}, \binits{T.E.}}:
\batitle{{UCSF Chimera} --- a visualization system for exploratory research and analysis}.
\bjtitle{Journal of Computational Chemistry}
\bvolume{25}(\bissue{13}),
\bfpage{1605}--\blpage{1612}
(\byear{2004})
\doiurl{10.1002/jcc.20084}
\end{barticle}
\endbibitem

\bibitem[\protect\citeauthoryear{{Chemical Computing Group ULC}}{2024}]{moe2024}
\begin{botherref}
\oauthor{\bsnm{{Chemical Computing Group ULC}}}:
Molecular Operating Environment ({MOE}).
\url{https://www.chemcomp.com/Products.htm},
Montreal, QC, Canada
(2024)
\end{botherref}
\endbibitem

\bibitem[\protect\citeauthoryear{Jo et~al.}{2008}]{jo2008}
\begin{barticle}
\bauthor{\bsnm{Jo}, \binits{S.}},
\bauthor{\bsnm{Kim}, \binits{T.}},
\bauthor{\bsnm{Iyer}, \binits{V.G.}},
\bauthor{\bsnm{Im}, \binits{W.}}:
\batitle{Charmm-gui: A web-based graphical user interface for charmm}.
\bjtitle{Journal of Computational Chemistry}
\bvolume{29}(\bissue{11}),
\bfpage{1859}--\blpage{1865}
(\byear{2008})
\doiurl{10.1002/jcc.20945}
\end{barticle}
\endbibitem

\bibitem[\protect\citeauthoryear{Lee et~al.}{2016}]{lee2016}
\begin{barticle}
\bauthor{\bsnm{Lee}, \binits{J.}},
\bauthor{\bsnm{Cheng}, \binits{X.}},
\bauthor{\bsnm{Swails}, \binits{J.M.}},
\bauthor{\bsnm{Yeom}, \binits{M.S.}},
\bauthor{\bsnm{Eastman}, \binits{P.K.}},
\bauthor{\bsnm{Lemkul}, \binits{J.A.}},
\bauthor{\bsnm{Wei}, \binits{S.}},
\bauthor{\bsnm{Buckner}, \binits{J.}},
\bauthor{\bsnm{Jeong}, \binits{J.C.}},
\bauthor{\bsnm{Qi}, \binits{Y.}},
\bauthor{\bsnm{Jo}, \binits{S.}},
\bauthor{\bsnm{Pande}, \binits{V.S.}},
\bauthor{\bsnm{Case}, \binits{D.A.}},
\bauthor{\bsnm{Brooks}, \binits{C.L.I.}},
\bauthor{\bsnm{MacKerell}, \binits{A.D.J.}},
\bauthor{\bsnm{Klauda}, \binits{J.B.}},
\bauthor{\bsnm{Im}, \binits{W.}}:
\batitle{Charmm-gui input generator for namd, gromacs, amber, openmm, and charmm/openmm simulations using the charmm36 additive force field}.
\bjtitle{Journal of Chemical Theory and Computation}
\bvolume{12}(\bissue{1}),
\bfpage{405}--\blpage{413}
(\byear{2016})
\doiurl{10.1021/acs.jctc.5b00935}
\end{barticle}
\endbibitem

\bibitem[\protect\citeauthoryear{Ernst et~al.}{2024}]{ernst2024}
\begin{barticle}
\bauthor{\bsnm{Ernst}, \binits{R.}},
\bauthor{\bsnm{Renne}, \binits{M.F.}},
\bauthor{\bsnm{Jain}, \binits{A.}},
\bauthor{\bsnm{Malsburg}, \binits{A.}}:
\batitle{Endoplasmic reticulum membrane homeostasis and the unfolded protein response}.
\bjtitle{Cold Spring Harbor Perspectives in Biology}
\bvolume{16}(\bissue{8}),
\bfpage{041400}
(\byear{2024})
\doiurl{10.1101/cshperspect.a041400}
\end{barticle}
\endbibitem

\bibitem[\protect\citeauthoryear{Jorgensen et~al.}{1983}]{jorgensen1983}
\begin{barticle}
\bauthor{\bsnm{Jorgensen}, \binits{W.L.}},
\bauthor{\bsnm{Chandrasekhar}, \binits{J.}},
\bauthor{\bsnm{Madura}, \binits{J.D.}},
\bauthor{\bsnm{Impey}, \binits{R.W.}},
\bauthor{\bsnm{Klein}, \binits{M.L.}}:
\batitle{Comparison of simple potential functions for simulating liquid water}.
\bjtitle{The Journal of Chemical Physics}
\bvolume{79}(\bissue{2}),
\bfpage{926}--\blpage{935}
(\byear{1983})
\doiurl{10.1063/1.445869}
\end{barticle}
\endbibitem

\bibitem[\protect\citeauthoryear{Van Der~Spoel et~al.}{2005}]{spoel2005}
\begin{barticle}
\bauthor{\bsnm{Van Der~Spoel}, \binits{D.}},
\bauthor{\bsnm{Lindahl}, \binits{E.}},
\bauthor{\bsnm{Hess}, \binits{B.}},
\bauthor{\bsnm{Groenhof}, \binits{G.}},
\bauthor{\bsnm{Mark}, \binits{A.E.}},
\bauthor{\bsnm{Berendsen}, \binits{H.J.C.}}:
\batitle{Gromacs: Fast, flexible, and free}.
\bjtitle{Journal of Computational Chemistry}
\bvolume{26}(\bissue{16}),
\bfpage{1701}--\blpage{1718}
(\byear{2005})
\doiurl{10.1002/jcc.20291}
\end{barticle}
\endbibitem

\bibitem[\protect\citeauthoryear{Abraham et~al.}{2015}]{abraham2015}
\begin{barticle}
\bauthor{\bsnm{Abraham}, \binits{M.J.}},
\bauthor{\bsnm{Murtola}, \binits{T.}},
\bauthor{\bsnm{Schulz}, \binits{R.}},
\bauthor{\bsnm{Páll}, \binits{S.}},
\bauthor{\bsnm{Smith}, \binits{J.C.}},
\bauthor{\bsnm{Hess}, \binits{B.}},
\bauthor{\bsnm{Lindahl}, \binits{E.}}:
\batitle{Gromacs: High performance molecular simulations through multi-level parallelism from laptops to supercomputers}.
\bjtitle{SoftwareX}
\bvolume{1-2},
\bfpage{19}--\blpage{25}
(\byear{2015})
\doiurl{10.1016/j.softx.2015.06.001}
\end{barticle}
\endbibitem

\bibitem[\protect\citeauthoryear{Brooks et~al.}{2009}]{brooks2009}
\begin{barticle}
\bauthor{\bsnm{Brooks}, \binits{B.R.}},
\bauthor{\bsnm{Brooks~III}, \binits{C.L.}},
\bauthor{\bsnm{Mackerell~Jr.}, \binits{A.D.}},
\bauthor{\bsnm{Nilsson}, \binits{L.}},
\bauthor{\bsnm{Petrella}, \binits{R.J.}},
\bauthor{\bsnm{Roux}, \binits{B.}},
\bauthor{\bsnm{Won}, \binits{Y.}},
\bauthor{\bsnm{Archontis}, \binits{G.}},
\bauthor{\bsnm{Bartels}, \binits{C.}},
\bauthor{\bsnm{Boresch}, \binits{S.}},
\bauthor{\bsnm{Caflisch}, \binits{A.}},
\bauthor{\bsnm{Caves}, \binits{L.}},
\bauthor{\bsnm{Cui}, \binits{Q.}},
\bauthor{\bsnm{Dinner}, \binits{A.R.}},
\bauthor{\bsnm{Feig}, \binits{M.}},
\bauthor{\bsnm{Fischer}, \binits{S.}},
\bauthor{\bsnm{Gao}, \binits{J.}},
\bauthor{\bsnm{Hodoscek}, \binits{M.}},
\bauthor{\bsnm{Im}, \binits{W.}},
\bauthor{\bsnm{Kuczera}, \binits{K.}},
\bauthor{\bsnm{Lazaridis}, \binits{T.}},
\bauthor{\bsnm{Ma}, \binits{J.}},
\bauthor{\bsnm{Ovchinnikov}, \binits{V.}},
\bauthor{\bsnm{Paci}, \binits{E.}},
\bauthor{\bsnm{Pastor}, \binits{R.W.}},
\bauthor{\bsnm{Post}, \binits{C.B.}},
\bauthor{\bsnm{Pu}, \binits{J.Z.}},
\bauthor{\bsnm{Schaefer}, \binits{M.}},
\bauthor{\bsnm{Tidor}, \binits{B.}},
\bauthor{\bsnm{Venable}, \binits{R.M.}},
\bauthor{\bsnm{Woodcock}, \binits{H.L.}},
\bauthor{\bsnm{Wu}, \binits{X.}},
\bauthor{\bsnm{Yang}, \binits{W.}},
\bauthor{\bsnm{York}, \binits{D.M.}},
\bauthor{\bsnm{Karplus}, \binits{M.}}:
\batitle{Charmm: The biomolecular simulation program}.
\bjtitle{Journal of Computational Chemistry}
\bvolume{30}(\bissue{10}),
\bfpage{1545}--\blpage{1614}
(\byear{2009})
\doiurl{10.1002/jcc.21287}
\end{barticle}
\endbibitem

\bibitem[\protect\citeauthoryear{Huang and MacKerell~Jr}{2013}]{huang2013}
\begin{barticle}
\bauthor{\bsnm{Huang}, \binits{J.}},
\bauthor{\bsnm{MacKerell~Jr}, \binits{A.D.}}:
\batitle{Charmm36 all-atom additive protein force field: Validation based on comparison to nmr data}.
\bjtitle{Journal of Computational Chemistry}
\bvolume{34}(\bissue{25}),
\bfpage{2135}--\blpage{2145}
(\byear{2013})
\doiurl{10.1002/jcc.23354}
\end{barticle}
\endbibitem

\bibitem[\protect\citeauthoryear{{GROMACS Development Team}}{2024}]{gromacs_manual_qmmm}
\begin{botherref}
\oauthor{\bsnm{{GROMACS Development Team}}}:
Hybrid Quantum-Classical simulations (QM/MM) with CP2K interface.
Accessed: 2026-03-02
(2024).
\url{https://manual.gromacs.org/current/reference-manual/special/qmmm.html}
\end{botherref}
\endbibitem

\bibitem[\protect\citeauthoryear{Kühne et~al.}{2020}]{kuhne2020}
\begin{botherref}
\oauthor{\bsnm{Kühne}, \binits{T.D.}},
\oauthor{\bsnm{Iannuzzi}, \binits{M.}},
\oauthor{\bsnm{Del~Ben}, \binits{M.}},
\oauthor{\bsnm{Rybkin}, \binits{V.V.}},
\oauthor{\bsnm{Seewald}, \binits{P.}},
\oauthor{\bsnm{Stein}, \binits{F.}},
\oauthor{\bsnm{Laino}, \binits{T.}},
\oauthor{\bsnm{Khaliullin}, \binits{R.Z.}},
\oauthor{\bsnm{Schütt}, \binits{O.}},
\oauthor{\bsnm{Schiffmann}, \binits{F.}},
\oauthor{\bsnm{Golze}, \binits{D.}},
\oauthor{\bsnm{Wilhelm}, \binits{J.}},
\oauthor{\bsnm{Chulkov}, \binits{S.}},
\oauthor{\bsnm{Bani-Hashemian}, \binits{M.H.}},
\oauthor{\bsnm{Weber}, \binits{V.}},
\oauthor{\bsnm{Borštnik}, \binits{U.}},
\oauthor{\bsnm{Taillefumier}, \binits{M.}},
\oauthor{\bsnm{Jakobovits}, \binits{A.S.}},
\oauthor{\bsnm{Lazzaro}, \binits{A.}},
\oauthor{\bsnm{Pabst}, \binits{H.}},
\oauthor{\bsnm{Müller}, \binits{T.}},
\oauthor{\bsnm{Schade}, \binits{R.}},
\oauthor{\bsnm{Guidon}, \binits{M.}},
\oauthor{\bsnm{Andermatt}, \binits{S.}},
\oauthor{\bsnm{Holmberg}, \binits{N.}},
\oauthor{\bsnm{Schenter}, \binits{G.K.}},
\oauthor{\bsnm{Hehn}, \binits{A.}},
\oauthor{\bsnm{Bussy}, \binits{A.}},
\oauthor{\bsnm{Belleflamme}, \binits{F.}},
\oauthor{\bsnm{Tabacchi}, \binits{G.}},
\oauthor{\bsnm{Glöß}, \binits{A.}},
\oauthor{\bsnm{Lass}, \binits{M.}},
\oauthor{\bsnm{Bethune}, \binits{I.}},
\oauthor{\bsnm{Mundy}, \binits{C.J.}},
\oauthor{\bsnm{Plessl}, \binits{C.}},
\oauthor{\bsnm{Watkins}, \binits{M.}},
\oauthor{\bsnm{VandeVondele}, \binits{J.}},
\oauthor{\bsnm{Krack}, \binits{M.}},
\oauthor{\bsnm{Hutter}, \binits{J.}}:
Cp2k: An electronic structure and molecular dynamics software package-quickstep: Efficient and accurate electronic structure calculations.
The Journal of Chemical Physics
\textbf{152}(19)
(2020)
\doiurl{10.1063/5.0007045}
\end{botherref}
\endbibitem

\bibitem[\protect\citeauthoryear{Perdew et~al.}{1996}]{perdew1996}
\begin{barticle}
\bauthor{\bsnm{Perdew}, \binits{J.P.}},
\bauthor{\bsnm{Burke}, \binits{K.}},
\bauthor{\bsnm{Ernzerhof}, \binits{M.}}:
\batitle{Generalized gradient approximation made simple}.
\bjtitle{Physical Review Letters}
\bvolume{77}(\bissue{18}),
\bfpage{3865}--\blpage{3868}
(\byear{1996})
\doiurl{10.1103/PhysRevLett.77.3865}
\end{barticle}
\endbibitem

\bibitem[\protect\citeauthoryear{VandeVondele and Hutter}{2007}]{vandevondele2007}
\begin{barticle}
\bauthor{\bsnm{VandeVondele}, \binits{J.}},
\bauthor{\bsnm{Hutter}, \binits{J.}}:
\batitle{Gaussian basis sets for accurate calculations on molecular systems in gas and condensed phases}.
\bjtitle{The Journal of Chemical Physics}
\bvolume{127}(\bissue{11}),
\bfpage{114105}
(\byear{2007})
\doiurl{10.1063/1.2770708}
\end{barticle}
\endbibitem

\bibitem[\protect\citeauthoryear{Becke}{1993}]{becke1993}
\begin{barticle}
\bauthor{\bsnm{Becke}, \binits{A.D.}}:
\batitle{Density-functional thermochemistry. {III}. the role of exact exchange}.
\bjtitle{The Journal of Chemical Physics}
\bvolume{98}(\bissue{7}),
\bfpage{5648}--\blpage{5652}
(\byear{1993})
\doiurl{10.1063/1.464913}
\end{barticle}
\endbibitem

\bibitem[\protect\citeauthoryear{Lee et~al.}{1988}]{lee1988}
\begin{barticle}
\bauthor{\bsnm{Lee}, \binits{C.}},
\bauthor{\bsnm{Yang}, \binits{W.}},
\bauthor{\bsnm{Parr}, \binits{R.G.}}:
\batitle{Development of the {Colle}-{Salvetti} correlation-energy formula into a functional of the electron density}.
\bjtitle{Physical Review B}
\bvolume{37}(\bissue{2}),
\bfpage{785}--\blpage{789}
(\byear{1988})
\doiurl{10.1103/PhysRevB.37.785}
\end{barticle}
\endbibitem

\bibitem[\protect\citeauthoryear{Stephens et~al.}{1994}]{stephens1994}
\begin{barticle}
\bauthor{\bsnm{Stephens}, \binits{P.J.}},
\bauthor{\bsnm{Devlin}, \binits{F.J.}},
\bauthor{\bsnm{Chabalowski}, \binits{C.F.}},
\bauthor{\bsnm{Frisch}, \binits{M.J.}}:
\batitle{Ab initio calculation of vibrational absorption and circular dichroism spectra using density functional force fields}.
\bjtitle{The Journal of Physical Chemistry}
\bvolume{98}(\bissue{45}),
\bfpage{11623}--\blpage{11627}
(\byear{1994})
\doiurl{10.1021/j100096a001}
\end{barticle}
\endbibitem

\bibitem[\protect\citeauthoryear{Guidon et~al.}{2010}]{guidon2010}
\begin{barticle}
\bauthor{\bsnm{Guidon}, \binits{M.}},
\bauthor{\bsnm{Hutter}, \binits{J.}},
\bauthor{\bsnm{VandeVondele}, \binits{J.}}:
\batitle{Auxiliary density matrix methods for hartree-fock exchange calculations}.
\bjtitle{Journal of Chemical Theory and Computation}
\bvolume{6}(\bissue{8}),
\bfpage{2348}--\blpage{2364}
(\byear{2010})
\doiurl{10.1021/ct1002225} .
\bcomment{PMID: 26613491}
\end{barticle}
\endbibitem

\bibitem[\protect\citeauthoryear{Laino et~al.}{2005}]{laino2005}
\begin{barticle}
\bauthor{\bsnm{Laino}, \binits{T.}},
\bauthor{\bsnm{Mohamed}, \binits{F.}},
\bauthor{\bsnm{Laio}, \binits{A.}},
\bauthor{\bsnm{Parrinello}, \binits{M.}}:
\batitle{An efficient real space multigrid qm/mm electrostatic coupling}.
\bjtitle{Journal of Chemical Theory and Computation}
\bvolume{1}(\bissue{6}),
\bfpage{1176}--\blpage{1184}
(\byear{2005})
\doiurl{10.1021/ct050123f} .
\bcomment{PMID: 26631661}
\end{barticle}
\endbibitem

\bibitem[\protect\citeauthoryear{Laino et~al.}{2006}]{laino2006}
\begin{barticle}
\bauthor{\bsnm{Laino}, \binits{T.}},
\bauthor{\bsnm{Mohamed}, \binits{F.}},
\bauthor{\bsnm{Laio}, \binits{A.}},
\bauthor{\bsnm{Parrinello}, \binits{M.}}:
\batitle{An efficient linear-scaling electrostatic coupling for treating periodic boundary conditions in qm/mm simulations}.
\bjtitle{Journal of Chemical Theory and Computation}
\bvolume{2}(\bissue{5}),
\bfpage{1370}--\blpage{1378}
(\byear{2006})
\doiurl{10.1021/ct6001169} .
\bcomment{PMID: 26626844}
\end{barticle}
\endbibitem

\bibitem[\protect\citeauthoryear{Pipek and Mezey}{1989}]{pipek1989}
\begin{barticle}
\bauthor{\bsnm{Pipek}, \binits{J.}},
\bauthor{\bsnm{Mezey}, \binits{P.G.}}:
\batitle{A fast intrinsic localization procedure applicable for ab initio and semiempirical linear combination of atomic orbital wave functions}.
\bjtitle{The Journal of Chemical Physics}
\bvolume{90}(\bissue{9}),
\bfpage{4916}--\blpage{4926}
(\byear{1989})
\doiurl{10.1063/1.456588}
\end{barticle}
\endbibitem

\bibitem[\protect\citeauthoryear{Löwdin}{1950}]{loewdin1950}
\begin{barticle}
\bauthor{\bsnm{Löwdin}, \binits{P.}}:
\batitle{On the non‐orthogonality problem connected with the use of atomic wave functions in the theory of molecules and crystals}.
\bjtitle{The Journal of Chemical Physics}
\bvolume{18}(\bissue{3}),
\bfpage{365}--\blpage{375}
(\byear{1950})
\doiurl{10.1063/1.1747632}
\end{barticle}
\endbibitem

\bibitem[\protect\citeauthoryear{Sun et~al.}{2018}]{sun2018pyscf}
\begin{barticle}
\bauthor{\bsnm{Sun}, \binits{Q.}},
\bauthor{\bsnm{Berkelbach}, \binits{T.C.}},
\bauthor{\bsnm{Blunt}, \binits{N.S.}},
\bauthor{\bsnm{Booth}, \binits{G.H.}},
\bauthor{\bsnm{Guo}, \binits{S.}},
\bauthor{\bsnm{Li}, \binits{Z.}},
\bauthor{\bsnm{Liu}, \binits{J.}},
\bauthor{\bsnm{McClain}, \binits{J.D.}},
\bauthor{\bsnm{Sayfutyarova}, \binits{E.R.}},
\bauthor{\bsnm{Sharma}, \binits{S.}}, \betal:
\batitle{Pyscf: the python-based simulations of chemistry framework}.
\bjtitle{Wiley Interdisciplinary Reviews: Computational Molecular Science}
\bvolume{8}(\bissue{1}),
\bfpage{1340}
(\byear{2018})
\doiurl{10.1002/wcms.1340}
\end{barticle}
\endbibitem

\bibitem[\protect\citeauthoryear{Li et~al.}{2024}]{li2024introducting}
\begin{botherref}
\oauthor{\bsnm{Li}, \binits{R.}},
\oauthor{\bsnm{Sun}, \binits{Q.}},
\oauthor{\bsnm{Zhang}, \binits{X.}},
\oauthor{\bsnm{Chan}, \binits{G.K.-L.}}:
Introducing GPU-acceleration into the Python-based Simulations of Chemistry Framework
(2024).
\url{https://arxiv.org/abs/2407.09700}
\end{botherref}
\endbibitem

\bibitem[\protect\citeauthoryear{Wu et~al.}{2024}]{wu2024enhancing}
\begin{botherref}
\oauthor{\bsnm{Wu}, \binits{X.}},
\oauthor{\bsnm{Sun}, \binits{Q.}},
\oauthor{\bsnm{Pu}, \binits{Z.}},
\oauthor{\bsnm{Zheng}, \binits{T.}},
\oauthor{\bsnm{Ma}, \binits{W.}},
\oauthor{\bsnm{Yan}, \binits{W.}},
\oauthor{\bsnm{Yu}, \binits{X.}},
\oauthor{\bsnm{Wu}, \binits{Z.}},
\oauthor{\bsnm{Huo}, \binits{M.}},
\oauthor{\bsnm{Li}, \binits{X.}},
\oauthor{\bsnm{Ren}, \binits{W.}},
\oauthor{\bsnm{Gong}, \binits{S.}},
\oauthor{\bsnm{Zhang}, \binits{Y.}},
\oauthor{\bsnm{Gao}, \binits{W.}}:
Enhancing GPU-acceleration in the Python-based Simulations of Chemistry Framework
(2024).
\url{https://arxiv.org/abs/2404.09452}
\end{botherref}
\endbibitem

\bibitem[\protect\citeauthoryear{Weigend and Ahlrichs}{2005}]{weigend2005}
\begin{barticle}
\bauthor{\bsnm{Weigend}, \binits{F.}},
\bauthor{\bsnm{Ahlrichs}, \binits{R.}}:
\batitle{Balanced basis sets of split valence, triple zeta valence and quadruple zeta valence quality for {H} to {Rn}: Design and assessment of accuracy}.
\bjtitle{Physical Chemistry Chemical Physics}
\bvolume{7}(\bissue{18}),
\bfpage{3297}--\blpage{3305}
(\byear{2005})
\doiurl{10.1039/b508541a}
\end{barticle}
\endbibitem

\bibitem[\protect\citeauthoryear{Mardirossian and Head-Gordon}{2016}]{mardirossian2016omegab97m}
\begin{botherref}
\oauthor{\bsnm{Mardirossian}, \binits{N.}},
\oauthor{\bsnm{Head-Gordon}, \binits{M.}}:
$\omega$b97m-v: A combinatorially optimized, range-separated hybrid, meta-gga density functional with vv10 nonlocal correlation.
The Journal of chemical physics
\textbf{144}(21)
(2016)
\doiurl{10.1063/1.4952647}
\end{botherref}
\endbibitem

\bibitem[\protect\citeauthoryear{Huzinaga and Cantu}{1971}]{huzinaga1971}
\begin{barticle}
\bauthor{\bsnm{Huzinaga}, \binits{S.}},
\bauthor{\bsnm{Cantu}, \binits{A.A.}}:
\batitle{Theory of separability of many-electron systems}.
\bjtitle{The Journal of Chemical Physics}
\bvolume{55}(\bissue{12}),
\bfpage{5543}--\blpage{5549}
(\byear{1971})
\doiurl{10.1063/1.1675720}
\end{barticle}
\endbibitem

\bibitem[\protect\citeauthoryear{Jordan and Wigner}{1928}]{jordan1928}
\begin{barticle}
\bauthor{\bsnm{Jordan}, \binits{P.}},
\bauthor{\bsnm{Wigner}, \binits{E.}}:
\batitle{{\"U}ber das {P}aulische {\"a}quivalenzverbot}.
\bjtitle{Zeitschrift f{\"u}r Physik}
\bvolume{47}(\bissue{9--10}),
\bfpage{631}--\blpage{651}
(\byear{1928})
\doiurl{10.1007/BF01331938}
\end{barticle}
\endbibitem

\bibitem[\protect\citeauthoryear{{Williams de la Bastida} et~al.}{2026}]{delabastida2026optimisedfermionqubitencodingsquantum}
\begin{barticle}
\bauthor{\bsnm{{Williams de la Bastida}}, \binits{M.I.}},
\bauthor{\bsnm{Bickley}, \binits{T.M.}},
\bauthor{\bsnm{Coveney}, \binits{P.V.}}:
\batitle{Optimised {{Fermion-Qubit Encodings}} for {{Quantum Simulation}} with {{Reduced Circuit Depth}}}.
\bjtitle{Quantum Science and Technology}
(\byear{2026})
\doiurl{10.1088/2058-9565/ae9b40}
\end{barticle}
\endbibitem

\bibitem[\protect\citeauthoryear{Piccinelli et~al.}{2025}]{piccinelli2025quantum}
\begin{botherref}
\oauthor{\bsnm{Piccinelli}, \binits{S.}},
\oauthor{\bsnm{Baiardi}, \binits{A.}},
\oauthor{\bsnm{Rossmannek}, \binits{M.}},
\oauthor{\bsnm{Vazquez}, \binits{A.C.}},
\oauthor{\bsnm{Tacchino}, \binits{F.}},
\oauthor{\bsnm{Mensa}, \binits{S.}},
\oauthor{\bsnm{Altamura}, \binits{E.}},
\oauthor{\bsnm{Alavi}, \binits{A.}},
\oauthor{\bsnm{Motta}, \binits{M.}},
\oauthor{\bsnm{Robledo-Moreno}, \binits{J.}}, et al.:
Quantum chemistry with provable convergence via randomized sample-based quantum diagonalization.
arXiv preprint
(2025)
{\href{https://arxiv.org/abs/2508.02578}{{2508.02578}}}
\end{botherref}
\endbibitem

\bibitem[\protect\citeauthoryear{Chung et~al.}{2025}]{chung2025qbeast}
\begin{botherref}
\oauthor{\bsnm{Chung}, \binits{M.}},
\oauthor{\bsnm{Gambo}, \binits{Y.}},
\oauthor{\bsnm{Mete}, \binits{B.}},
\oauthor{\bsnm{To}, \binits{X.-T.M.}},
\oauthor{\bsnm{Krötz}, \binits{F.}},
\oauthor{\bsnm{Staudacher}, \binits{K.}},
\oauthor{\bsnm{Letras}, \binits{M.}},
\oauthor{\bsnm{Deng}, \binits{X.}},
\oauthor{\bsnm{Vavilala}, \binits{M.}},
\oauthor{\bsnm{Raoofy}, \binits{A.}},
\oauthor{\bsnm{Echavarria}, \binits{J.}},
\oauthor{\bsnm{Iapichino}, \binits{L.}},
\oauthor{\bsnm{Schulz}, \binits{L.}},
\oauthor{\bsnm{Weidendorfer}, \binits{J.}},
\oauthor{\bsnm{Schulz}, \binits{M.}}:
Q-BEAST: A Practical Course on Experimental Evaluation and Characterization of Quantum Computing Systems
(2025).
\url{https://arxiv.org/abs/2508.14084}
\end{botherref}
\endbibitem

\bibitem[\protect\citeauthoryear{Burgholzer et~al.}{2026}]{burgholzer2026}
\begin{bchapter}
\bauthor{\bsnm{Burgholzer}, \binits{L.}},
\bauthor{\bsnm{Echavarria}, \binits{J.}},
\bauthor{\bsnm{Hopf}, \binits{P.}},
\bauthor{\bsnm{Stade}, \binits{Y.}},
\bauthor{\bsnm{Rovara}, \binits{D.}},
\bauthor{\bsnm{Schmid}, \binits{L.}},
\bauthor{\bsnm{Kaya}, \binits{E.}},
\bauthor{\bsnm{Mete}, \binits{B.}},
\bauthor{\bsnm{Farooqi}, \binits{M.N.}},
\bauthor{\bsnm{Chung}, \binits{M.}},
\bauthor{\bsnm{De~Pascale}, \binits{M.}},
\bauthor{\bsnm{Schulz}, \binits{L.}},
\bauthor{\bsnm{Schulz}, \binits{M.}},
\bauthor{\bsnm{Wille}, \binits{R.}}:
\bctitle{The munich quantum software stack: Connecting end users, integrating diverse quantum technologies, accelerating hpc}.
In: \bbtitle{Proceedings of the Supercomputing Asia and International Conference on High Performance Computing in Asia Pacific Region}.
\bsertitle{SCA/HPCAsia '26},
pp. \bfpage{55}--\blpage{67}.
\bpublisher{Association for Computing Machinery},
\blocation{New York, NY, USA}
(\byear{2026}).
\doiurl{10.1145/3773656.3773669}
\end{bchapter}
\endbibitem

\bibitem[\protect\citeauthoryear{Yoo et~al.}{2003}]{yoo2003}
\begin{bchapter}
\bauthor{\bsnm{Yoo}, \binits{A.B.}},
\bauthor{\bsnm{Jette}, \binits{M.A.}},
\bauthor{\bsnm{Grondona}, \binits{M.}}:
\bctitle{Slurm: Simple linux utility for resource management}.
In: \beditor{\bsnm{Feitelson}, \binits{D.}},
\beditor{\bsnm{Rudolph}, \binits{L.}},
\beditor{\bsnm{Schwiegelshohn}, \binits{U.}} (eds.)
\bbtitle{Job Scheduling Strategies for Parallel Processing},
vol. \bseriesno{2862},
pp. \bfpage{44}--\blpage{60}.
\bpublisher{Springer},
\blocation{Berlin, Heidelberg}
(\byear{2003}).
\doiurl{10.1007/10968987_3}
\end{bchapter}
\endbibitem

\bibitem[\protect\citeauthoryear{Farooqi et~al.}{2026}]{farooqi2026enahpcqcworkflow}
\begin{botherref}
\oauthor{\bsnm{Farooqi}, \binits{M.N.}},
\oauthor{\bsnm{Chung}, \binits{M.}},
\oauthor{\bsnm{Mete}, \binits{B.}},
\oauthor{\bsnm{Mansfield}, \binits{E.}},
\oauthor{\bsnm{Hoffmann}, \binits{B.}},
\oauthor{\bsnm{Mattsson}, \binits{T.}},
\oauthor{\bsnm{Schulz}, \binits{L.}},
\oauthor{\bsnm{Echavarria}, \binits{J.}}:
Enabling Hybrid HPCQC Workflows with a Heterogeneous Software Stack
(2026).
\url{https://arxiv.org/abs/2608.14827}
\end{botherref}
\endbibitem

\bibitem[\protect\citeauthoryear{{NVIDIA Corporation}}{2024}]{cudaq}
\begin{botherref}
\oauthor{\bsnm{{NVIDIA Corporation}}}:
{NVIDIA CUDA-Q}.
\url{https://nvidia.github.io/cuda-quantum/}.
Accessed: 2025-11-04
(2024)
\end{botherref}
\endbibitem

\bibitem[\protect\citeauthoryear{Orlov et~al.}{2026}]{orlov2026data}
\begin{botherref}
\oauthor{\bsnm{Orlov}, \binits{A.}},
\oauthor{\bsnm{Ralli}, \binits{A.}},
\oauthor{\bsnm{Weaving}, \binits{T.}},
\oauthor{\bsnm{Bastida}, \binits{M.}},
\oauthor{\bsnm{Martin}, \binits{R.}},
\oauthor{\bsnm{Chandani}, \binits{Z.}},
\oauthor{\bsnm{Chung}, \binits{M.}},
\oauthor{\bsnm{{\v{S}}imkovic~IV}, \binits{F.}},
\oauthor{\bsnm{Sriluckshmy}, \binits{P.V.}},
\oauthor{\bsnm{Iapichino}, \binits{L.}},
\oauthor{\bsnm{Mendl}, \binits{C.B.}},
\oauthor{\bsnm{Townsend-Nicholson}, \binits{A.}},
\oauthor{\bsnm{Coveney}, \binits{P.V.}}:
Data associated with ``Predicting reaction barrier heights in adhesion GPCR autoproteolysis with a multiscale classical-quantum computing pipeline''.
Zenodo
(2026).
\doiurl{10.5281/zenodo.21417282} .
\url{https://doi.org/10.5281/zenodo.21417282}
\end{botherref}
\endbibitem

\bibitem[\protect\citeauthoryear{{UCL Centre for Computational Science}}{2026}]{gpcr2026github}
\begin{botherref}
\oauthor{\bsnm{{UCL Centre for Computational Science}}}:
GPCR.
\url{https://github.com/UCL-CCS/GPCR}.
GitHub repository
(2026)
\end{botherref}
\endbibitem

\end{thebibliography}



\begin{thebibliography}{79}
\ifx \bisbn   \undefined \def \bisbn  #1{ISBN #1}\fi
\ifx \binits  \undefined \def \binits#1{#1}\fi
\ifx \bauthor  \undefined \def \bauthor#1{#1}\fi
\ifx \batitle  \undefined \def \batitle#1{#1}\fi
\ifx \bjtitle  \undefined \def \bjtitle#1{#1}\fi
\ifx \bvolume  \undefined \def \bvolume#1{\textbf{#1}}\fi
\ifx \byear  \undefined \def \byear#1{#1}\fi
\ifx \bissue  \undefined \def \bissue#1{#1}\fi
\ifx \bfpage  \undefined \def \bfpage#1{#1}\fi
\ifx \blpage  \undefined \def \blpage #1{#1}\fi
\ifx \burl  \undefined \def \burl#1{\textsf{#1}}\fi
\ifx \doiurl  \undefined \def \doiurl#1{\url{https://doi.org/#1}}\fi
\ifx \betal  \undefined \def \betal{\textit{et al.}}\fi
\ifx \binstitute  \undefined \def \binstitute#1{#1}\fi
\ifx \binstitutionaled  \undefined \def \binstitutionaled#1{#1}\fi
\ifx \bctitle  \undefined \def \bctitle#1{#1}\fi
\ifx \beditor  \undefined \def \beditor#1{#1}\fi
\ifx \bpublisher  \undefined \def \bpublisher#1{#1}\fi
\ifx \bbtitle  \undefined \def \bbtitle#1{#1}\fi
\ifx \bedition  \undefined \def \bedition#1{#1}\fi
\ifx \bseriesno  \undefined \def \bseriesno#1{#1}\fi
\ifx \blocation  \undefined \def \blocation#1{#1}\fi
\ifx \bsertitle  \undefined \def \bsertitle#1{#1}\fi
\ifx \bsnm \undefined \def \bsnm#1{#1}\fi
\ifx \bsuffix \undefined \def \bsuffix#1{#1}\fi
\ifx \bparticle \undefined \def \bparticle#1{#1}\fi
\ifx \barticle \undefined \def \barticle#1{#1}\fi
\bibcommenthead
\ifx \bconfdate \undefined \def \bconfdate #1{#1}\fi
\ifx \botherref \undefined \def \botherref #1{#1}\fi
\ifx \url \undefined \def \url#1{\textsf{#1}}\fi
\ifx \bchapter \undefined \def \bchapter#1{#1}\fi
\ifx \bbook \undefined \def \bbook#1{#1}\fi
\ifx \bcomment \undefined \def \bcomment#1{#1}\fi
\ifx \oauthor \undefined \def \oauthor#1{#1}\fi
\ifx \citeauthoryear \undefined \def \citeauthoryear#1{#1}\fi
\ifx \endbibitem  \undefined \def \endbibitem {}\fi
\ifx \bconflocation  \undefined \def \bconflocation#1{#1}\fi
\ifx \arxivurl  \undefined \def \arxivurl#1{\textsf{#1}}\fi
\csname PreBibitemsHook\endcsname

\bibitem[\protect\citeauthoryear{{GPCRdb}}{2026a}]{gpcrdb_adgrg1_active}
\begin{botherref}
\oauthor{\bsnm{{GPCRdb}}}:
{ADGRG1} ({GPR56}) human active state homology model.
Accessed: 2026-02-18
(2026).
\url{https://gpcrdb.org/structure/homology_models/agrg1_human_active}
\end{botherref}
\endbibitem

\bibitem[\protect\citeauthoryear{{GPCRdb}}{2026b}]{gpcrdb_adgrg1_inactive}
\begin{botherref}
\oauthor{\bsnm{{GPCRdb}}}:
{ADGRG1} ({GPR56}) human inactive state homology model.
Accessed: 2026-02-18
(2026).
\url{https://gpcrdb.org/structure/homology_models/agrg1_human_inactive}
\end{botherref}
\endbibitem

\bibitem[\protect\citeauthoryear{Jumper et~al.}{2021}]{jumper2021}
\begin{barticle}
\bauthor{\bsnm{Jumper}, \binits{J.}},
\bauthor{\bsnm{Evans}, \binits{R.}},
\bauthor{\bsnm{Pritzel}, \binits{A.}},
\bauthor{\bsnm{Green}, \binits{T.}},
\bauthor{\bsnm{Figurnov}, \binits{M.}},
\bauthor{\bsnm{Ronneberger}, \binits{O.}},
\bauthor{\bsnm{Tunyasuvunakool}, \binits{K.}},
\bauthor{\bsnm{Bates}, \binits{R.}},
\bauthor{\bsnm{{\v{Z}}{\'i}dek}, \binits{A.}},
\bauthor{\bsnm{Potapenko}, \binits{A.}},
\bauthor{\bsnm{Bridgland}, \binits{A.}},
\bauthor{\bsnm{Meyer}, \binits{C.}},
\bauthor{\bsnm{Kohl}, \binits{S.A.A.}},
\bauthor{\bsnm{Ballard}, \binits{A.J.}},
\bauthor{\bsnm{Cowie}, \binits{A.}},
\bauthor{\bsnm{Romera-Paredes}, \binits{B.}},
\bauthor{\bsnm{Nikolov}, \binits{S.}},
\bauthor{\bsnm{Jain}, \binits{R.}},
\bauthor{\bsnm{Adler}, \binits{J.}},
\bauthor{\bsnm{Back}, \binits{T.}},
\bauthor{\bsnm{Petersen}, \binits{S.}},
\bauthor{\bsnm{Reiman}, \binits{D.}},
\bauthor{\bsnm{Clancy}, \binits{E.}},
\bauthor{\bsnm{Zielinski}, \binits{M.}},
\bauthor{\bsnm{Steinegger}, \binits{M.}},
\bauthor{\bsnm{Pacholska}, \binits{M.}},
\bauthor{\bsnm{Berghammer}, \binits{T.}},
\bauthor{\bsnm{Bodenstein}, \binits{S.}},
\bauthor{\bsnm{Silver}, \binits{D.}},
\bauthor{\bsnm{Vinyals}, \binits{O.}},
\bauthor{\bsnm{Senior}, \binits{A.W.}},
\bauthor{\bsnm{Kavukcuoglu}, \binits{K.}},
\bauthor{\bsnm{Kohli}, \binits{P.}},
\bauthor{\bsnm{Hassabis}, \binits{D.}}:
\batitle{Highly accurate protein structure prediction with alphafold}.
\bjtitle{Nature}
\bvolume{596}(\bissue{7873}),
\bfpage{583}--\blpage{589}
(\byear{2021})
\doiurl{10.1038/s41586-021-03819-2}
\end{barticle}
\endbibitem

\bibitem[\protect\citeauthoryear{Varadi et~al.}{2021}]{varadi2021}
\begin{barticle}
\bauthor{\bsnm{Varadi}, \binits{M.}},
\bauthor{\bsnm{Anyango}, \binits{S.}},
\bauthor{\bsnm{Deshpande}, \binits{M.}},
\bauthor{\bsnm{Nair}, \binits{S.}},
\bauthor{\bsnm{Natassia}, \binits{C.}},
\bauthor{\bsnm{Yordanova}, \binits{G.}},
\bauthor{\bsnm{Yuan}, \binits{D.}},
\bauthor{\bsnm{Stroe}, \binits{O.}},
\bauthor{\bsnm{Wood}, \binits{G.}},
\bauthor{\bsnm{Laydon}, \binits{A.}},
\bauthor{\bsnm{Žídek}, \binits{A.}},
\bauthor{\bsnm{Green}, \binits{T.}},
\bauthor{\bsnm{Tunyasuvunakool}, \binits{K.}},
\bauthor{\bsnm{Petersen}, \binits{S.}},
\bauthor{\bsnm{Jumper}, \binits{J.}},
\bauthor{\bsnm{Clancy}, \binits{E.}},
\bauthor{\bsnm{Green}, \binits{R.}},
\bauthor{\bsnm{Vora}, \binits{A.}},
\bauthor{\bsnm{Lutfi}, \binits{M.}},
\bauthor{\bsnm{Figurnov}, \binits{M.}},
\bauthor{\bsnm{Cowie}, \binits{A.}},
\bauthor{\bsnm{Hobbs}, \binits{N.}},
\bauthor{\bsnm{Kohli}, \binits{P.}},
\bauthor{\bsnm{Kleywegt}, \binits{G.}},
\bauthor{\bsnm{Birney}, \binits{E.}},
\bauthor{\bsnm{Hassabis}, \binits{D.}},
\bauthor{\bsnm{Velankar}, \binits{S.}}:
\batitle{Alphafold protein structure database: massively expanding the structural coverage of protein-sequence space with high-accuracy models}.
\bjtitle{Nucleic Acids Research}
\bvolume{50}(\bissue{D1}),
\bfpage{439}--\blpage{444}
(\byear{2021})
\doiurl{10.1093/nar/gkab1061}
\end{barticle}
\endbibitem

\bibitem[\protect\citeauthoryear{Jo et~al.}{2008}]{jo2008}
\begin{barticle}
\bauthor{\bsnm{Jo}, \binits{S.}},
\bauthor{\bsnm{Kim}, \binits{T.}},
\bauthor{\bsnm{Iyer}, \binits{V.G.}},
\bauthor{\bsnm{Im}, \binits{W.}}:
\batitle{Charmm-gui: A web-based graphical user interface for charmm}.
\bjtitle{Journal of Computational Chemistry}
\bvolume{29}(\bissue{11}),
\bfpage{1859}--\blpage{1865}
(\byear{2008})
\doiurl{10.1002/jcc.20945}
\end{barticle}
\endbibitem

\bibitem[\protect\citeauthoryear{Lee et~al.}{2016}]{lee2016}
\begin{barticle}
\bauthor{\bsnm{Lee}, \binits{J.}},
\bauthor{\bsnm{Cheng}, \binits{X.}},
\bauthor{\bsnm{Swails}, \binits{J.M.}},
\bauthor{\bsnm{Yeom}, \binits{M.S.}},
\bauthor{\bsnm{Eastman}, \binits{P.K.}},
\bauthor{\bsnm{Lemkul}, \binits{J.A.}},
\bauthor{\bsnm{Wei}, \binits{S.}},
\bauthor{\bsnm{Buckner}, \binits{J.}},
\bauthor{\bsnm{Jeong}, \binits{J.C.}},
\bauthor{\bsnm{Qi}, \binits{Y.}},
\bauthor{\bsnm{Jo}, \binits{S.}},
\bauthor{\bsnm{Pande}, \binits{V.S.}},
\bauthor{\bsnm{Case}, \binits{D.A.}},
\bauthor{\bsnm{Brooks}, \binits{C.L.I.}},
\bauthor{\bsnm{MacKerell}, \binits{A.D.J.}},
\bauthor{\bsnm{Klauda}, \binits{J.B.}},
\bauthor{\bsnm{Im}, \binits{W.}}:
\batitle{Charmm-gui input generator for namd, gromacs, amber, openmm, and charmm/openmm simulations using the charmm36 additive force field}.
\bjtitle{Journal of Chemical Theory and Computation}
\bvolume{12}(\bissue{1}),
\bfpage{405}--\blpage{413}
(\byear{2016})
\doiurl{10.1021/acs.jctc.5b00935}
\end{barticle}
\endbibitem

\bibitem[\protect\citeauthoryear{Hess et~al.}{1997}]{hess1997}
\begin{barticle}
\bauthor{\bsnm{Hess}, \binits{B.}},
\bauthor{\bsnm{Bekker}, \binits{H.}},
\bauthor{\bsnm{Berendsen}, \binits{H.J.C.}},
\bauthor{\bsnm{Fraaije}, \binits{J.G.E.M.}}:
\batitle{{LINCS}: A linear constraint solver for molecular simulations}.
\bjtitle{Journal of Computational Chemistry}
\bvolume{18}(\bissue{12}),
\bfpage{1463}--\blpage{1472}
(\byear{1997})
\doiurl{10.1002/(SICI)1096-987X(199709)18:12<1463::AID-JCC4>3.0.CO;2-}
\end{barticle}
\endbibitem

\bibitem[\protect\citeauthoryear{Darden et~al.}{1993}]{darden1993}
\begin{barticle}
\bauthor{\bsnm{Darden}, \binits{T.}},
\bauthor{\bsnm{York}, \binits{D.}},
\bauthor{\bsnm{Pedersen}, \binits{L.}}:
\batitle{Particle mesh {Ewald}: An {N}$\cdot$log({N}) method for {Ewald} sums in large systems}.
\bjtitle{The Journal of Chemical Physics}
\bvolume{98}(\bissue{12}),
\bfpage{10089}--\blpage{10092}
(\byear{1993})
\doiurl{10.1063/1.464397}
\end{barticle}
\endbibitem

\bibitem[\protect\citeauthoryear{Bernetti and Bussi}{2020}]{bernetti2020}
\begin{barticle}
\bauthor{\bsnm{Bernetti}, \binits{M.}},
\bauthor{\bsnm{Bussi}, \binits{G.}}:
\batitle{Pressure control using stochastic cell rescaling}.
\bjtitle{The Journal of Chemical Physics}
\bvolume{153}(\bissue{11}),
\bfpage{114107}
(\byear{2020})
\doiurl{10.1063/5.0020514}
\end{barticle}
\endbibitem

\bibitem[\protect\citeauthoryear{Seufert et~al.}{2025}]{seufert2025}
\begin{barticle}
\bauthor{\bsnm{Seufert}, \binits{F.}},
\bauthor{\bsnm{Chung}, \binits{Y.K.}},
\bauthor{\bsnm{Schick}, \binits{R.}},
\bauthor{\bsnm{Pohl}, \binits{F.}},
\bauthor{\bsnm{Str{\"a}ter}, \binits{N.}},
\bauthor{\bsnm{Langenhan}, \binits{T.}},
\bauthor{\bsnm{Batebi}, \binits{H.}},
\bauthor{\bsnm{Hildebrand}, \binits{P.W.}}:
\batitle{Mechanistic insights into adhesion {GPCR} autoproteolysis by a multiscale computational approach}.
\bjtitle{Research Square (preprint)}
(\byear{2025})
\doiurl{10.21203/rs.3.rs-7121662/v1}
\end{barticle}
\endbibitem

\bibitem[\protect\citeauthoryear{Laino et~al.}{2005}]{laino2005}
\begin{barticle}
\bauthor{\bsnm{Laino}, \binits{T.}},
\bauthor{\bsnm{Mohamed}, \binits{F.}},
\bauthor{\bsnm{Laio}, \binits{A.}},
\bauthor{\bsnm{Parrinello}, \binits{M.}}:
\batitle{An efficient real space multigrid qm/mm electrostatic coupling}.
\bjtitle{Journal of Chemical Theory and Computation}
\bvolume{1}(\bissue{6}),
\bfpage{1176}--\blpage{1184}
(\byear{2005})
\doiurl{10.1021/ct050123f} .
\bcomment{PMID: 26631661}
\end{barticle}
\endbibitem

\bibitem[\protect\citeauthoryear{Laino et~al.}{2006}]{laino2006}
\begin{barticle}
\bauthor{\bsnm{Laino}, \binits{T.}},
\bauthor{\bsnm{Mohamed}, \binits{F.}},
\bauthor{\bsnm{Laio}, \binits{A.}},
\bauthor{\bsnm{Parrinello}, \binits{M.}}:
\batitle{An efficient linear-scaling electrostatic coupling for treating periodic boundary conditions in qm/mm simulations}.
\bjtitle{Journal of Chemical Theory and Computation}
\bvolume{2}(\bissue{5}),
\bfpage{1370}--\blpage{1378}
(\byear{2006})
\doiurl{10.1021/ct6001169} .
\bcomment{PMID: 26626844}
\end{barticle}
\endbibitem

\bibitem[\protect\citeauthoryear{Manby et~al.}{2012}]{manby2012}
\begin{barticle}
\bauthor{\bsnm{Manby}, \binits{F.R.}},
\bauthor{\bsnm{Stella}, \binits{M.}},
\bauthor{\bsnm{Goodpaster}, \binits{J.D.}},
\bauthor{\bsnm{Miller~III}, \binits{T.F.}}:
\batitle{A simple, exact density-functional-theory embedding scheme}.
\bjtitle{Journal of Chemical Theory and Computation}
\bvolume{8}(\bissue{8}),
\bfpage{2564}--\blpage{2568}
(\byear{2012})
\doiurl{10.1021/ct300544e}
\end{barticle}
\endbibitem

\bibitem[\protect\citeauthoryear{Goodpaster et~al.}{2014}]{goodpaster2014}
\begin{barticle}
\bauthor{\bsnm{Goodpaster}, \binits{J.D.}},
\bauthor{\bsnm{Barnes}, \binits{T.A.}},
\bauthor{\bsnm{Manby}, \binits{F.R.}},
\bauthor{\bsnm{Miller~III}, \binits{T.F.}}:
\batitle{Accurate and systematically improvable density functional theory embedding for correlated wavefunctions}.
\bjtitle{The Journal of Chemical Physics}
\bvolume{140}(\bissue{18}),
\bfpage{18}--\blpage{507}
(\byear{2014})
\doiurl{10.1063/1.4864040}
\end{barticle}
\endbibitem

\bibitem[\protect\citeauthoryear{Ralli et~al.}{2024}]{ralli2024}
\begin{barticle}
\bauthor{\bsnm{Ralli}, \binits{A.}},
\bauthor{\bsnm{Bastida}, \binits{M.}},
\bauthor{\bsnm{Coveney}, \binits{P.V.}}:
\batitle{Scalable approach to quantum simulation via projection-based embedding}.
\bjtitle{Physical Review A}
\bvolume{109},
\bfpage{022418}
(\byear{2024})
\doiurl{10.1103/PhysRevA.109.022418}
\end{barticle}
\endbibitem

\bibitem[\protect\citeauthoryear{Pipek and Mezey}{1989}]{pipek1989}
\begin{barticle}
\bauthor{\bsnm{Pipek}, \binits{J.}},
\bauthor{\bsnm{Mezey}, \binits{P.G.}}:
\batitle{A fast intrinsic localization procedure applicable for ab initio and semiempirical linear combination of atomic orbital wave functions}.
\bjtitle{The Journal of Chemical Physics}
\bvolume{90}(\bissue{9}),
\bfpage{4916}--\blpage{4926}
(\byear{1989})
\doiurl{10.1063/1.456588}
\end{barticle}
\endbibitem

\bibitem[\protect\citeauthoryear{Löwdin}{1950}]{loewdin1950}
\begin{barticle}
\bauthor{\bsnm{Löwdin}, \binits{P.}}:
\batitle{On the non‐orthogonality problem connected with the use of atomic wave functions in the theory of molecules and crystals}.
\bjtitle{The Journal of Chemical Physics}
\bvolume{18}(\bissue{3}),
\bfpage{365}--\blpage{375}
(\byear{1950})
\doiurl{10.1063/1.1747632}
\end{barticle}
\endbibitem

\bibitem[\protect\citeauthoryear{Sun et~al.}{2018}]{sun2018pyscf}
\begin{barticle}
\bauthor{\bsnm{Sun}, \binits{Q.}},
\bauthor{\bsnm{Berkelbach}, \binits{T.C.}},
\bauthor{\bsnm{Blunt}, \binits{N.S.}},
\bauthor{\bsnm{Booth}, \binits{G.H.}},
\bauthor{\bsnm{Guo}, \binits{S.}},
\bauthor{\bsnm{Li}, \binits{Z.}},
\bauthor{\bsnm{Liu}, \binits{J.}},
\bauthor{\bsnm{McClain}, \binits{J.D.}},
\bauthor{\bsnm{Sayfutyarova}, \binits{E.R.}},
\bauthor{\bsnm{Sharma}, \binits{S.}}, \betal:
\batitle{Pyscf: the python-based simulations of chemistry framework}.
\bjtitle{Wiley Interdisciplinary Reviews: Computational Molecular Science}
\bvolume{8}(\bissue{1}),
\bfpage{1340}
(\byear{2018})
\doiurl{10.1002/wcms.1340}
\end{barticle}
\endbibitem

\bibitem[\protect\citeauthoryear{Orlov et~al.}{2026}]{orlov2026data}
\begin{botherref}
\oauthor{\bsnm{Orlov}, \binits{A.}},
\oauthor{\bsnm{Ralli}, \binits{A.}},
\oauthor{\bsnm{Weaving}, \binits{T.}},
\oauthor{\bsnm{Bastida}, \binits{M.}},
\oauthor{\bsnm{Martin}, \binits{R.}},
\oauthor{\bsnm{Chandani}, \binits{Z.}},
\oauthor{\bsnm{Chung}, \binits{M.}},
\oauthor{\bsnm{{\v{S}}imkovic~IV}, \binits{F.}},
\oauthor{\bsnm{Sriluckshmy}, \binits{P.V.}},
\oauthor{\bsnm{Iapichino}, \binits{L.}},
\oauthor{\bsnm{Mendl}, \binits{C.B.}},
\oauthor{\bsnm{Townsend-Nicholson}, \binits{A.}},
\oauthor{\bsnm{Coveney}, \binits{P.V.}}:
Data associated with ``Predicting reaction barrier heights in adhesion GPCR autoproteolysis with a multiscale classical-quantum computing pipeline''.
Zenodo
(2026).
\doiurl{10.5281/zenodo.21417282} .
\url{https://doi.org/10.5281/zenodo.21417282}
\end{botherref}
\endbibitem

\bibitem[\protect\citeauthoryear{Jordan and Wigner}{1928}]{jordan1928}
\begin{barticle}
\bauthor{\bsnm{Jordan}, \binits{P.}},
\bauthor{\bsnm{Wigner}, \binits{E.}}:
\batitle{{\"U}ber das {P}aulische {\"a}quivalenzverbot}.
\bjtitle{Zeitschrift f{\"u}r Physik}
\bvolume{47}(\bissue{9--10}),
\bfpage{631}--\blpage{651}
(\byear{1928})
\doiurl{10.1007/BF01331938}
\end{barticle}
\endbibitem

\bibitem[\protect\citeauthoryear{{Williams de la Bastida} et~al.}{2026}]{delabastida2026optimisedfermionqubitencodingsquantum}
\begin{barticle}
\bauthor{\bsnm{{Williams de la Bastida}}, \binits{M.I.}},
\bauthor{\bsnm{Bickley}, \binits{T.M.}},
\bauthor{\bsnm{Coveney}, \binits{P.V.}}:
\batitle{Optimised {{Fermion-Qubit Encodings}} for {{Quantum Simulation}} with {{Reduced Circuit Depth}}}.
\bjtitle{Quantum Science and Technology}
(\byear{2026})
\doiurl{10.1088/2058-9565/ae9b40}
\end{barticle}
\endbibitem

\bibitem[\protect\citeauthoryear{Williams de~la Bastida and Bickley}{2026}]{topphatt}
\begin{botherref}
\oauthor{\bsnm{Bastida}, \binits{M.}},
\oauthor{\bsnm{Bickley}, \binits{T.}}:
UCL-{CCS}/ferrmion
(2026).
\url{https://github.com/UCL-CCS/ferrmion}
\end{botherref}
\endbibitem

\bibitem[\protect\citeauthoryear{Lanczos}{1950}]{lanczos1950iteration}
\begin{barticle}
\bauthor{\bsnm{Lanczos}, \binits{C.}}:
\batitle{An iteration method for the solution of the eigenvalue problem of linear differential and integral operators}.
\bjtitle{Journal of research of the National Bureau of Standards}
\bvolume{45}(\bissue{4}),
\bfpage{255}--\blpage{282}
(\byear{1950})
\doiurl{10.6028/jres.045.026}
\end{barticle}
\endbibitem

\bibitem[\protect\citeauthoryear{Davidson}{1975}]{DAVIDSON197587}
\begin{barticle}
\bauthor{\bsnm{Davidson}, \binits{E.R.}}:
\batitle{The iterative calculation of a few of the lowest eigenvalues and corresponding eigenvectors of large real-symmetric matrices}.
\bjtitle{Journal of Computational Physics}
\bvolume{17}(\bissue{1}),
\bfpage{87}--\blpage{94}
(\byear{1975})
\doiurl{10.1016/0021-9991(75)90065-0}
\end{barticle}
\endbibitem

\bibitem[\protect\citeauthoryear{Parrish and McMahon}{2019}]{parrish2019filter}
\begin{botherref}
\oauthor{\bsnm{Parrish}, \binits{R.M.}},
\oauthor{\bsnm{McMahon}, \binits{P.L.}}:
Quantum filter diagonalization: Quantum eigendecomposition without full quantum phase estimation.
arXiv preprint
(2019)
{\href{https://arxiv.org/abs/1909.08925}{{1909.08925}}}
\end{botherref}
\endbibitem

\bibitem[\protect\citeauthoryear{Yoshioka et~al.}{2025}]{yoshioka2025krylov}
\begin{barticle}
\bauthor{\bsnm{Yoshioka}, \binits{N.}},
\bauthor{\bsnm{Amico}, \binits{M.}},
\bauthor{\bsnm{Kirby}, \binits{W.}},
\bauthor{\bsnm{Jurcevic}, \binits{P.}},
\bauthor{\bsnm{Dutt}, \binits{A.}},
\bauthor{\bsnm{Fuller}, \binits{B.}},
\bauthor{\bsnm{Garion}, \binits{S.}},
\bauthor{\bsnm{Haas}, \binits{H.}},
\bauthor{\bsnm{Hamamura}, \binits{I.}},
\bauthor{\bsnm{Ivrii}, \binits{A.}}, \betal:
\batitle{Krylov diagonalization of large many-body hamiltonians on a quantum processor}.
\bjtitle{Nature Communications}
\bvolume{16}(\bissue{1}),
\bfpage{5014}
(\byear{2025})
\doiurl{10.1038/s41467-025-59716-z}
\end{barticle}
\endbibitem

\bibitem[\protect\citeauthoryear{Yu et~al.}{2025}]{Yu2025}
\begin{botherref}
\oauthor{\bsnm{Yu}, \binits{J.}},
\oauthor{\bsnm{Moreno}, \binits{J.R.}},
\oauthor{\bsnm{Iosue}, \binits{J.}},
\oauthor{\bsnm{Bertels}, \binits{L.}},
\oauthor{\bsnm{Claudino}, \binits{D.}},
\oauthor{\bsnm{Fuller}, \binits{B.}},
\oauthor{\bsnm{Groszkowski}, \binits{P.}},
\oauthor{\bsnm{Humble}, \binits{T.S.}},
\oauthor{\bsnm{Jurcevic}, \binits{P.}},
\oauthor{\bsnm{Kirby}, \binits{W.}},
\oauthor{\bsnm{Maier}, \binits{T.A.}},
\oauthor{\bsnm{Motta}, \binits{M.}},
\oauthor{\bsnm{Pokharel}, \binits{B.}},
\oauthor{\bsnm{Seif}, \binits{A.}},
\oauthor{\bsnm{Shehata}, \binits{A.}},
\oauthor{\bsnm{Sung}, \binits{K.J.}},
\oauthor{\bsnm{Tran}, \binits{M.C.}},
\oauthor{\bsnm{Tripathi}, \binits{V.}},
\oauthor{\bsnm{Mezzacapo}, \binits{A.}},
\oauthor{\bsnm{Sharma}, \binits{K.}}:
Sample-based {Krylov} quantum diagonalization.
arXiv preprint
(2025)
{\href{https://arxiv.org/abs/arXiv:2501.09702}{{arXiv:2501.09702}}}
\end{botherref}
\endbibitem

\bibitem[\protect\citeauthoryear{Piccinelli et~al.}{2025}]{piccinelli2025quantum}
\begin{botherref}
\oauthor{\bsnm{Piccinelli}, \binits{S.}},
\oauthor{\bsnm{Baiardi}, \binits{A.}},
\oauthor{\bsnm{Rossmannek}, \binits{M.}},
\oauthor{\bsnm{Vazquez}, \binits{A.C.}},
\oauthor{\bsnm{Tacchino}, \binits{F.}},
\oauthor{\bsnm{Mensa}, \binits{S.}},
\oauthor{\bsnm{Altamura}, \binits{E.}},
\oauthor{\bsnm{Alavi}, \binits{A.}},
\oauthor{\bsnm{Motta}, \binits{M.}},
\oauthor{\bsnm{Robledo-Moreno}, \binits{J.}}, et al.:
Quantum chemistry with provable convergence via randomized sample-based quantum diagonalization.
arXiv preprint
(2025)
{\href{https://arxiv.org/abs/2508.02578}{{2508.02578}}}
\end{botherref}
\endbibitem

\bibitem[\protect\citeauthoryear{Matsuzawa and Kurashige}{2020}]{matsuzawa2020jastrow}
\begin{barticle}
\bauthor{\bsnm{Matsuzawa}, \binits{Y.}},
\bauthor{\bsnm{Kurashige}, \binits{Y.}}:
\batitle{Jastrow-type decomposition in quantum chemistry for low-depth quantum circuits}.
\bjtitle{Journal of chemical theory and computation}
\bvolume{16}(\bissue{2}),
\bfpage{944}--\blpage{952}
(\byear{2020})
\doiurl{10.1021/acs.jctc.9b00963}
\end{barticle}
\endbibitem

\bibitem[\protect\citeauthoryear{Motta et~al.}{2023}]{motta2023bridging}
\begin{barticle}
\bauthor{\bsnm{Motta}, \binits{M.}},
\bauthor{\bsnm{Sung}, \binits{K.J.}},
\bauthor{\bsnm{Whaley}, \binits{K.B.}},
\bauthor{\bsnm{Head-Gordon}, \binits{M.}},
\bauthor{\bsnm{Shee}, \binits{J.}}:
\batitle{Bridging physical intuition and hardware efficiency for correlated electronic states: the local unitary cluster jastrow ansatz for electronic structure}.
\bjtitle{Chemical Science}
\bvolume{14}(\bissue{40}),
\bfpage{11213}--\blpage{11227}
(\byear{2023})
\doiurl{10.1039/D3SC02516K}
\end{barticle}
\endbibitem

\bibitem[\protect\citeauthoryear{Robledo-Moreno et~al.}{2025}]{robledo2025}
\begin{barticle}
\bauthor{\bsnm{Robledo-Moreno}, \binits{J.}},
\bauthor{\bsnm{Motta}, \binits{M.}},
\bauthor{\bsnm{Haas}, \binits{H.}},
\bauthor{\bsnm{Javadi-Abhari}, \binits{A.}},
\bauthor{\bsnm{Jurcevic}, \binits{P.}},
\bauthor{\bsnm{Kirby}, \binits{W.}},
\bauthor{\bsnm{Martiel}, \binits{S.}},
\bauthor{\bsnm{Sharma}, \binits{K.}},
\bauthor{\bsnm{Sharma}, \binits{S.}},
\bauthor{\bsnm{Shirakawa}, \binits{T.}},
\bauthor{\bsnm{Sitdikov}, \binits{I.}},
\bauthor{\bsnm{Sun}, \binits{R.-Y.}},
\bauthor{\bsnm{Sung}, \binits{K.J.}},
\bauthor{\bsnm{Takita}, \binits{M.}},
\bauthor{\bsnm{Tran}, \binits{M.C.}},
\bauthor{\bsnm{Yunoki}, \binits{S.}},
\bauthor{\bsnm{Mezzacapo}, \binits{A.}}:
\batitle{Chemistry beyond the scale of exact diagonalization on a quantum-centric supercomputer}.
\bjtitle{Science Advances}
\bvolume{11}(\bissue{25}),
\bfpage{9991}
(\byear{2025})
\doiurl{10.1126/sciadv.adu9991}
\end{barticle}
\endbibitem

\bibitem[\protect\citeauthoryear{Kaliakin et~al.}{2025}]{kaliakin2025accurate}
\begin{barticle}
\bauthor{\bsnm{Kaliakin}, \binits{D.}},
\bauthor{\bsnm{Shajan}, \binits{A.}},
\bauthor{\bsnm{Liang}, \binits{F.}},
\bauthor{\bsnm{Robledo~Moreno}, \binits{J.}},
\bauthor{\bsnm{Li}, \binits{Z.}},
\bauthor{\bsnm{Mitra}, \binits{A.}},
\bauthor{\bsnm{Motta}, \binits{M.}},
\bauthor{\bsnm{Johnson}, \binits{C.}},
\bauthor{\bsnm{Saki}, \binits{A.A.}},
\bauthor{\bsnm{Das}, \binits{S.}}, \betal:
\batitle{Accurate quantum-centric simulations of intermolecular interactions}.
\bjtitle{Communications Physics}
\bvolume{8}(\bissue{1}),
\bfpage{396}
(\byear{2025})
\doiurl{10.1038/s42005-025-02305-9}
\end{barticle}
\endbibitem

\bibitem[\protect\citeauthoryear{Barison et~al.}{2025}]{barison2025quantum}
\begin{barticle}
\bauthor{\bsnm{Barison}, \binits{S.}},
\bauthor{\bsnm{Moreno}, \binits{J.R.}},
\bauthor{\bsnm{Motta}, \binits{M.}}:
\batitle{Quantum-centric computation of molecular excited states with extended sample-based quantum diagonalization}.
\bjtitle{Quantum Science and Technology}
\bvolume{10}(\bissue{2}),
\bfpage{025034}
(\byear{2025})
\doiurl{10.1088/2058-9565/adb781}
\end{barticle}
\endbibitem

\bibitem[\protect\citeauthoryear{Duriez et~al.}{2026}]{duriez2025computing}
\begin{barticle}
\bauthor{\bsnm{Duriez}, \binits{A.}},
\bauthor{\bsnm{Carvalho}, \binits{P.C.}},
\bauthor{\bsnm{Barroca}, \binits{M.A.}},
\bauthor{\bsnm{Zipoli}, \binits{F.}},
\bauthor{\bsnm{Jaderberg}, \binits{B.}},
\bauthor{\bsnm{Ferreira}, \binits{R.N.B.}},
\bauthor{\bsnm{Sharma}, \binits{K.}},
\bauthor{\bsnm{Mezzacapo}, \binits{A.}},
\bauthor{\bsnm{Wunsch}, \binits{B.}},
\bauthor{\bsnm{Steiner}, \binits{M.}}:
\batitle{Computing band gaps of periodic materials via sample-based quantum diagonalization}.
\bjtitle{npj Computational Materials}
\bvolume{12},
\bfpage{94}
(\byear{2026})
\doiurl{10.1038/s41524-026-02059-0}
{\href{https://arxiv.org/abs/2503.10901}{{2503.10901}}}
\end{barticle}
\endbibitem

\bibitem[\protect\citeauthoryear{Barroca et~al.}{2025}]{barroca2025surface}
\begin{botherref}
\oauthor{\bsnm{Barroca}, \binits{M.A.}},
\oauthor{\bsnm{Gujarati}, \binits{T.}},
\oauthor{\bsnm{Sharma}, \binits{V.}},
\oauthor{\bsnm{Ferreira}, \binits{R.N.B.}},
\oauthor{\bsnm{Na}, \binits{Y.-H.}},
\oauthor{\bsnm{Giammona}, \binits{M.}},
\oauthor{\bsnm{Mezzacapo}, \binits{A.}},
\oauthor{\bsnm{Wunsch}, \binits{B.}},
\oauthor{\bsnm{Steiner}, \binits{M.}}:
Surface reaction simulations for battery materials through sample-based quantum diagonalization and local embedding.
arXiv preprint
(2025)
{\href{https://arxiv.org/abs/2503.10923}{{2503.10923}}}
\end{botherref}
\endbibitem

\bibitem[\protect\citeauthoryear{Nogaki et~al.}{2025}]{nogaki2025symmetry}
\begin{botherref}
\oauthor{\bsnm{Nogaki}, \binits{K.}},
\oauthor{\bsnm{Backes}, \binits{S.}},
\oauthor{\bsnm{Shirakawa}, \binits{T.}},
\oauthor{\bsnm{Yunoki}, \binits{S.}},
\oauthor{\bsnm{Arita}, \binits{R.}}:
Symmetry-adapted sample-based quantum diagonalization: Application to lattice model.
arXiv preprint
(2025)
{\href{https://arxiv.org/abs/2505.00914}{{2505.00914}}}
\end{botherref}
\endbibitem

\bibitem[\protect\citeauthoryear{Liepuoniute et~al.}{2025}]{liepuoniute2024quantum}
\begin{barticle}
\bauthor{\bsnm{Liepuoniute}, \binits{I.}},
\bauthor{\bsnm{Doney}, \binits{K.D.}},
\bauthor{\bsnm{Robledo~Moreno}, \binits{J.}},
\bauthor{\bsnm{Job}, \binits{J.A.}},
\bauthor{\bsnm{Friend}, \binits{W.S.}},
\bauthor{\bsnm{Jones}, \binits{G.O.}}:
\batitle{Quantum-centric computational study of methylene singlet and triplet states}.
\bjtitle{Journal of Chemical Theory and Computation}
\bvolume{21}(\bissue{10}),
\bfpage{5062}--\blpage{5070}
(\byear{2025})
\doiurl{10.1021/acs.jctc.5c00075}
\end{barticle}
\endbibitem

\bibitem[\protect\citeauthoryear{Kaliakin et~al.}{2025}]{kaliakin2025implicit}
\begin{barticle}
\bauthor{\bsnm{Kaliakin}, \binits{D.}},
\bauthor{\bsnm{Shajan}, \binits{A.}},
\bauthor{\bsnm{Liang}, \binits{F.}},
\bauthor{\bsnm{Merz~Jr}, \binits{K.M.}}:
\batitle{Implicit solvent sample-based quantum diagonalization}.
\bjtitle{The Journal of Physical Chemistry B}
\bvolume{129}(\bissue{23}),
\bfpage{5788}--\blpage{5796}
(\byear{2025})
\doiurl{10.1021/acs.jpcb.5c01030}
\end{barticle}
\endbibitem

\bibitem[\protect\citeauthoryear{Bazayeva et~al.}{2025}]{bazayeva2025quantum}
\begin{botherref}
\oauthor{\bsnm{Bazayeva}, \binits{M.}},
\oauthor{\bsnm{Li}, \binits{Z.}},
\oauthor{\bsnm{Kaliakin}, \binits{D.}},
\oauthor{\bsnm{Liang}, \binits{F.}},
\oauthor{\bsnm{Shajan}, \binits{A.}},
\oauthor{\bsnm{Das}, \binits{S.}},
\oauthor{\bsnm{Merz~Jr}, \binits{K.M.}}:
Quantum-centric alchemical free energy calculations.
arXiv preprint
(2025)
{\href{https://arxiv.org/abs/2506.20825}{{2506.20825}}}
\end{botherref}
\endbibitem

\bibitem[\protect\citeauthoryear{Sriluckshmy et~al.}{2025}]{sriluckshmy2025quantum}
\begin{botherref}
\oauthor{\bsnm{Sriluckshmy}, \binits{P.}},
\oauthor{\bsnm{Jamet}, \binits{F.}},
\oauthor{\bsnm{{\v{S}}imkovic~IV}, \binits{F.}}:
Quantum assisted ghost gutzwiller ansatz.
arXiv preprint
(2025)
{\href{https://arxiv.org/abs/2506.21431}{{2506.21431}}}
\end{botherref}
\endbibitem

\bibitem[\protect\citeauthoryear{Shajan et~al.}{2025}]{shajan2025toward}
\begin{barticle}
\bauthor{\bsnm{Shajan}, \binits{A.}},
\bauthor{\bsnm{Kaliakin}, \binits{D.}},
\bauthor{\bsnm{Mitra}, \binits{A.}},
\bauthor{\bsnm{Robledo~Moreno}, \binits{J.}},
\bauthor{\bsnm{Li}, \binits{Z.}},
\bauthor{\bsnm{Motta}, \binits{M.}},
\bauthor{\bsnm{Johnson}, \binits{C.}},
\bauthor{\bsnm{Saki}, \binits{A.A.}},
\bauthor{\bsnm{Das}, \binits{S.}},
\bauthor{\bsnm{Sitdikov}, \binits{I.}}, \betal:
\batitle{Toward quantum-centric simulations of extended molecules: Sample-based quantum diagonalization enhanced with density matrix embedding theory}.
\bjtitle{Journal of Chemical Theory and Computation}
\bvolume{21}(\bissue{14}),
\bfpage{6801}--\blpage{6810}
(\byear{2025})
\doiurl{10.1021/acs.jctc.5c00114}
\end{barticle}
\endbibitem

\bibitem[\protect\citeauthoryear{Smith et~al.}{2025}]{smith2025quantum}
\begin{botherref}
\oauthor{\bsnm{Smith}, \binits{T.}},
\oauthor{\bsnm{Gujarati}, \binits{T.P.}},
\oauthor{\bsnm{Motta}, \binits{M.}},
\oauthor{\bsnm{Link}, \binits{B.}},
\oauthor{\bsnm{Liepuoniute}, \binits{I.}},
\oauthor{\bsnm{Friedhoff}, \binits{T.}},
\oauthor{\bsnm{Nishimura}, \binits{H.}},
\oauthor{\bsnm{Nguyen}, \binits{N.}},
\oauthor{\bsnm{Williams}, \binits{K.S.}},
\oauthor{\bsnm{Moreno}, \binits{J.R.}}, et al.:
Quantum-centric simulation of hydrogen abstraction by sample-based quantum diagonalization and entanglement forging.
arXiv preprint
(2025)
{\href{https://arxiv.org/abs/2508.08229}{{2508.08229}}}
\end{botherref}
\endbibitem

\bibitem[\protect\citeauthoryear{Danilov et~al.}{2025}]{danilov2025enhancing}
\begin{barticle}
\bauthor{\bsnm{Danilov}, \binits{D.}},
\bauthor{\bsnm{Robledo-Moreno}, \binits{J.}},
\bauthor{\bsnm{Sung}, \binits{K.J.}},
\bauthor{\bsnm{Motta}, \binits{M.}},
\bauthor{\bsnm{Shee}, \binits{J.}}:
\batitle{Enhancing the accuracy and efficiency of sample-based quantum diagonalization with phaseless auxiliary-field quantum monte carlo}.
\bjtitle{Journal of Chemical Theory and Computation}
\bvolume{21}(\bissue{22}),
\bfpage{11585}--\blpage{11594}
(\byear{2025})
\doiurl{10.1021/acs.jctc.5c01407}
\end{barticle}
\endbibitem

\bibitem[\protect\citeauthoryear{Shirakawa et~al.}{2025}]{shirakawa2025closed}
\begin{botherref}
\oauthor{\bsnm{Shirakawa}, \binits{T.}},
\oauthor{\bsnm{Robledo-Moreno}, \binits{J.}},
\oauthor{\bsnm{Itoko}, \binits{T.}},
\oauthor{\bsnm{Tripathi}, \binits{V.}},
\oauthor{\bsnm{Ueda}, \binits{K.}},
\oauthor{\bsnm{Kawashima}, \binits{Y.}},
\oauthor{\bsnm{Broers}, \binits{L.}},
\oauthor{\bsnm{Kirby}, \binits{W.}},
\oauthor{\bsnm{Pathak}, \binits{H.}},
\oauthor{\bsnm{Paik}, \binits{H.}}, et al.:
Closed-loop calculations of electronic structure on a quantum processor and a classical supercomputer at full scale.
arXiv preprint
(2025)
{\href{https://arxiv.org/abs/2511.00224}{{2511.00224}}}
\end{botherref}
\endbibitem

\bibitem[\protect\citeauthoryear{McFarthing et~al.}{2026}]{mcfarthing2026noise}
\begin{botherref}
\oauthor{\bsnm{McFarthing}, \binits{S.}},
\oauthor{\bsnm{Pellow-Jarman}, \binits{A.}},
\oauthor{\bsnm{Petruccione}, \binits{F.}}:
Noise-resilient quantum chemistry with half the qubits.
arXiv preprint
(2026)
{\href{https://arxiv.org/abs/2602.01165}{{2602.01165}}}
\end{botherref}
\endbibitem

\bibitem[\protect\citeauthoryear{Park et~al.}{2026}]{park2026cluster}
\begin{botherref}
\oauthor{\bsnm{Park}, \binits{B.}},
\oauthor{\bsnm{Kang}, \binits{S.}},
\oauthor{\bsnm{Seo}, \binits{J.}},
\oauthor{\bsnm{Baek}, \binits{J.}},
\oauthor{\bsnm{Ahn}, \binits{D.}},
\oauthor{\bsnm{Jeong}, \binits{K.}}:
Cluster-adaptive sample-based quantum diagonalization for strongly correlated systems.
arXiv preprint
(2026)
{\href{https://arxiv.org/abs/2603.09346}{{2603.09346}}}
\end{botherref}
\endbibitem

\bibitem[\protect\citeauthoryear{Li et~al.}{2026}]{li2026protein}
\begin{botherref}
\oauthor{\bsnm{Li}, \binits{Z.}},
\oauthor{\bsnm{Bazayeva}, \binits{M.}},
\oauthor{\bsnm{Pellegrini}, \binits{T.}},
\oauthor{\bsnm{Bhowmik}, \binits{S.}},
\oauthor{\bsnm{Das}, \binits{S.}},
\oauthor{\bsnm{Kaliakin}, \binits{D.}},
\oauthor{\bsnm{Liang}, \binits{F.}},
\oauthor{\bsnm{Shajan}, \binits{A.}},
\oauthor{\bsnm{Merz~Jr}, \binits{K.M.}}:
Protein-ligand free energy perturbation on quantum hardware.
arXiv preprint
(2026)
{\href{https://arxiv.org/abs/2604.09857}{{2604.09857}}}
\end{botherref}
\endbibitem

\bibitem[\protect\citeauthoryear{Shajan et~al.}{2026}]{shajan2026molecular}
\begin{barticle}
\bauthor{\bsnm{Shajan}, \binits{A.}},
\bauthor{\bsnm{Kaliakin}, \binits{D.}},
\bauthor{\bsnm{Liang}, \binits{F.}},
\bauthor{\bsnm{Pellegrini}, \binits{T.}},
\bauthor{\bsnm{Doga}, \binits{H.}},
\bauthor{\bsnm{Bhowmik}, \binits{S.}},
\bauthor{\bsnm{Das}, \binits{S.}},
\bauthor{\bsnm{Mezzacapo}, \binits{A.}},
\bauthor{\bsnm{Motta}, \binits{M.}},
\bauthor{\bsnm{Merz~Jr}, \binits{K.M.}}:
\batitle{Molecular quantum computations on a protein}.
\bjtitle{Journal of Chemical Theory and Computation}
(\byear{2026})
\doiurl{10.1021/acs.jctc.6c00364}
\end{barticle}
\endbibitem

\bibitem[\protect\citeauthoryear{{Anurag K. S. V.} et~al.}{2026}]{anurag2026towards}
\begin{botherref}
\oauthor{\bsnm{{Anurag K. S. V.}}},
\oauthor{\bsnm{Patra}, \binits{A.K.}},
\oauthor{\bsnm{Mukherjee}, \binits{M.}},
\oauthor{\bsnm{Shukla}, \binits{A.}},
\oauthor{\bsnm{{Sai Shankar P.}}},
\oauthor{\bsnm{Bhat}, \binits{R.}},
\oauthor{\bsnm{{Radhika T. S. L.}}},
\oauthor{\bsnm{{Jaiganesh G.}}}:
Towards chemically accurate and scalable quantum simulations on {IQM} quantum hardware: A quantum-{HPC} hybrid approach.
arXiv preprint
(2026)
{\href{https://arxiv.org/abs/2604.01983}{{2604.01983}}}
\end{botherref}
\endbibitem

\bibitem[\protect\citeauthoryear{Hafid et~al.}{2025}]{hafid2025hardness}
\begin{botherref}
\oauthor{\bsnm{Hafid}, \binits{A.}},
\oauthor{\bsnm{Iwakiri}, \binits{H.}},
\oauthor{\bsnm{Tsubouchi}, \binits{K.}},
\oauthor{\bsnm{Yoshioka}, \binits{N.}},
\oauthor{\bsnm{Kohda}, \binits{M.}}:
Hardness of classically sampling quantum chemistry circuits.
arXiv preprint
(2025)
{\href{https://arxiv.org/abs/2504.12893}{{2504.12893}}}
\end{botherref}
\endbibitem

\bibitem[\protect\citeauthoryear{Rudolph and Tindall}{2025}]{rudolph2025simulating}
\begin{botherref}
\oauthor{\bsnm{Rudolph}, \binits{M.S.}},
\oauthor{\bsnm{Tindall}, \binits{J.}}:
Simulating and sampling from quantum circuits with {2D} tensor networks.
arXiv preprint
(2025)
{\href{https://arxiv.org/abs/2507.11424}{{2507.11424}}}
\end{botherref}
\endbibitem

\bibitem[\protect\citeauthoryear{Yoffe et~al.}{2024}]{yoffe2024qubit}
\begin{barticle}
\bauthor{\bsnm{Yoffe}, \binits{D.}},
\bauthor{\bsnm{Entin}, \binits{N.}},
\bauthor{\bsnm{Natan}, \binits{A.}},
\bauthor{\bsnm{Makmal}, \binits{A.}}:
\batitle{A qubit-efficient variational selected configuration-interaction method}.
\bjtitle{Quantum Science and Technology}
\bvolume{10}(\bissue{1}),
\bfpage{015020}
(\byear{2024})
\doiurl{10.1088/2058-9565/ad7d32}
\end{barticle}
\endbibitem

\bibitem[\protect\citeauthoryear{N{\"u}tzel et~al.}{2025}]{nutzel2025solving}
\begin{barticle}
\bauthor{\bsnm{N{\"u}tzel}, \binits{L.}},
\bauthor{\bsnm{Gresch}, \binits{A.}},
\bauthor{\bsnm{Hehn}, \binits{L.}},
\bauthor{\bsnm{Marti}, \binits{L.}},
\bauthor{\bsnm{Freund}, \binits{R.}},
\bauthor{\bsnm{Steiner}, \binits{A.}},
\bauthor{\bsnm{Marciniak}, \binits{C.D.}},
\bauthor{\bsnm{Eckstein}, \binits{T.}},
\bauthor{\bsnm{Stockinger}, \binits{N.}},
\bauthor{\bsnm{Wolf}, \binits{S.}}, \betal:
\batitle{Solving an industrially relevant quantum chemistry problem on quantum hardware}.
\bjtitle{Quantum Science and Technology}
\bvolume{10}(\bissue{1}),
\bfpage{015066}
(\byear{2025})
\doiurl{10.1088/2058-9565/ad9ed3}
\end{barticle}
\endbibitem

\bibitem[\protect\citeauthoryear{Bauer et~al.}{2025}]{bauer2025efficient}
\begin{barticle}
\bauthor{\bsnm{Bauer}, \binits{N.}},
\bauthor{\bsnm{Yeter-Aydeniz}, \binits{K.}},
\bauthor{\bsnm{Siopsis}, \binits{G.}}:
\batitle{Efficient quantum chemistry calculations on noisy quantum hardware}.
\bjtitle{Advanced Quantum Technologies}
\bvolume{8}(\bissue{11}),
\bfpage{00248}
(\byear{2025})
\doiurl{10.1002/qute.202500248}
\end{barticle}
\endbibitem

\bibitem[\protect\citeauthoryear{Ohgoe et~al.}{2025}]{ohgoe2025quantum}
\begin{barticle}
\bauthor{\bsnm{Ohgoe}, \binits{T.}},
\bauthor{\bsnm{Iwakiri}, \binits{H.}},
\bauthor{\bsnm{Ichikawa}, \binits{K.}},
\bauthor{\bsnm{Koh}, \binits{S.}},
\bauthor{\bsnm{Kohda}, \binits{M.}}:
\batitle{Quantum computation of a quasiparticle band structure with the quantum-selected configuration interaction}.
\bjtitle{Journal of the Physical Society of Japan}
\bvolume{94}(\bissue{11}),
\bfpage{114002}
(\byear{2025})
\doiurl{10.7566/JPSJ.94.114002}
\end{barticle}
\endbibitem

\bibitem[\protect\citeauthoryear{Shirai et~al.}{2025}]{shirai2025enhancing}
\begin{barticle}
\bauthor{\bsnm{Shirai}, \binits{S.}},
\bauthor{\bsnm{Tseng}, \binits{S.-Y.}},
\bauthor{\bsnm{Iwakiri}, \binits{H.}},
\bauthor{\bsnm{Horiba}, \binits{T.}},
\bauthor{\bsnm{Hirai}, \binits{H.}},
\bauthor{\bsnm{Koh}, \binits{S.}}:
\batitle{Enhancing accuracy of quantum-selected configuration interaction calculations using multireference perturbation theory: Application to aromatic molecules}.
\bjtitle{ACS omega}
\bvolume{10}(\bissue{35}),
\bfpage{39736}--\blpage{39750}
(\byear{2025})
\doiurl{10.1021/acsomega.5c03371}
\end{barticle}
\endbibitem

\bibitem[\protect\citeauthoryear{Pellow-Jarman et~al.}{2025}]{pellow2025hivqe}
\begin{botherref}
\oauthor{\bsnm{Pellow-Jarman}, \binits{A.}},
\oauthor{\bsnm{McFarthing}, \binits{S.}},
\oauthor{\bsnm{Kang}, \binits{D.H.}},
\oauthor{\bsnm{Yoo}, \binits{P.}},
\oauthor{\bsnm{Elala}, \binits{E.E.}},
\oauthor{\bsnm{Pellow-Jarman}, \binits{R.}},
\oauthor{\bsnm{Nakliang}, \binits{P.M.}},
\oauthor{\bsnm{Kim}, \binits{J.}},
\oauthor{\bsnm{Rhee}, \binits{J.-K.K.}}:
Hivqe: Handover iterative variational quantum eigensolver for efficient quantum chemistry calculations.
arXiv preprint
(2025)
{\href{https://arxiv.org/abs/2503.06292}{{2503.06292}}}
\end{botherref}
\endbibitem

\bibitem[\protect\citeauthoryear{Bickley et~al.}{2025}]{bickley2025}
\begin{barticle}
\bauthor{\bsnm{Bickley}, \binits{T.M.}},
\bauthor{\bsnm{Mingare}, \binits{A.}},
\bauthor{\bsnm{Weaving}, \binits{T.}},
\bauthor{\bsnm{Bastida}, \binits{M.}},
\bauthor{\bsnm{Wan}, \binits{S.}},
\bauthor{\bsnm{Nibbi}, \binits{M.}},
\bauthor{\bsnm{Seitz}, \binits{P.}},
\bauthor{\bsnm{Ralli}, \binits{A.}},
\bauthor{\bsnm{Love}, \binits{P.J.}},
\bauthor{\bsnm{Chung}, \binits{M.}},
\bauthor{\bsnm{Hernández~Vera}, \binits{M.}},
\bauthor{\bsnm{Schulz}, \binits{L.}},
\bauthor{\bsnm{Coveney}, \binits{P.V.}}:
\batitle{Extending quantum computing through subspace{,} embedding and classical molecular dynamics techniques}.
\bjtitle{Digital Discovery}
\bvolume{4},
\bfpage{3427}--\blpage{3444}
(\byear{2025})
\doiurl{10.1039/D5DD00225G}
\end{barticle}
\endbibitem

\bibitem[\protect\citeauthoryear{Kemmoku et~al.}{2026}]{kemmoku2026generative}
\begin{botherref}
\oauthor{\bsnm{Kemmoku}, \binits{R.}},
\oauthor{\bsnm{Gao}, \binits{Q.}},
\oauthor{\bsnm{Kanno}, \binits{S.}},
\oauthor{\bsnm{Keithley}, \binits{K.}},
\oauthor{\bsnm{Hamamura}, \binits{I.}},
\oauthor{\bsnm{Yamamoto}, \binits{N.}},
\oauthor{\bsnm{Nakaji}, \binits{K.}}:
Generative circuit design for quantum-selected configuration interaction.
arXiv preprint
(2026)
{\href{https://arxiv.org/abs/2604.09756}{{2604.09756}}}
\end{botherref}
\endbibitem

\bibitem[\protect\citeauthoryear{Chen et~al.}{2026}]{chen2026neural}
\begin{barticle}
\bauthor{\bsnm{Chen}, \binits{Y.-C.}},
\bauthor{\bsnm{Wu}, \binits{R.}},
\bauthor{\bsnm{Cheng}, \binits{M.H.}},
\bauthor{\bsnm{Hsieh}, \binits{M.-H.}}:
\batitle{Neural network assisted fermionic compression encoding: A lossy quantum-selected configuration interaction framework for resource-efficient ground-state simulations}.
\bjtitle{Physical Review Research}
\bvolume{8}(\bissue{1}),
\bfpage{013128}
(\byear{2026})
\doiurl{10.1103/9x8b-4j41}
\end{barticle}
\endbibitem

\bibitem[\protect\citeauthoryear{Sugisaki et~al.}{2025}]{sugisaki2024hamiltonian}
\begin{barticle}
\bauthor{\bsnm{Sugisaki}, \binits{K.}},
\bauthor{\bsnm{Kanno}, \binits{S.}},
\bauthor{\bsnm{Itoko}, \binits{T.}},
\bauthor{\bsnm{Sakuma}, \binits{R.}},
\bauthor{\bsnm{Yamamoto}, \binits{N.}}:
\batitle{Hamiltonian simulation-based quantum-selected configuration interaction for large-scale electronic structure calculations with a quantum computer}.
\bjtitle{Physical Chemistry Chemical Physics}
\bvolume{27}(\bissue{38}),
\bfpage{20869}--\blpage{20884}
(\byear{2025})
\doiurl{10.1039/D5CP02202A}
\end{barticle}
\endbibitem

\bibitem[\protect\citeauthoryear{Mikkelsen and Nakagawa}{2025}]{mikkelsen2024quantum}
\begin{barticle}
\bauthor{\bsnm{Mikkelsen}, \binits{M.}},
\bauthor{\bsnm{Nakagawa}, \binits{Y.O.}}:
\batitle{Quantum-selected configuration interaction with time-evolved state}.
\bjtitle{Physical Review Research}
\bvolume{7}(\bissue{4}),
\bfpage{043043}
(\byear{2025})
\doiurl{10.1103/75pv-hbrx}
\end{barticle}
\endbibitem

\bibitem[\protect\citeauthoryear{Oumarou et~al.}{2025}]{oumarou2025molecular}
\begin{barticle}
\bauthor{\bsnm{Oumarou}, \binits{O.}},
\bauthor{\bsnm{Ollitrault}, \binits{P.J.}},
\bauthor{\bsnm{Cortes}, \binits{C.L.}},
\bauthor{\bsnm{Scheurer}, \binits{M.}},
\bauthor{\bsnm{Parrish}, \binits{R.M.}},
\bauthor{\bsnm{Gogolin}, \binits{C.}}:
\batitle{Molecular properties from quantum {Krylov} subspace diagonalization}.
\bjtitle{Journal of Chemical Theory and Computation}
\bvolume{21}(\bissue{9}),
\bfpage{4543}--\blpage{4552}
(\byear{2025})
\doiurl{10.1021/acs.jctc.5c00194}
\end{barticle}
\endbibitem

\bibitem[\protect\citeauthoryear{Graves et~al.}{2026}]{graves2026resource}
\begin{botherref}
\oauthor{\bsnm{Graves}, \binits{V.}},
\oauthor{\bsnm{Hlatshwayo}, \binits{M.Q.}},
\oauthor{\bsnm{Kapourniotis}, \binits{T.}},
\oauthor{\bsnm{Georgopoulos}, \binits{K.}}:
Resource-efficient quantum algorithms for selected hamiltonian subspace diagonalization.
arXiv preprint
(2026)
{\href{https://arxiv.org/abs/2603.13160}{{2603.13160}}}
\end{botherref}
\endbibitem

\bibitem[\protect\citeauthoryear{Weaving et~al.}{2026}]{weaving2026selected}
\begin{barticle}
\bauthor{\bsnm{Weaving}, \binits{T.}},
\bauthor{\bsnm{Mingare}, \binits{A.}},
\bauthor{\bsnm{Ralli}, \binits{A.}},
\bauthor{\bsnm{Coveney}, \binits{P.V.}}:
\batitle{Selected configuration interaction using time-evolved population statistics}.
\bjtitle{Journal of Chemical Theory and Computation}
\bvolume{22}(\bissue{9}),
\bfpage{4315}--\blpage{4328}
(\byear{2026})
\doiurl{10.1021/acs.jctc.5c01994}
\end{barticle}
\endbibitem

\bibitem[\protect\citeauthoryear{Campbell}{2019}]{campbell2019}
\begin{barticle}
\bauthor{\bsnm{Campbell}, \binits{E.}}:
\batitle{Random compiler for fast {H}amiltonian simulation}.
\bjtitle{Physical Review Letters}
\bvolume{123},
\bfpage{070503}
(\byear{2019})
\doiurl{10.1103/PhysRevLett.123.070503}
\end{barticle}
\endbibitem

\bibitem[\protect\citeauthoryear{Ouyang et~al.}{2020}]{ouyang2020compilation}
\begin{barticle}
\bauthor{\bsnm{Ouyang}, \binits{Y.}},
\bauthor{\bsnm{White}, \binits{D.R.}},
\bauthor{\bsnm{Campbell}, \binits{E.T.}}:
\batitle{Compilation by stochastic hamiltonian sparsification}.
\bjtitle{Quantum}
\bvolume{4},
\bfpage{235}
(\byear{2020})
\doiurl{10.22331/q-2020-02-27-235}
\end{barticle}
\endbibitem

\bibitem[\protect\citeauthoryear{Alexeev et~al.}{2024}]{alexeev2024quantum}
\begin{barticle}
\bauthor{\bsnm{Alexeev}, \binits{Y.}},
\bauthor{\bsnm{Amsler}, \binits{M.}},
\bauthor{\bsnm{Barroca}, \binits{M.A.}},
\bauthor{\bsnm{Bassini}, \binits{S.}},
\bauthor{\bsnm{Battelle}, \binits{T.}},
\bauthor{\bsnm{Camps}, \binits{D.}},
\bauthor{\bsnm{Casanova}, \binits{D.}},
\bauthor{\bsnm{Choi}, \binits{Y.J.}},
\bauthor{\bsnm{Chong}, \binits{F.T.}},
\bauthor{\bsnm{Chung}, \binits{C.}}, \betal:
\batitle{Quantum-centric supercomputing for materials science: A perspective on challenges and future directions}.
\bjtitle{Future Generation Computer Systems}
\bvolume{160},
\bfpage{666}--\blpage{710}
(\byear{2024})
\doiurl{10.1016/j.future.2024.04.060}
\end{barticle}
\endbibitem

\bibitem[\protect\citeauthoryear{Suzuki}{1976}]{suzuki1976generalized}
\begin{barticle}
\bauthor{\bsnm{Suzuki}, \binits{M.}}:
\batitle{Generalized trotter's formula and systematic approximants of exponential operators and inner derivations with applications to many-body problems}.
\bjtitle{Communications in Mathematical Physics}
\bvolume{51}(\bissue{2}),
\bfpage{183}--\blpage{190}
(\byear{1976})
\doiurl{10.1007/BF01609348}
\end{barticle}
\endbibitem

\bibitem[\protect\citeauthoryear{Holmes et~al.}{2016}]{holmes2016heat}
\begin{barticle}
\bauthor{\bsnm{Holmes}, \binits{A.A.}},
\bauthor{\bsnm{Tubman}, \binits{N.M.}},
\bauthor{\bsnm{Umrigar}, \binits{C.}}:
\batitle{Heat-bath configuration interaction: An efficient selected configuration interaction algorithm inspired by heat-bath sampling}.
\bjtitle{Journal of chemical theory and computation}
\bvolume{12}(\bissue{8}),
\bfpage{3674}--\blpage{3680}
(\byear{2016})
\doiurl{10.1021/acs.jctc.6b00407}
\end{barticle}
\endbibitem

\bibitem[\protect\citeauthoryear{Sharma et~al.}{2017}]{sharma2017semistochastic}
\begin{barticle}
\bauthor{\bsnm{Sharma}, \binits{S.}},
\bauthor{\bsnm{Holmes}, \binits{A.A.}},
\bauthor{\bsnm{Jeanmairet}, \binits{G.}},
\bauthor{\bsnm{Alavi}, \binits{A.}},
\bauthor{\bsnm{Umrigar}, \binits{C.J.}}:
\batitle{Semistochastic heat-bath configuration interaction method: Selected configuration interaction with semistochastic perturbation theory}.
\bjtitle{Journal of chemical theory and computation}
\bvolume{13}(\bissue{4}),
\bfpage{1595}--\blpage{1604}
(\byear{2017})
\doiurl{10.1021/acs.jctc.6b01028}
\end{barticle}
\endbibitem

\bibitem[\protect\citeauthoryear{Song et~al.}{2020}]{song2020multi}
\begin{barticle}
\bauthor{\bsnm{Song}, \binits{Y.}},
\bauthor{\bsnm{Cheng}, \binits{Y.}},
\bauthor{\bsnm{Ma}, \binits{Y.}},
\bauthor{\bsnm{Ma}, \binits{H.}}:
\batitle{Multi-reference {Epstein}--{Nesbet} perturbation theory with density matrix renormalization group reference wavefunction}.
\bjtitle{Electronic Structure}
\bvolume{2}(\bissue{1}),
\bfpage{014002}
(\byear{2020})
\doiurl{10.1088/2516-1075/ab72db}
\end{barticle}
\endbibitem

\bibitem[\protect\citeauthoryear{Burgholzer et~al.}{2026}]{burgholzer2026}
\begin{bchapter}
\bauthor{\bsnm{Burgholzer}, \binits{L.}},
\bauthor{\bsnm{Echavarria}, \binits{J.}},
\bauthor{\bsnm{Hopf}, \binits{P.}},
\bauthor{\bsnm{Stade}, \binits{Y.}},
\bauthor{\bsnm{Rovara}, \binits{D.}},
\bauthor{\bsnm{Schmid}, \binits{L.}},
\bauthor{\bsnm{Kaya}, \binits{E.}},
\bauthor{\bsnm{Mete}, \binits{B.}},
\bauthor{\bsnm{Farooqi}, \binits{M.N.}},
\bauthor{\bsnm{Chung}, \binits{M.}},
\bauthor{\bsnm{De~Pascale}, \binits{M.}},
\bauthor{\bsnm{Schulz}, \binits{L.}},
\bauthor{\bsnm{Schulz}, \binits{M.}},
\bauthor{\bsnm{Wille}, \binits{R.}}:
\bctitle{The munich quantum software stack: Connecting end users, integrating diverse quantum technologies, accelerating hpc}.
In: \bbtitle{Proceedings of the Supercomputing Asia and International Conference on High Performance Computing in Asia Pacific Region}.
\bsertitle{SCA/HPCAsia '26},
pp. \bfpage{55}--\blpage{67}.
\bpublisher{Association for Computing Machinery},
\blocation{New York, NY, USA}
(\byear{2026}).
\doiurl{10.1145/3773656.3773669}
\end{bchapter}
\endbibitem

\bibitem[\protect\citeauthoryear{Yoo et~al.}{2003}]{yoo2003}
\begin{bchapter}
\bauthor{\bsnm{Yoo}, \binits{A.B.}},
\bauthor{\bsnm{Jette}, \binits{M.A.}},
\bauthor{\bsnm{Grondona}, \binits{M.}}:
\bctitle{Slurm: Simple linux utility for resource management}.
In: \beditor{\bsnm{Feitelson}, \binits{D.}},
\beditor{\bsnm{Rudolph}, \binits{L.}},
\beditor{\bsnm{Schwiegelshohn}, \binits{U.}} (eds.)
\bbtitle{Job Scheduling Strategies for Parallel Processing},
vol. \bseriesno{2862},
pp. \bfpage{44}--\blpage{60}.
\bpublisher{Springer},
\blocation{Berlin, Heidelberg}
(\byear{2003}).
\doiurl{10.1007/10968987_3}
\end{bchapter}
\endbibitem

\bibitem[\protect\citeauthoryear{Farooqi et~al.}{2026}]{farooqi2026enahpcqcworkflow}
\begin{botherref}
\oauthor{\bsnm{Farooqi}, \binits{M.N.}},
\oauthor{\bsnm{Chung}, \binits{M.}},
\oauthor{\bsnm{Mete}, \binits{B.}},
\oauthor{\bsnm{Mansfield}, \binits{E.}},
\oauthor{\bsnm{Hoffmann}, \binits{B.}},
\oauthor{\bsnm{Mattsson}, \binits{T.}},
\oauthor{\bsnm{Schulz}, \binits{L.}},
\oauthor{\bsnm{Echavarria}, \binits{J.}}:
Enabling Hybrid HPCQC Workflows with a Heterogeneous Software Stack
(2026).
\url{https://arxiv.org/abs/2608.14827}
\end{botherref}
\endbibitem

\bibitem[\protect\citeauthoryear{MacArthur et~al.}{2017}]{macarthur2017mpiinfiniband}
\begin{barticle}
\bauthor{\bsnm{MacArthur}, \binits{P.}},
\bauthor{\bsnm{Liu}, \binits{Q.}},
\bauthor{\bsnm{Russell}, \binits{R.D.}},
\bauthor{\bsnm{Mizero}, \binits{F.}},
\bauthor{\bsnm{Veeraraghavan}, \binits{M.}},
\bauthor{\bsnm{Dennis}, \binits{J.M.}}:
\batitle{An integrated tutorial on infiniband, verbs, and mpi}.
\bjtitle{IEEE Communications Surveys \& Tutorials}
\bvolume{19}(\bissue{4}),
\bfpage{2894}--\blpage{2926}
(\byear{2017})
\doiurl{10.1109/COMST.2017.2746083}
\end{barticle}
\endbibitem

\bibitem[\protect\citeauthoryear{{BayernKI Consortium}}{2025}]{bayernki2025}
\begin{botherref}
\oauthor{\bsnm{{BayernKI Consortium}}}:
BayernKI: High-Performance Computing Infrastructure for Academic AI Research in Bavaria.
\url{https://www.ki-in-bayern.de/}.
Jointly operated by Leibniz Supercomputing Centre (LRZ) and Erlangen National High Performance Computing Center (NHR@FAU)
(2025)
\end{botherref}
\endbibitem

\bibitem[\protect\citeauthoryear{Boyle}{2024}]{boyle2024nvidia}
\begin{botherref}
\oauthor{\bsnm{Boyle}, \binits{C.}}:
NVIDIA Eos Revealed: Peek Into Operations of a Top 10 Supercomputer.
NVIDIA Blog.
Accessed: May 2026
(2024).
\url{https://blogs.nvidia.com/blog/eos/}
\end{botherref}
\endbibitem

\bibitem[\protect\citeauthoryear{Calzona et~al.}{2026}]{iqm-error-reduction-tools}
\begin{botherref}
\oauthor{\bsnm{Calzona}, \binits{A.}},
\oauthor{\bsnm{Guimarães}, \binits{J.D.}},
\oauthor{\bsnm{Hosseinkhani}, \binits{A.}},
\oauthor{\bsnm{Pina-Canelles}, \binits{V.}},
\oauthor{\bsnm{Rath}, \binits{A.}},
\oauthor{\bsnm{Thapa}, \binits{M.}}:
IQM Error Reduction Tools
(2026).
\url{https://github.com/iqm-finland/sdk/tree/main/src/4.6.3/iqm-error-reduction-tools}
\end{botherref}
\endbibitem

\end{thebibliography}
